\documentclass[fleqn,usenatbib]{mnras}
\usepackage{newtxtext,newtxmath}
\usepackage{graphicx}
\usepackage[T1]{fontenc}
\usepackage{ae,aecompl}
\usepackage[version=3]{mhchem}
\usepackage{amsmath}
\usepackage{hyperref}
\usepackage{booktabs}   
\usepackage{multirow}
\usepackage{comment}    

\usepackage{makecell}

\newcommand{\angstrom}{\mbox{\normalfont\AA}}

\graphicspath{{./images/}}

\title[]{Modelling the cosmological Lyman-Werner background radiation field in the Early Universe}

\author[A. Incatasciato et al.]{
Andrea Incatasciato,$^{1}$\thanks{Contact e-mail: \href{mailto:andrea@roe.ac.uk}{andrea@roe.ac.uk}}
Sadegh Khochfar$^{1}$
and Jose O{\~n}orbe$^{2}$
\\
$^{1}$Institute for Astronomy, University of Edinburgh, Royal Observatory, Blackford Hill, Edinburgh EH9 3HJ, United Kingdom\\
$^{2}$Facultad de Físicas, Universidad de Sevilla, Avda. Reina Mercedes s/n. Campus Reina Mercedes. E-41012, Seville, Spain
}

\date{Accepted XXX. Received YYY; in original form ZZZ}

\pubyear{2023}

\begin{document}
\label{firstpage}
\pagerange{\pageref{firstpage}--\pageref{lastpage}}
\maketitle

\begin{abstract}

The Lyman-Werner (LW) radiation field is a key ingredient in the chemo-thermal evolution of gas in the Early Universe, as it dissociates \ce{H_2} molecules, the primary cooling channel in an environment devoid of metals and dust.
Despite its important role, it is still not implemented in cosmological simulations on a regular basis, in contrast to the ionising UV background. This is in part due to uncertainty in the source modelling, their spectra and abundance, as well as the detailed physics involved in the propagation of the photons and their interactions with the molecules.
The goal of this work is to produce an accurate model of the LW radiation field at $z\geq6$, by post-processing the physics-rich high-resolution FiBY simulation. Our novelties include updated cross sections for \ce{H_2}, \ce{H^-} and \ce{H^+_2} chemical species, IGM absorption by neutral Hydrogen and various spectral models for Population III and Population II stars.
With our fiducial set of parameters, we show that the mean LW intensity steadily increases by three orders of magnitude from $z\sim23$ to $z\sim6$, while spatial inhomogeneities originate from massive star-forming galaxies that dominate the photon budget up to a distance of $\sim 100$ proper kpc. Our model can be easily applied to other simulations or semi-analytical models as an external radiation field that regulates the formation of stars and massive black hole seeds in high-$z$ low-mass halos.

\end{abstract}


\section{Introduction}
\label{sec:introduction}

Molecular Hydrogen (\ce{H_2}) is a key ingredient of the early-universe chemistry, as it represents the main cooling channel of pristine gas at $\mathrm{T}<10^4$ K \citep{Saslaw:1967,Peebles:1968}. Light primordial elements such as Hydrogen and Helium are efficient coolants in their atomic form only above that temperature. On the other hand, heavier elements (collectively referred to as \textit{metals}) do not form during the Big Bang Nucleosynthesis and are a product of the evolution and explosion of stars \citep{Kobayashi:2020}, either in isolation or in binary systems; hence cooling due to metal-line transitions \citep{Smith:2008}, C-, F-, and O-based molecules and dust grains \citep{Hirashita:2002} starts dominating the energy budget of the interstellar medium (ISM) only after the first chemical enrichment episodes.

The abundance of \ce{H_2} (and secondarily of other simple molecules, e.g. \ce{HD} and \ce{HeH^+}) strongly influences the thermo-dynamical evolution of the gas that condenses in the first mini-halos forming at redshift $z \leq 30$ (see e.g. \citealt{Abel:2000}, or \citealt{Galli:2013} for a review).
Molecular cooling allows the gas to reach temperatures as low as $\sim 200$ K, condense to high densities and form the first Population III (PopIII) stars \citep{Haiman:1996,Tegmark:1997}.
Analytical models, 1D and 3D simulations all show that the compressional heating that develops while gas falls into dark matter halos is efficiently dissipated with a central \ce{H_2} fractional abundance of at least $10^{-5}-10^{-4}$ \citep{Abel:2000,Machacek:2001,Yoshida:2006,Latif:2019}. This sets a clear consensus about the initial phase of metal-free PopIII star formation episodes, while models diverge on the final outcome of this process (the multiplicity and the Initial Mass Function - IMF - of PopIII stars), due to differences in the spatial and mass resolution, and in the treatment of accretion, gas chemistry and turbulence. (see \citealt{Bromm:2004} for a review, or e.g. \citealt{Hirano:2015,Chiaki:2022} and \citealt{Latif:2022a} for more recent discussions).

Nevertheless, PopIII stars are generally thought to be massive and hot \citep{Bromm:1999,Abel:2002} and are predicted to emit a copious amount of energetic photons during their very short lifetime \citep{Schaerer:2002}. They explode as violent supernovae, leaving black hole remnants with masses $\sim 10-100 \ \mathrm{M}_\odot$ \citep{Fryer:2001,Madau:2001} and enriching the universe with metals \citep{Heger:2002}, that pave the way for the formation of the first proto-galaxies made of metal-poor Population II (PopII) stars \citep{Bromm:2003}.

Due to their peculiar features, PopIII stars represent also the most important source of Lyman-Werner (LW) photons at the Cosmic Dawn \citep[e.g.][]{Haiman:2000,Agarwal:2012}.
The LW radiation lies within the soft-UV part of the electromagnetic spectrum (its range is commonly indicated as $11-13.6 \ \mathrm{eV}$, or $911-1150 \ \angstrom$) and is able to efficiently dissociate \ce{H_2} through the two-step Solomon process \citep{Solomon:1965,Stecher:1967}. \ce{H_2} formation can be also prevented with the detachment of \ce{H^-} and the dissociation of \ce{H^+_2}, due to NIR-VIS-NUV photons with a few to $\sim 10$ eV \citep{Glover:2015a,Glover:2015b}. \ce{H^-} and \ce{H^+_2} indeed represent the two main \ce{H_2} formation channels in the ISM at moderate densities and devoid of dust grains. Radiation above the Lyman limit, that in principle would be able to directly dissociate \ce{H_2} molecules, is rapidly absorbed by atomic H in the diffuse intergalactic medium (IGM), that is still completely neutral at this stage. LW photons, on the contrary, have a very long mean free path ($\sim 100 \ \mathrm{cMpc}$, \citealt{Haiman:2000,Ahn:2009}), as they can only be absorbed when redshifted to the exact frequencies of the atomic Lyman transitions. \ce{H_2} molecules, instead, are not dense enough in the IGM to play any role in this context. This leads to the definition of a spatially nearly-homogeneous background, whose intensity at the Lyman limit at $z\sim25-10$ is often bracketed by $J_{21}\sim10^{-3}$ and $J_{21}\sim 10^2$, where $J_{21}$ is the Lyman-Werner background (LWB) intensity normalised to $10^{-21} \ \mathrm{erg} \ \mathrm{s}^{-1} \ \mathrm{sr}^{-1} \ \mathrm{Hz}^{-1} \ \mathrm{cm}^{-2}$ \citep[e.g.][]{Haiman:1997,Machacek:2001,Ahn:2009,Trenti:2009,Johnson:2013}.

The build-up of a homogeneous LWB during the formation of the first cosmological structures has important implications on the PopIII star formation \citep{Haiman:2000}, as it makes molecular cooling inefficient in low-mass halos. Without \ce{H_2} molecules, star formation is delayed until dark matter halos reach virial temperatures of $T_\mathrm{vir} \sim 10^4$ K, when atomic H cooling becomes efficient and the collapse can start \citep{Haiman:1997}.
Recently, many theoretical efforts have been focused on trying to quantify this effect. The interplay between a time-varying LWB and PopIII star formation has been explored with cosmological hydro-dynamical simulations \citep{Wise:2012a,Johnson:2013}, that are designed to accurately capture the highly non linear evolution of cosmic structures, and with semi-analytical/semi-numerical models \citep{Haiman:2000,Ahn:2009,Trenti:2009,Agarwal:2012,Visbal:2020,Qin:2020}, that on the other hand require a certain number of approximations and a priori assumptions, but allow a fast parameter exploration. In addition, \cite{Latif:2019,Schauer:2021,Kulkarni:2021} employed high-resolution small-scale cosmological simulations to explore the minimum halo mass required for PopIII star formation in molecular cooling halos under a range of constant LWB intensities.

The modelling of the LW radiation is usually approximated, due to the technical complexity of the calculation from first principles \citep{Abel:1997,Wolcott-Green:2017} and the lack of constraints on the spectra of the stellar populations responsible for the LW emission \citep{Bromm:2004}. Often, only very young PopIII and PopII stars are considered in the radiative budget, stellar evolution is neglected and the emissivity is kept constant \citep{Greif:2006}. A fully self-consistent treatment of the closed loop between star formation and the growth of a LWB, that exerts a negative feedback on the subsequent star formation episodes, is also made difficult by the computational cost of radiative transfer methods over large cosmological volumes \citep{Johnson:2013}.

Another matter of debate is the importance of the LW radiation in the context of the Direct Collapse Black Hole (DCBH) scenario \citep{Begelman:2006,Lodato:2006,Dijkstra:2008,Agarwal:2012}, that represents one of the most promising formation channels of the initial seeds of the supermassive black holes observed at $z>6$ \citep{Fan:2006,Mortlock:2011,Banados:2018}. Halos illuminated by high LW intensity, such as small star-less satellites of massive high-redshift galaxies, where the radiation from the neighbouring galaxies prevails by orders of magnitude over the large-scale background, have been proposed as birthplaces of black holes with initial masses of $10^{4-6} \ \mathrm{M_\odot}$ (\citealt{Agarwal:2014,Wise:2019,Agarwal:2019,Lupi:2021}, see also \citealt{Fernandez:2014,Bonoli:2014}).

A critical value $J_{21,\mathrm{crit}}$ of LW intensity is usually assumed to express the minimum level of radiation needed to efficiently prevent \ce{H_2} molecular cooling, the first-order requirement of the DCBH scenario together with a pristine chemical composition. In the last few years many studies have explored its feasibility with 1D and 3D hydro-dynamical simulations that employ non-equilibrium chemistry, high spatial and temporal resolution and in some cases a self-consistent treatment of the radiative feedback from the central object \citep{Omukai:2008,Shang:2010,Regan:2014a,Luo:2018}. However, a consensus on the value of $J_{21,\mathrm{crit}}$ is still lacking. Recent studies \cite[e.g.][]{Sugimura:2014,Latif:2015,Wolcott-Green:2017} have shown that, if the interstellar radiation field is modelled as a black-body, $J_{21,\mathrm{crit}}$ can vary by many orders of magnitude (from 10 to $10^5$), depending on the assumed black-body temperature, usually $10^5$ K ($10^4$ K) if PopIII (PopII) stars dominate the radiation field. \cite{Agarwal:2015} have highlighted that considering the evolution of the spectral shape across the lifetime of a stellar population has an important impact on $J_{21,\mathrm{crit}}$, especially when also long-lived low-mass stars are included. Furthermore \cite{Glover:2015a,Glover:2015b,Agarwal:2016,Sugimura:2016,Luo:2020} all proposed that the chemical network employed in the simulations should also include \ce{H^-} detachment and \ce{H^+_2} dissociation, to provide a more accurate estimate of the \ce{H_2} formation rate. Further degrees of freedom include the \ce{H_2} self-shielding treatment, that in the optically-thick regime can reduce the effect of the LW photons by up to three orders of magnitude and strongly depends on the accuracy of the calculation \citep{Draine:1996,Wolcott-Green:2011,Hartwig:2015a,Wolcott-Green:2019}, and the impact of additional fields, such as X-rays or cosmic rays, that can increase the fraction of free electrons, thus facilitating the formation of \ce{H_2} \citep{Inayoshi:2011,Inayoshi:2015,Regan:2016,Glover:2016,Park:2021}. These uncertainties lie on top of other aspects, such as unresolved fragmentation and long-term sustainability of the mass accretion flow, whose role the scientific community still has to have a final say on \citep[e.g.][]{Ge:2017,Bhowmick:2022}.

With this work we tackle some of the current limitations of the studies on the effect of the LW radiation on the formation of stars and black hole seeds in the Early Universe. In particular, we aim at showing how the LWB and the other associated photochemical rates can be accurately modeled given a star formation history, that can be either derived from a simulation or a semi-analytical model. We also study the spatial inhomogeneities of the radiation field \citep{Haiman:2000,Ahn:2009,Dijkstra:2014}. To do so we post-process the First Billion Year (FiBY) suite of cosmological simulations. We describe the FiBY project in Sec.~\ref{sec:fiby}, together with the methods employed in the post-processing algorithm. We keep an agnostic approach  with regards to the IMF of  PopIII and PopII stars (Sec.~\ref{sec:SEDs}), in order to show the intrinsic uncertainty due to the current lack of constraints on the stellar models. We outline our code that accurately calculates the photochemical rates in Sec.~\ref{sec:rates}, while deferring to a companion paper (\textcolor{blue}{Incatasciato et al., in prep}) for a in depth comparison of the specific methods to calculate the \ce{H_2} dissociation rate given a stellar spectrum. IGM absorption is described in Sec.~\ref{sec:fmod}. Our homogeneous LWB model is outlined in Sec.~\ref{sec:LWB}, while its spatial inhomogeneities are investigated in Sec.~\ref{sec:histogs}. Finally, our considerations on the impact of the LWB on PopIII star formation are reported in Sec.~\ref{sec:minhmass}. We then complete the paper with further discussions and our conclusions in Sec.~\ref{sec:conclusion}.

\section{Methods}
\label{sec:methods}

In this work we use the simulations of the First Billion Year (FiBY) project to evaluate the evolution of the LW background\footnote{Here and in the following, when we refer to the LW background, we implicitly consider not only the photons responsible for the \ce{H_2} dissociation, but also the ones relevant for the \ce{H^-} detachment and the \ce{H^+_2} dissociation.} at $z\geq6$. The FiBY suite is described in Section~\ref{sec:fiby}. To obtain an estimate of the LW background we sample the radiation field by choosing random points (\textit{observers}) within the simulation box. For each observer we sum the radiation emitted by all the sources, taking into account various stellar models (described in Section~\ref{sec:SEDs}), a detailed calculation of the photo-chemical rates, including recent updates to take into account molecular level populations (Section~\ref{sec:rates}) and the absorption by the IGM (Section~\ref{sec:fmod}). The number of points used in each snapshot is selected such as the mean and the standard deviation of $J_{21}$ converge to percent level and corresponds to $\mathcal{O}(10^4)$. We repeat this exercise for each snapshot available at $z\geq6$.

\subsection{FiBY}
\label{sec:fiby}

The FiBY project \citep{Johnson:2013,Paardekooper:2013,Agarwal:2014,Paardekooper:2015,Cullen:2017,Phipps:2020} includes a set of high-resolution and physics-rich cosmological simulations of the early universe. The simulations were run with a modified version of the \textsc{gadget-3} code \citep{Springel:2001,Springel:2005}, already employed for the OWLS project \citep{Schaye:2010}. The code was updated further to include the relevant physical processes and stellar models required for a better modelling of the formation of the first stars and proto-galaxies at $z \sim 30-6$. Substructures  within  the  simulations are identified  with  the \textsc{subfind} algorithm \citep{Springel:2001} and merger trees are generated with the method described in \cite{Neistein:2012}.

We refer the reader to the original FiBY papers and the references therein for a detailed description of the sub-grid models and provide here only a brief summary. For completeness and convenience of the reader, we also summarise the key parameters of all the runs (e.g. box size and mass resolution) used in this work in Table~\ref{Table:FiBYruns}. All the simulations were run using the following cosmological parameters, consistent with those reported by the Wilkinson Microwave Anisotropy Probe (WMAP) team in \citet{Komatsu:2009}: $\Omega_m=0.265$, $\Omega_b=0.0448$, $\Omega_{\Lambda}=0.735$, $H_0=71$ km s$^{-1}$ Mpc$^{-1}$ and $\sigma_8=0.81$. The same cosmological parameters are assumed throughout this work, unless otherwise stated.

Collisionless dark matter particles and SPH gas particles are the two main constituents of the simulated volumes. The thermodynamical evolution of the gas particles is described with the usual atomic cooling due to H and He, but also with metal line cooling (C, N, O, Ne, Mg, Si, S, Ca and Fe, \citealt{Wiersma:2009}) and \ce{H_2} and \ce{HD} non-equilibrium chemistry \citep{Abel:1997,Yoshida:2006}. The multi-phase interstellar medium (ISM) is modelled with an effective equation of state (EOS), following \cite{Schaye:2008}, explicitly designed to yield star formation rates consistent with the observed Schmidt–Kennicutt law \citep{Schmidt:1959,Kennicutt:1998}. The density threshold for the effective EOS is $10 \ \mathrm{cm}^3$, that represents also the threshold for the star formation. Depending on the metallicity of the star forming gas, collisionless particles representing  metal-free Population III or metal-poor Population II stars are spawned. Each stellar particle is treated as a single stellar population with a perfectly-sampled IMF. PopIII stars are assigned a \cite{Salpeter:1955} IMF with stellar masses in the range 21-500 $\mathrm{M}_\odot$, consistent with the top-heavy IMF predicted e.g. by \cite{Bromm:2004}, while PopII stars have a \cite{Chabrier:2003} IMF that extends down to sub-solar masses. The critical metallicity to distinguish PopIII and PopII stellar particles is  $Z_\mathrm{crit}=10^{-4} \ \mathrm{Z}_\odot$ \citep{Maio:2011}\footnote{The specific choice of the critical metallicity does not strongly impact our results, as metal pollution proceeds quickly and increases the metallicity of the interstellar medium (ISM) to large values in the hosting and neighbouring halos \citep{Maio:2010,Maio:2011,Smith:2015}.}, where $\mathrm{Z}_\odot=0.02$. Massive stars explode as supernovae at the end of their lives. Metal enrichment due to core-collapse (CCSNe, $8 \ \mathrm{M}_\odot < M_\star < 100 \ \mathrm{M}_\odot$) and pair-instability (PISNe, $140 \ \mathrm{M}_\odot < M_\star < 260 \ \mathrm{M}_\odot$) supernovae follows \cite{Heger:2002,Heger:2010}. The thermal energy due to the explosions is stochastically injected to the neighbouring particles following the scheme described by \cite{DallaVecchia:2012}. The cosmic reionisation is modelled with a time-dependent and spatially-uniform UV radiation background \citep{Haardt:2001}, while high-density gas is shielded against the UVB as proposed by \cite{Nagamine:2010}. One specific run (FiBY\_LW, see Table~\ref{Table:FiBYruns}) includes also an on-the-fly LW background, that comprises both a homogeneous component dependent on the cosmic star formation rate \citep{Greif:2006} and the contribution from the local sources \citep{Johnson:2013}. In this run, the \ce{H_2} self-shielding follows the prescriptions by \cite{Wolcott-Green:2011}.

\begin{table*}
\centering
\begin{tabular}{ |c|c|c|c|c|c|c|c| } 
 \hline
 Name & Boxlength [cMpc/h] (cMpc) & N$^{1/3}$ & M$_\mathrm{dm}$ [M$_\odot$/h] & M$_\mathrm{g}$ [M$_\odot$/h] & z$_\mathrm{end}$ & LW background & M$_\mathrm{halo,min}$ [M$_\odot$] \\
 \hline\hline
 \textbf{FiBY\_XL} & 22.72 (32) & 684 & $2.24\times10^6$ & $4.56\times10^5$ & 4 & N & $1.6\times10^8$ \\
 \hline
 \textbf{FiBY\_L} & 11.36 (16) & 684 & $2.80\times10^5$ & $5.70\times10^4$ & 4 & N & $2\times10^7$ \\
 \hline
 \textbf{FiBY\_M} & 5.68 (8) & 684 & $3.50\times10^4$ & $7.12\times10^3$ & 6 & N & $2.5\times10^6$ \\
 \hline
 \textbf{FiBY\_S} & 2.84 (4) & 684 & $4.37\times10^3$ & $8.90\times10^2$ & 6 & N & $3\times10^5$ \\
 \hline
 \textbf{FiBY\_LW} & 2.84 (4) & 684 & $4.37\times10^3$ & $8.90\times10^2$ & 6 & Y & $3\times10^5$ \\
 \hline
\end{tabular}
\caption{Compilation of the different FiBY simulations used in this work. We report the box size in the second column, where the corresponding value in brackets is the size when assuming $\mathrm{h}=0.71$. Columns 3, 4, and 5 show the number of particles of each component (dark matter and gas) and their (initial) mass. The final redshift reached is listed in Column 6, while the next one indicates whether the LW background is self-consistently calculated starting from the star formation rate and the local contribution of young stars (see Section 2.2 of \citealt{Johnson:2013}). The last Column reports a rough estimate of the mass of the smallest halos resolved (with at least 50 dark matter particles).}
\label{Table:FiBYruns}
\end{table*}

The main sources of LW photons in the early universe are PopIII and PopII stars. The simulations predict high-redshift UV-luminosity function and a star formation main sequence that are in good agreement with observational constraints \citep{Cullen:2017}, as well as an overall star formation rate density ($\rho_\mathrm{SFR}$) that is in fair agreement with observational bounds \citep{Johnson:2013}. This gives us confidence that to first order stars form within the simulations at the right rate and in the right objects.

\begin{figure}
    \centering
    \includegraphics[width=\linewidth]{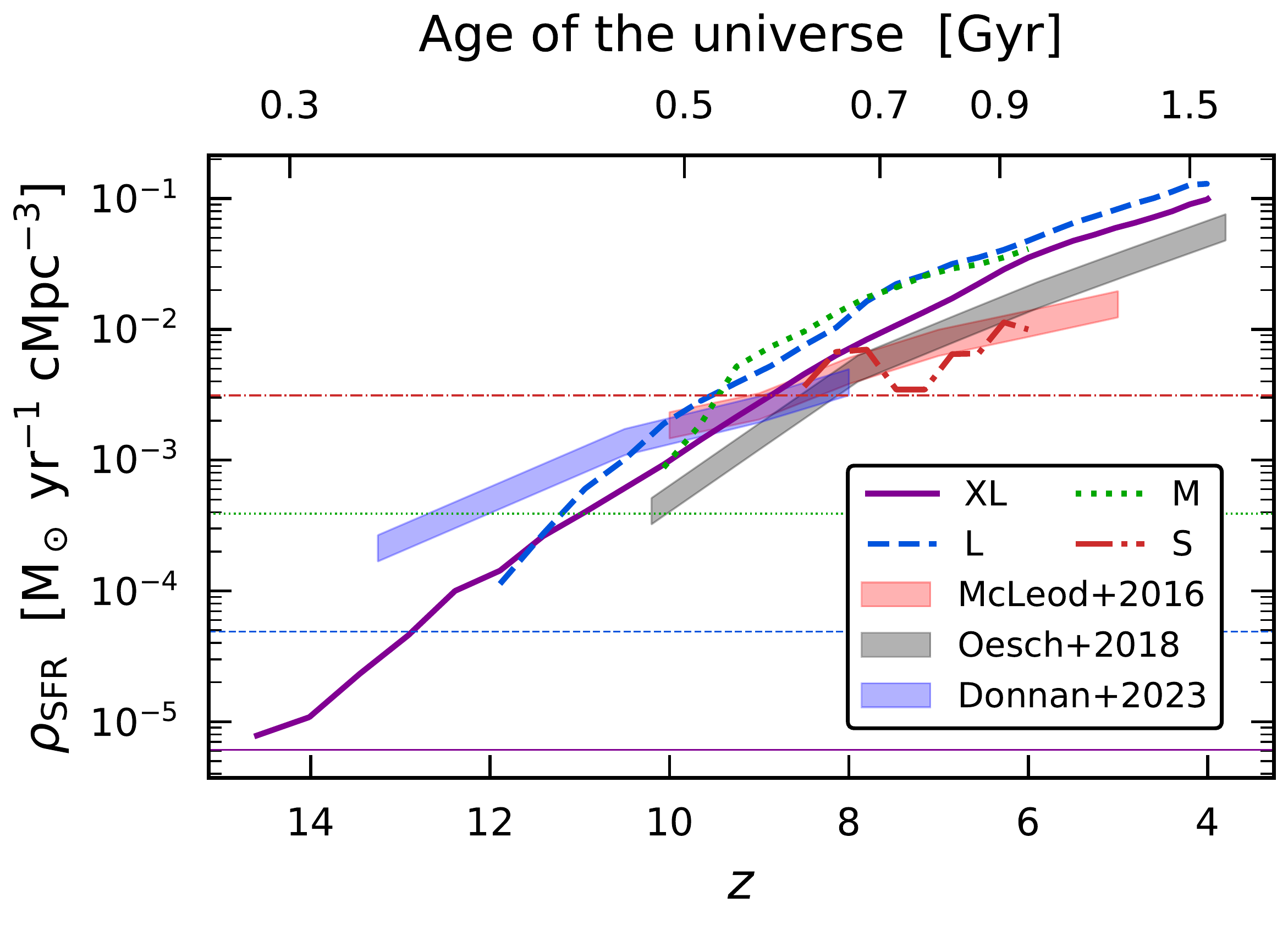}
    \vspace{-0.6cm}
    \caption{The star formation rate density ($\rho_\mathrm{SFR}$) in the following FiBY simulations: XL (solid purple), L (dashed blue), M (dotted green) and S (dash-dotted red line). Coloured shaded areas indicate the $\rho_\mathrm{SFR}$ derived from deep HST and JWST+COSMOS observations: \protect\citet[$z=5-7$]{Bowler:2015}, \protect\citet[$z=8$]{McLure:2013}, \protect\citet[$z=9$]{McLeod:2015} and \protect\citet[$z=10$]{McLeod:2016} are in red, \protect\citet[$z=4-8$]{Bouwens:2016}, \protect\citet[$z=9$]{Oesch:2014} and \protect\citet[$z=10$]{Oesch:2018} in grey and \protect\cite{Donnan:2023} in blue.
    To ensure a consistency between observations and simulations, we estimate the FiBY $\rho_\mathrm{SFR}$ only from galaxies with $\mathrm{SFR}\gtrsim0.3 \ \mathrm{M_\odot} \ \mathrm{yr}^{-1}$, corresponding to the usual integration limit of $M_\mathrm{{UV}}=-17$ in the observed UV luminosity function. The horizontal lines indicate the minimum value that can be predicted by FiBY due to the limited volume of each simulation.
    The shaded regions quantify the uncertainties due to the assumptions on the stellar IMF: the higher (lower) values are for a \protect\citealt{Salpeter:1955} (\protect\citealt{Chabrier:2003}) IMF, that give a slightly different UV luminosity per stellar mass, with a correction factor of 0.63 as suggested by \protect\cite{Madau:2014}.}
    \vspace{-0.2cm}
    \label{fig:SFRD_SFRthresh}
\end{figure}

For the sake of completeness we show in Fig.~\ref{fig:SFRD_SFRthresh} $\rho_\mathrm{SFR}$ of the XL (solid purple), L (dashed blue), M (dotted green) and S (dash-dotted red) FiBY simulations, superimposed over the one obtained with deep HST observations at $z\sim4-10$ from two collaborations, in grey \citep{Oesch:2014,Bouwens:2016,Oesch:2018} and in red \citep{McLure:2013,Bowler:2015,McLeod:2015,McLeod:2016}, and recent constraints from ground (COSMOS/UltraVISTA) and JWST NIRCam photometry \citep{Donnan:2023} at $z\sim8-15$ in blue. We make use of the shaded regions to highlight the uncertainties due to the underlying assumption on the stellar IMF, where the higher (lower) values are for a \citealt{Salpeter:1955} \citep{Chabrier:2003} IMF, that have different UV luminosity per stellar mass conversion factors \citep{Madau:2014}. To ensure a consistency between observations and simulations, we include only galaxies with $\mathrm{SFR} \gtrsim \mathrm{SFR_{min}}=0.3 \ \mathrm{M_\odot} \ \mathrm{yr}^{-1}$, that corresponds to the integration limit of the UV luminosity function down to $M_\mathrm{UV}=-17$ as in \cite{Oesch:2018} and \cite{Donnan:2023}.

The simulations employ the same number of particles to evolve the dark matter and baryonic density fields ($684^3$ each) within cubic volumes of different sizes (from 32 cMpc to 4 cMpc). They therefore investigate different sections of the halo mass function. In particular, only the M and S boxes properly resolve \ce{H_2}-cooling halos with $M_\mathrm{h}<10^{7-8} \ \mathrm{M}_\odot$ (last column in Table~\ref{Table:FiBYruns}), but lack the rarer massive galaxies due to the limited volume. On the other hand, the larger L and XL boxes focus on atomic-cooling halos and include a wide range of cosmic environments, hence they contain massive galaxies above the observational limit of $\mathrm{SFR_{min}}$ from earlier times ($z\sim11-14$, exactly the redshift range that is currently being studied for the first time with JWST, \citealt{Donnan:2023,Harikane:2023}). The minimum $\rho_\mathrm{SFR}$ that can be estimated from each simulation (corresponding to only one galaxy with $\mathrm{SFR}=\mathrm{SFR_{min}}$ in the entire volume) is shown with the horizontal thin lines.

The $\rho_\mathrm{SFR}$ in FiBY shows a reasonable convergence between the XL, L and M volumes, and an evolution with redshift that is consistent with observations by \cite{Oesch:2018} and collaborators. However, the slightly different absolute values suggest that FiBY might overproduce stars in massive galaxies at $z\lesssim10$. Nevertheless, deviations on this level cannot be too surprising, given the uncertainties both on the observational side (dust correction, incompleteness) and in simulations (resolution, LW radiation, stellar feedback, only to name a few, see e.g. \citealt{Vogelsberger:2020}). The shallower evolution found by \cite{McLeod:2016} and \cite{Donnan:2023} indeed demonstrates that observations still do not provide a unique solution to this problem. Finally, the S box shows a peculiar evolution at $z=6-9$ that matches \cite{McLeod:2016}, despite being very close to the limits set by the small simulated volume. In conclusion, the FiBY simulations produce a fairly realistic high-$z$ universe and we consider it a useful tool to model the evolution of the LWB in the pre-Reionisation Era.

\subsection{Stellar emission}
\label{sec:SEDs}

We use 9 different models for the spectral energy distribution (SED) of the stars.
For PopIII and PopII stars we employ the models described in Table~\ref{Table:SEDpop3} and Table~\ref{Table:SEDpop2} respectively, calculated with the publicly available stellar population synthesis (SPS) codes  Yggdrasil \citep{Zackrisson:2011}, Slug2 \citep{daSilva:2012,daSilva:2014} and BPASS \citep{Stanway:2018}. Yggdrasil uses models for PopIII stars from \cite{Schaerer:2002} and \cite{Raiter:2010} and provides pre-computed SEDs for a very top-heavy IMF (\citealt{Salpeter:1955} between 50 and 500 M$_\odot$) and a more moderate one (lognormal with characteristic mass equal to 10 M$_\odot$). Slug2, instead, allows the user to calculate stellar SEDs with a wide variety of IMFs and evolutionary tracks (e.g. Geneva \citealt{Eggenberger:2008}, Padova \citealt{Bressan:1993} and MIST \citealt{Dotter:2016}), and atmosphere models resembling the Starburst99 SPS code of \cite{Leitherer:1999}.\footnote{Another important feature of Slug2 is to allow a stochastic sampling of the IMF. We do not make use of it in this work.} BPASS provides a large set of pre-computed SEDs with an in-depth treatment of stellar binary systems. The minimum available metallicity is $5\times10^{-4}$ Z$_\odot$, hence we use BPASS models only for metal-poor PopII stars.

To include the approximation of the stellar spectra commonly assumed in the literature, we additionally consider two black-body (BB) spectra with $T_\mathrm{rad}=10^5$ K and $10^4$ K for PopIII and PopII stars respectively.
The normalisation of these spectra is chosen such that the number of emitted photons in the LW range per stellar baryon $\eta_\mathrm{LW}$ is $2\times 10^4$ and $4\times 10^3$ respectively, as adopted in \cite{Greif:2006} and \cite{Johnson:2013}. When using the BB spectra we do not consider stars older than 5 Myr, in order to match the model used in the FiBY \citep{Johnson:2013}.

PopIII and PopII stars often coexist in simulated high-$z$ galaxies. The total LWB is therefore calculated as the sum of the contributions from these two distinct stellar populations, where we consider 5 out of the 20 possible combinations:
\begin{itemize}
    \item \textbf{FID}: \textbf{PopIII\_Ygg2} + \textbf{PopII\_BPASS\_Chab}, this is our \textquotesingle fiducial\textquotesingle \ choice; see bottom panel of Fig.~\ref{fig:example_spectra} for an example of the SEDs 1 Myr after the star formation episode;
    \item \textbf{TH}: \textbf{PopIII\_Ygg1} + \textbf{PopII\_BPASS\_TH}, with top-heavy IMFs;
    \item \textbf{BH}: \textbf{PopIII\_Ygg2} + \textbf{PopII\_BPASS\_BH}, with bottom-heavy IMFs;
    \item \textbf{SLUG}: \textbf{PopIII\_Slug} + \textbf{PopII\_Slug}, where both SEDs are calculated with the Slug2 SPS code;
    \item \textbf{BB}: \textbf{PopIII\_BB5} + \textbf{PopII\_BB4}, with single-temperature black-body spectra.
\end{itemize}
In particular \textquotesingle TH\textquotesingle \ and \textquotesingle BH\textquotesingle \ should bracket the level of uncertainty introduced by the choice of IMF, where the contribution from high-mass and low-mass stars respectively is enhanced with respect to our fiducial model and to all the other combinations neglected in this work.

\begin{figure}
    \centering
    \includegraphics[width=\linewidth]{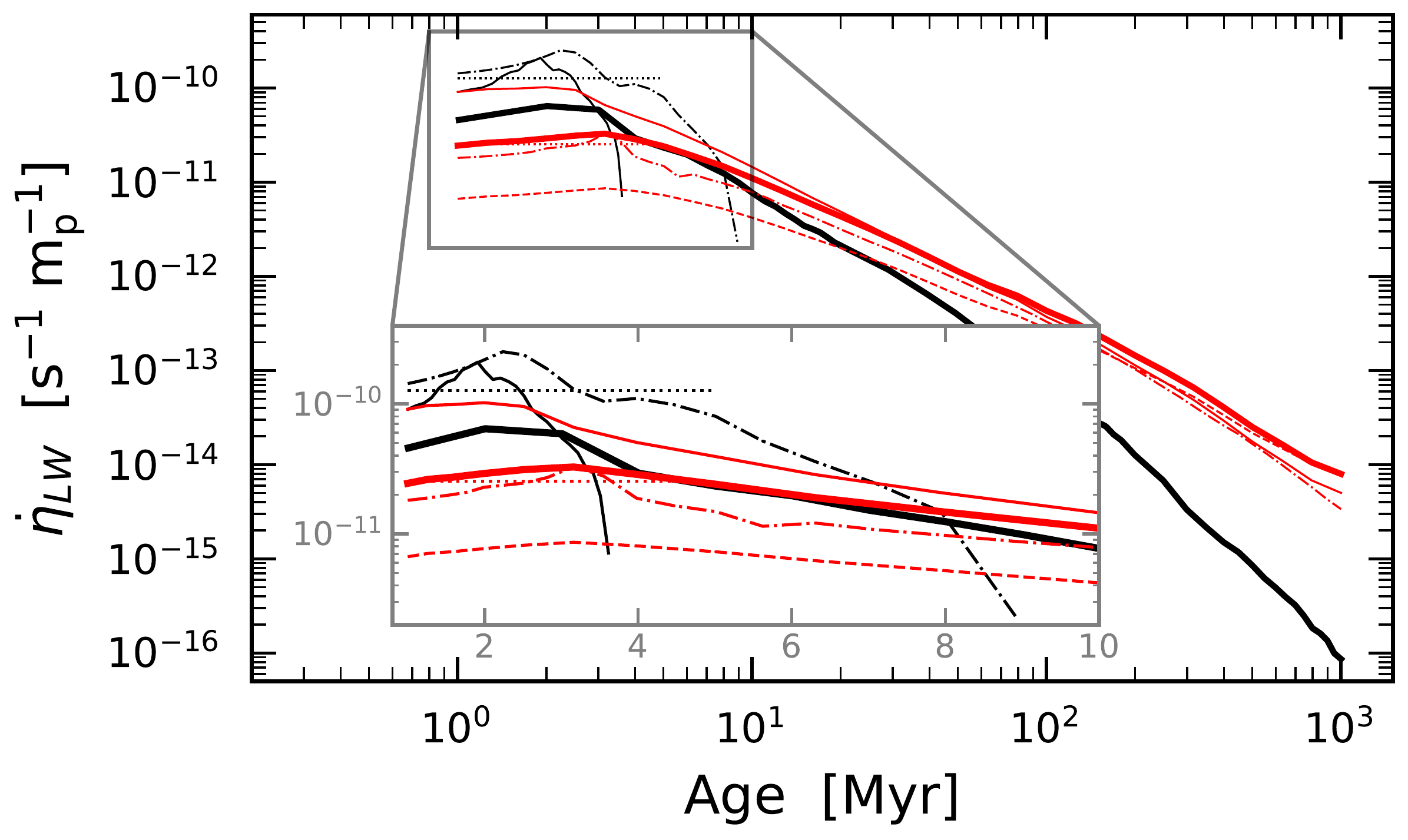}
    \vspace{-0.6cm}
    \caption{LW photon emission rate per stellar baryon for each SED used in this work. Models for PopIII and PopII stars are shown in black and red respectively: with reference to the definitions in the main text and Tables~\ref{Table:SEDpop3}-\ref{Table:SEDpop2}, the \textquotesingle FID\textquotesingle \ (thick solid), the \textquotesingle TH\textquotesingle \ (thin solid), the \textquotesingle SLUG\textquotesingle \ (dot-dashed), the \textquotesingle BB\textquotesingle \ (dashed) and finally the BPASS bottom-heavy IMF (red dashed line).}
    \vspace{-0.2cm}
    \label{fig:SEDs_LW_emission}
\end{figure}

We show in Fig.~\ref{fig:SEDs_LW_emission} the emission rate per stellar baryon of LW photons in the range 11-13.6 eV for each SED used in this work. Black and red lines indicate PopIII and PopII models respectively. Our fiducial choice for PopIII stars (black thick solid line) is conservative, as it could have been predicted from the IMF, since the other PopIII models emit $\sim5$ times more LW photons in the early stages, but die off very quickly after 5-10 Myr. The differences in the IMFs for PopII stars can be noticed in the first 10 Myr, where the number of high-mass stars determines a factor of 4-5 higher (lower) LW emission of BPASS\_TH (BPASS\_BH) with respect to the fiducial case (red thin solid, dashed and thick solid lines respectively), while they all show pretty similar evolution at later times.
PopII\_BB is hardly distinguishable from the fiducial SED, but it shows a totally different shape when it comes to low energy photons that determine the \ce{H^+_2} dissociation and \ce{H^-} detachment rates (see bottom panel in Fig.~\ref{fig:example_spectra}).

Once the radiation is emitted by a star, it takes a not-negligible time to reach an observer at a given comoving distance $r_\mathrm{com}$. By reverting Eq.8 of \cite{Ahn:2009}, in the high-redshift limit
\begin{equation}
    z_\mathrm{em} = \left((1+z_\mathrm{obs})^{-1/2}-\frac{H_0\Omega_\mathrm{m}^{1/2}r_\mathrm{com}}{2c}\right)^{-2}-1
\end{equation}
expresses the redshift of emission of a photon observed at $z_\mathrm{obs}$ (here the redshift of a given simulation snapshot).
We hence account for the light time travel $\Delta t = t(z_\mathrm{obs}) - t(z_\mathrm{em})$ when choosing the age of the emitted spectrum of each stellar particle, where $t(z)$ is the age of the universe at a given redshift. The time resolution of the spectra generated with the SPS codes listed in Tables~\ref{Table:SEDpop3}-\ref{Table:SEDpop2} allows us to properly follow the spectral evolution of a stellar population.
The physical motion of the stars is instead negligible in this context.

\begin{table*}
\centering
\begin{tabular}{ |c|c|c|c|c|c| } 
 \hline
 Name & IMF & Parameters & Metallicity [Z$_\odot$] & SPS code & References \\ 
 \hline\hline
 \textbf{PopIII\_Ygg1} & \cite{Salpeter:1955} & $M_\mathrm{min} = 50$, $M_\mathrm{max} = 500$ & 0 & Yggdrasil & \cite{Schaerer:2002,Zackrisson:2011} \\
 \hline
 \textbf{PopIII\_Ygg2} & Lognormal & \makecell{$M_\mathrm{min} = 1$, $M_\mathrm{max} = 500$ \\ $M_\mathrm{c} = 10$, $\sigma=1$} & 0 & Yggdrasil & \cite{Raiter:2010,Zackrisson:2011} \\
 \hline
 \textbf{PopIII\_Slug} & \cite{Salpeter:1955} & $M_\mathrm{min} = 21$, $M_\mathrm{max} = 300$ & $10^{-4}$ & Slug2 & \cite{daSilva:2012,daSilva:2014} \\
 \hline
\end{tabular}
\caption{SEDs for PopIII stars, where the masses are in M$_\odot$. Nebular emission and extinction are neglected. The Slug2 spectra are calculated with the MISTv1.0 \citep{Dotter:2016,Choi:2016} non-rotating stellar tracks and are generated with logarithmic timesteps of 0.05 dex from 1 Myr to 1 Gyr, in order to accurately resolve the rapid evolution of young stellar populations. The presented  metallicity is the minimum available for these stellar tracks.}
\label{Table:SEDpop3}
\end{table*}

\begin{table*}
\centering
\begin{tabular}{ |c|c|c|c|c|c|c| } 
 \hline
 Name & IMF & Parameters & Metallicity [Z$_\odot$] & SPS code & References \\ 
 \hline\hline
 \textbf{PopII\_BPASS\_TH} & Double power-law & \makecell{$\alpha_1=-1.3$, $\alpha_2=-2$ \\ $M_\mathrm{t} = 0.5$, $M_\mathrm{max} = 300$} & $5\times10^{-4}$ & BPASS & \cite{Stanway:2018} \\
 \hline
 \textbf{PopII\_BPASS\_Chab} & \cite{Chabrier:2003} & $M_\mathrm{t} = 1$, $M_\mathrm{max} = 100$ & $5\times10^{-4}$ & BPASS & \cite{Stanway:2018} \\
 \hline
 \textbf{PopII\_BPASS\_BH} & Double power-law & \makecell{$\alpha_1=-1.3$, $\alpha_2=-2.7$ \\ $M_\mathrm{t} = 0.5$, $M_\mathrm{max} = 100$} & $5\times10^{-4}$ & BPASS & \cite{Stanway:2018} \\
 \hline
 \textbf{PopII\_Slug} & \cite{Chabrier:2003} & $M_\mathrm{t} = 1$, $M_\mathrm{max} = 120$ & $10^{-3}$ & Slug2 & \cite{daSilva:2012,daSilva:2014} \\
 \hline
\end{tabular}
\caption{Same as Table~\ref{Table:SEDpop3}, but for for PopII stars. $\alpha_1$ and $\alpha_2$ are the low-mass and the high-mass slopes respectively and the masses are in M$_\odot$. The BPASS SEDs include binaries as according to \protect\cite{Stanway:2018} and have $M_\mathrm{min} = 0.1$ M$_\odot$. Nebular emission and extinction are neglected.}
\label{Table:SEDpop2}
\end{table*}

\subsection{Optically-thin photochemical rates}
\label{sec:rates}

Three photo-reactions need to be taken into account to accurately evaluate the formation and destruction of \ce{H_2} in pristine gas \citep[e.g.][]{Glover:2015a}:
\begin{subequations}
\label{eq:photoreacts}
\begin{align}
\label{eq:photoH2}
    \ce{H_2} + \gamma &\rightarrow \ce{H^\textasteriskcentered_2} \rightarrow \ce{H} + \ce{H}   &h\nu \gtrsim  6.7 \ \mathrm{eV}\\
\label{eq:photoHM}
	\ce{H^-} + \gamma &\rightarrow \ce{H} + \ce{e^-} &h\nu \gtrsim 0.76 \ \mathrm{eV}\\
\label{eq:photoH2p}
    \ce{H^+_2} + \gamma &\rightarrow \ce{H} + \ce{H^+} &h\nu \gtrsim 0.1 \ \mathrm{eV}
\end{align}
\end{subequations}
Reaction \ref{eq:photoH2} represents the indirect dissociation of \ce{H_2} molecules by LW photons via the two-step Solomon process \citep{Solomon:1965}, while reactions \ref{eq:photoHM} and \ref{eq:photoH2p} are the detachment of \ce{H^-} and the dissociation of \ce{H^+_2} respectively; these two chemical species are the main catalysts that lead to the formation of \ce{H_2} at moderate densities in a gas devoid of metals and dust.
The full frequency-dependent computation of these rates requires the following integration over the relevant range of photon energies:
\begin{equation}
\label{eq:intspecxsigma}
    k \ [\mathrm{s}^{-1}] = \int^{\nu_\mathrm{max}}_{\nu_\mathrm{min}} \frac{4\pi J_\nu \sigma(\nu)}{h\nu} \mathrm{d}\nu
\end{equation}
where $J_\nu$ is the radiation intensity in erg s$^{-1}$ sr$^{-1}$ Hz$^{-1}$ cm$^{-2}$, $\sigma_\nu$ is the frequency-dependent cross section in cm$^{2}$ and $\nu_\mathrm{max}$ is the Lyman limit that corresponds to the ionisation energy of Hydrogen atoms at 13.6 eV. As commonly assumed, photons above this threshold are neglected, as they are quickly absorbed by the ISM in the proximity of the source. We stress here that for the dissociation of molecules in Reactions~\ref{eq:photoH2} and \ref{eq:photoH2p} the minimum energies required are lower than the threshold energies usually adopted in the literature ($\sim11$ eV and $\sim2.65$ eV respectively, see e.g. \citealt{Abel:1997,Glover:2015a}). The latter are valid when only the ground state roto-vibrational level of the respective molecule is taken into account; however, we include in our model also the appropriate population of excited levels, that have lower bounding energies, hence the lower threshold energies. A detailed discussion on the molecular level populations is deferred to a companion paper \textcolor{blue}{(Incatasciato et al., in prep.)}, while here we limit the description of the rates calculation to a more general level.

\subsubsection{\ce{H_2} photodissociation rate}
\label{sec:H2rate}

The indirect photodissociation of \ce{H_2} molecules takes place through the Solomon process \citep{Solomon:1965}: a molecule in the roto-vibrational level with quantum numbers ($v, J$) of the electronic ground state $\mathrm{X}^1\Sigma^+_\mathrm{g}$ is excited to a ($v', J'$) state of the $\mathrm{B}^1\Sigma^+_\mathrm{u}$ or $\mathrm{C}^1\Sigma^{+/-}_\mathrm{u}$ electronic level, due to the absorption of a Lyman or Werner photon respectively. A fraction ($\sim 15\%$, on average) of the excited molecules then decay into the vibrational continuum of the ground state, resulting in its subsequent dissociation \citep{Abgrall:2000}.

To compute the optically-thin dissociation rate we follow the approach described in \cite{Draine:1996,Abel:1997,Wolcott-Green:2011},
for the continuum limit: the \textquotesingle effective cross section\textquotesingle \ can be calculated with
\begin{equation}
\label{eq:effsigma}
    \sigma(\nu)=C
    \sum_{v, J}\left[\sum_{v', J'}\left(\sum_\mathrm{i\in\mathrm{LW}} f_\mathrm{osc,i} \mathrm{V}(\nu-\nu_{0,\mathrm{i}}) f_{\mathrm{diss},v',J'}\right) N_\mathrm{X}(v, J)\right]
\end{equation}
where $C=\frac{\pi e^2}{4\pi m_\mathrm{e}c\epsilon_0}$ \citep{Corney:1977}\footnote{$e$ is the electron charge, $m_\mathrm{e}$ is the electron mass, $c$ is the speed of light in vacuum and $\epsilon_0$ is the electric constant.} and the summation runs over all the possible LW transitions, excited and ground state levels. $\mathrm{V}(\nu-\nu_{0,\mathrm{i}})$ is the Voigt line profile of the i-th transition between the ground state level ($v, J$) and the excited state level ($v', J'$), whose width takes into account both the natural damping coefficient and the thermal broadening; $f_\mathrm{osc,i}$ is the transition oscillator strength, $f_{\mathrm{diss},v',J'}$ is the fraction of molecules that dissociate after the excitation and $N_\mathrm{X}(v, J)$ is the fraction of molecules initially in the level ($v, J$).

The LW transitions are taken from the databases of \cite{Ubachs:2019} and \cite{Salumbides:2015}, where the transition frequency $\nu_0$, the oscillator strength $f_\mathrm{osc}$ and the natural damping coefficient $\Gamma$ are reported for each transition. These two datasets are complementary and are updated versions of the widely-used database by \cite{Abgrall:1993a,Abgrall:1993b,Abgrall:1993c}. The fraction $f_{\mathrm{diss},v',J'}$ of excited molecules ($v', J'$) that dissociate is instead derived from \cite{Abgrall:2000} as $A_\mathrm{c}/A_\mathrm{t}$, where $A_\mathrm{c}$ is the probability of decay to the vibrational continuum and $A_\mathrm{t}$ is the total probability of decay of an electronically excited state B/C. We also include the data from \cite{Abgrall:1997} \footnote{\ce{H_2} continuum emission probabilities at \url{https://molat.obspm.fr}} to derive the mean kinetic energy of the products of the dissociation (two H atoms), that in turn allows to estimate the average heating rate due to \ce{H_2} photodissociation. We find $\sim 0.4$ eV per dissociated molecule, similar to \citealt{Black:1977}, but with some variations of the order of 30\% depending on the gas temperature and density and the shape of the incident spectrum.

We also take into account how the roto-vibrational levels of the electronic ground state $\mathrm{X}^1\Sigma^+_\mathrm{g}$ are populated, for a given combination of gas temperature and density. These levels can be excited and de-excited both due to collisions or the absorption/emission of photons. When these processes are frequent enough (above a certain density threshold) the local thermo-dynamical equilibrium (LTE) is reached and the level population follows the Boltzmann distribution.
The LTE density threshold for \ce{H_2} molecules lies between $10^3$ and $10^6$ cm$^{-3}$ depending on the gas temperature. At lower densities only the first rotational levels of the ground vibrational level ($v=0$, $J=0-3$) are populated. At intermediate densities, we interpolate between the non-LTE ($k_{\ce{H_2},0}$) and the LTE case ($k_{\ce{H_2},\mathrm{LTE}}$) as in \cite{Glover:2015a}:
\begin{equation}
\label{eq:interpLTE}
    k_{\ce{H_2}} = k_{\ce{H_2},\mathrm{LTE}}\left(\frac{k_{\ce{H_2},0}}{k_{\ce{H_2},\mathrm{LTE}}}\right)^\alpha
\end{equation}
where $\alpha=(1+n/n_\mathrm{crit}(T_\mathrm{gas}))^{-1}$.
Interested readers will find a detailed description of $n_\mathrm{crit}$ in \textcolor{blue}{Incatasciato et al. (in prep.)}.

In this work we do not vary the gas temperature (set at $10^3 \ \mathrm{K}$) and density (set at $10^2 \ \mathrm{cm}^{-3}$), that is well within the non-LTE limit, $n_\mathrm{crit}$ being approximately 3 orders of magnitude higher at $10^3 \ \mathrm{K}$. This choice of gas temperature and density ensures that the $\ce{H_2}$ and $\ce{H^+_2}$ dissociation rates are representative of the initial stages of collapse of a gas cloud in a low-metallicity environment \citep{Omukai:2005}. The effect of the LW radiation during the subsequent evolution at densities $n\gtrsim10^4 \ \mathrm{cm}^{-3}$ would instead involve other physical processes, such as the \ce{H_2} self-shielding \citep{Wolcott-Green:2011,Hartwig:2015a,Wolcott-Green:2019}, that are beyond the scope of this work.

\subsubsection{\ce{H^+_2} photodissociation}
\label{sec:H2prate}

Updated state-resolved cross sections for the \ce{H^+_2} photodissociation and the inverse process (radiative association) are available in the literature \citep[e.g.,][]{Babb:2015,Zammit:2017} for photons with energies as high as 40 eV.
We choose to use the data from \cite{Zammit:2017,Zammit:2018} as they are including the cross sections for all the 423 roto-vibrational levels of the electronic ground state of \ce{H^+_2}. This makes the calculation more reliable in the LTE limit at high temperatures ($T_\mathrm{gas} \sim 10^3-10^4$ K). They also take into account transitions to 23 different electronic excited states, while \cite{Babb:2015} considers only the first excited state 2p$\sigma_\mathrm{u}$: this is less crucial for the purpose of our work, as the total cross section is essentially due to transitions to the continuum of the first excited state, other than at energies $\gtrsim 12$ eV ($\lambda < 1000 \ \angstrom$) where the contribution from the other states is noticeable, and we neglect photons above the Hydrogen ionisation limit.

We again follow the approach of \cite{Glover:2015a} as in Eq.~\ref{eq:interpLTE} to interpolate between the non-LTE and the LTE rates. For the non-LTE limit we assume that all the molecules are in the roto-vibrational level with the lowest energy ($v=0$, $J=0$) \citep[see e.g.][]{Shapiro:1987,Latif:2015,Glover:2015a}. The LTE limit is assumed for a gas density above the critical value $n_\mathrm{crit}$, that is determined as in \cite{Glover:2015a} Section B1.2, assuming that H atoms and free electrons are the most important collisional partners.

\subsubsection{\ce{H^-} photodetachment}
\label{sec:HMrate}

Several cross sections are available in the literature for this process: \cite{Shapiro:1987,John:1988,Chuzhoy:2007,McLaughlin:2017}. We choose the latter, that for the first time includes the resonances at 11 eV. This gives an increase of $\sim 20\%$ on the \ce{H^-} detachment rate for energetic spectra (e.g. a black-body spectrum with $T_\mathrm{rad}=10^5$ K, \citealt{Glover:2015b}).

\subsection{IGM optical depth}
\label{sec:fmod}

\begin{figure}
    \centering
    \includegraphics[width=\linewidth]{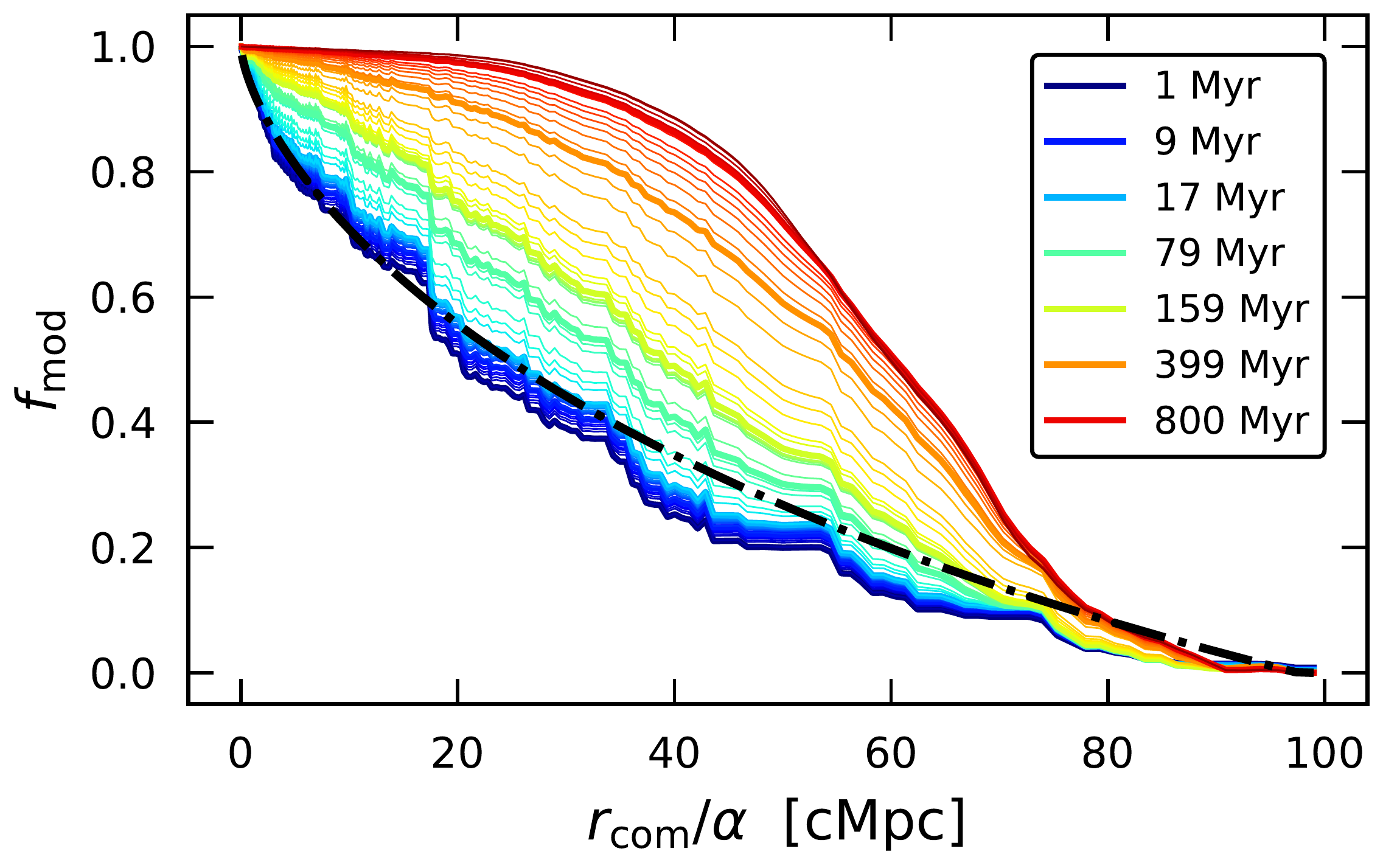}
    \vspace{-0.6cm}
    \caption{\ce{H_2} modulation factor assuming a PopIII\_Ygg2 stellar population at different ages.
    The black dot-dashed line is the \protect\cite{Ahn:2009} fit, that reproduces fairly well $f_{\mathrm{mod}}$ for young stars ($<10$ Myr, in blue), but fails to match it for increasing ages (green, yellow and red solid lines).}
    \vspace{-0.2cm}
    \label{fig:fmod_Ygg2_evol}
\end{figure}

\begin{figure}
    \centering
    \includegraphics[width=\linewidth]{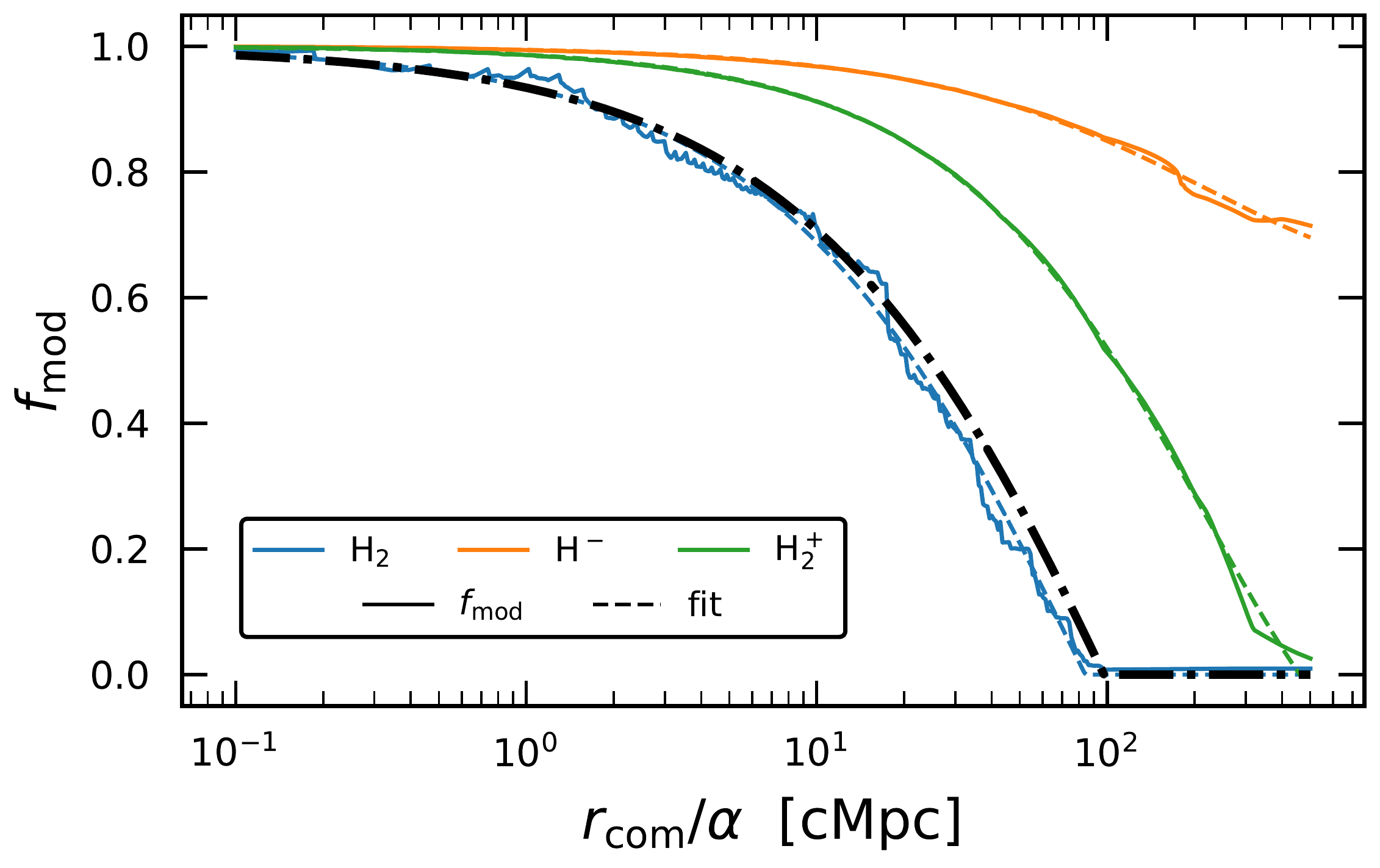}
    \vspace{-0.6cm}
    \caption{Modulation factors (solid lines) and the corresponding fits (dashed lines) for the three reactions and a 1-Myr-old PopIII\_Ygg2 SED. The optimal parameters valid for this particular stellar model are listed in Table~\ref{Table:fmod_fit_params}. As in Fig.~\ref{fig:fmod_Ygg2_evol} the black dot-dashed line is the \protect\cite{Ahn:2009} fit for \ce{H_2}. It can be noted here that $f_{\mathrm{mod}}$ for \ce{H_2} is slightly non-monotonic, unlike what is proposed by \protect\cite{Ahn:2009} and \protect\cite{Fialkov:2013}, due to the presence of LW transitions at energies lower than the Lyman-$\alpha$ transition, that do not enter any absorption window. This, however, represents only a second-order effect, while the trend first shown by \protect\cite{Ahn:2009} is confirmed and is valid for young stellar populations in general (see the text for the relative discussion). For the same reason $f_{\mathrm{mod}}$ for \ce{H_2} doesn't reach exactly zero at the LW horizon, but values as low as $10^{-3}-10^{-4}$ that can be approximated with zero.}
    \vspace{-0.2cm}
    \label{fig:fmod_fit_3reac}
\end{figure}

\cite{Haiman:2000} show that the LW radiation can be efficiently absorbed by the diffuse neutral gas of the IGM. In the pre-reionisation universe, in fact, the Hydrogen optical depth in the Lyman lines (energy range $10.2-13.6$ eV) is very high ($\tau\sim10^6$ at $z\sim20$); hence, LW photons are absorbed by H atoms as soon as they are cosmologically redshifted into the closest atomic Lyman transition. As suggested by \cite{Haiman:2000} and \cite{Ahn:2009} the contribution of \ce{H_2} molecules to the IGM optical depth is subdominant and can be neglected due to its low abundance in the diffuse gas.

If we consider a LW photon emitted at redshift $z_\mathrm{em}$ with energy $h\nu_\mathrm{em}$ and its closest Lyman line with energy $h\nu_\mathrm{line}$, the maximum distance at which the photon can be observed corresponds to a minimum redshift $z_\mathrm{obs}$ expressed as
\begin{equation}
    \frac{1+z_\mathrm{obs}}{1+z_\mathrm{em}} = \frac{\nu_\mathrm{line}}{\nu_\mathrm{em}}.
\end{equation}
Under the assumption of a homogeneous LW background, this formula leads to the definition of the \textquotesingle sawtooth modulation\textquotesingle \ (see Fig.1 in \citealt{Haiman:2000} and \citealt{Ahn:2009}, for a flat emitted spectrum).

However, this works aims at studying the LW background beyond the homogeneous universe approximation. We need to consider the so-called \textquotesingle picket-fence\textquotesingle \ modulation factor by \cite{Ahn:2009}, that describes (from the point of view of a single source) the fraction of unabsorbed spectrum in the LW energy range at a comoving distance $r_\mathrm{com}$:
\begin{equation}
\label{eq:fmod}
    f_{\mathrm{mod}} = \mathrm{Max}(0,A\exp[-(r_\mathrm{com}/B\alpha)^C]-D)
\end{equation}
where $A=1.7$, $B=116.29$, $C=0.68$, $D=0.7$, $r_\mathrm{com}$ is in (comoving) Mpc and
\begin{equation}
\label{eq:fmod_alpha}
    \alpha = \left(\frac{h}{0.7}\right)^{-1} \left(\frac{\Omega_\mathrm{m}}{0.27}\right)^{-1/2} \left(\frac{1+z_\mathrm{em}}{21}\right)^{-1/2}
\end{equation}
contains the dependency on the cosmological parameters and the redshift of emission. The parameters in the expression were estimated by \cite{Ahn:2009} considering a flat spectrum in the energy interval $11.5-13.6$ eV. From Eq.~\ref{eq:fmod} the maximum distance that a LW photon can travel is $R_\mathrm{LW} \simeq 97\alpha$ cMpc. This defines a \textquotesingle LW horizon\textquotesingle, that represents the largest volume that needs to be considered in order to evaluate a self-consistent LW background.

\cite{Fialkov:2013} showed that a more accurate evaluation of $f_{\mathrm{mod}}$ is obtained if the full frequency-dependent calculation is performed without simplifying assumptions, such as the LW transitions being uniformly distributed in frequency or a flat incident spectrum.
We build further on this, by recalculating the fit of Eq.~\ref{eq:fmod} for each SED mentioned in Section~\ref{sec:SEDs} and for each of the three photochemical rates described in Section~\ref{sec:rates}. In particular we calculate here for the first time the modulation factor for $k_\mathrm{\ce{H^-}}$ and $k_\mathrm{\ce{H^+_2}}$, that have a much larger \textquotesingle horizon\textquotesingle, as the corresponding threshold energies are lower than than $\mathrm{E_{Ly\alpha}}=10.2$ eV.

Given a stellar spectrum and the assumption of a universe at the mean density, we perform the full frequency-dependent rate calculation for the transmitted spectrum at distances from $0.1\alpha$ cMpc to $500\alpha$ cMpc. The result is then fitted using the same functional form as in Eq.~\ref{eq:fmod}, where we fix $D=A-1$, but we re-evaluate the other parameters for each SED at each stellar age. By automatically incorporating the appropriate cross section, in this work (as also in \citealt{Fialkov:2013}) $f_{\mathrm{mod}}$ represents the true correction factor to the photochemical rates\footnote{We neglect here the resonant photons that are absorbed by the neutral IGM and re-emitted at lower frequencies. The dataset of LW transitions adopted in this work is more extended than the one in \cite{Fialkov:2013} and some low-energy photons might still excite \ce{H_2} molecules through LW transitions at energies < 10.2 eV. However, our choice is still a reasonable assumption, given that those transitions are not important outside the LTE regime of dense gas (defined as above a critical density of $n\sim 10^4 \ \mathrm{cm}^{-3}$).} and not just the fraction of unabsorbed spectrum as in \cite{Ahn:2009}.

Fig.~\ref{fig:fmod_Ygg2_evol} shows how $f_{\mathrm{mod}}$ for the \ce{H_2} dissociation rate changes if spectra with different shapes are assumed. In particular we use PopIII\_Ygg2 SEDs at different ages, from young energetic spectra (in blue, < 10 Myr) to intermediate and old stellar populations (green, yellow and red solid lines). The dot-dashed black line represents the fit from \cite{Ahn:2009}, that matches reasonably well only the modulation factor for young stellar populations at $r_\mathrm{com} < 40\alpha \ \mathrm{cMpc}$. At larger distances the fit overestimates it, even though less than in \cite{Fialkov:2013}, as in our larger dataset of LW transitions some at $\sim11$ eV do not end up in any absorption window until very large distances, hence they contribute to the dissociation rate. For older stellar populations, instead, $f_{\mathrm{mod}}$ evolves more and more slowly with the distance, as hard photons at $\sim13$ eV, that are absorbed closer to the emitting source due to the high density of Lyman lines, have a minor impact to the total \ce{H_2} dissociation rate. We have verified that the same trend is found for all the other SEDs we use in this work, for both PopIII and PopII stars. Assuming \cite{Ahn:2009} fitting function regardless of the stellar age would hence lead to an underestimation of the contribution of old stars to the LW background by a factor of $2-3$.

In Fig.~\ref{fig:fmod_fit_3reac} we show $f_{\mathrm{mod}}$ for the \ce{H_2} dissociation, the \ce{H^-} detachment and the \ce{H^+_2} dissociation (blue, orange and green solid lines respectively). The dashed lines show the relative fits, with the black dot-dashed line being the \cite{Ahn:2009} fit for \ce{H_2}. The SED used in this illustrative example is the one for a 1-Myr-old PopIII\_Ygg2 stellar population and the corresponding fitting parameters are reported in Table~\ref{Table:fmod_fit_params}. As already discussed above, the \cite{Ahn:2009} fit closely describes the \ce{H_2} $f_{\mathrm{mod}}$ for the radiation emitted by young stars (our fitting parameters are only slightly different) and the LW horizon at $\sim 100$ cMpc is still valid and independent from the spectral shape (see Fig.~\ref{fig:fmod_Ygg2_evol}). $f_{\mathrm{mod}}$ for the other two reactions, instead, decreases much more slowly with the distance and never actually reaches zero, as photons below the Lyman $\alpha$ line are not absorbed by the neutral Hydrogen. This in principle would imply that the volume employed for the calculation of the background is not finite. However, we choose to limit it to a sphere of radius 500$\alpha$ cMpc. We motivate our strategy in the next section.

\begin{table}
    \centering
    \renewcommand{\arraystretch}{1.5}
    \setlength{\heavyrulewidth}{1.5pt}
    \setlength{\abovetopsep}{4pt}
    \begin{tabular}{ cccc }
        \toprule
        Rate & $A$ & $B$ & $C$ \\
        \midrule
        $\ce{H_2}$ & 1.548 & 79.733 & 0.719 \\
        $\ce{H^-}$ & 0.357 & 217.445 & 0.776 \\
        $\ce{H^+_2}$ & 1.188 & 220.907 & 0.831 \\
        \midrule
        $\ce{H_2}$ - \cite{Ahn:2009} & 1.7 & 116.29 & 0.68 \\
        \bottomrule
    \end{tabular}
    \caption{Optimal parameters for the fit shown in Fig.~\ref{fig:fmod_fit_3reac}. The parameters are defined as in Eq.~\ref{eq:fmod} and we have fixed $D=A-1$.}
\label{Table:fmod_fit_params}
\end{table}

\subsubsection{Cosmological volume needed for the LWB evaluation}
\label{sec:copies}

\begin{figure}
    \centering
    \includegraphics[width=\linewidth]{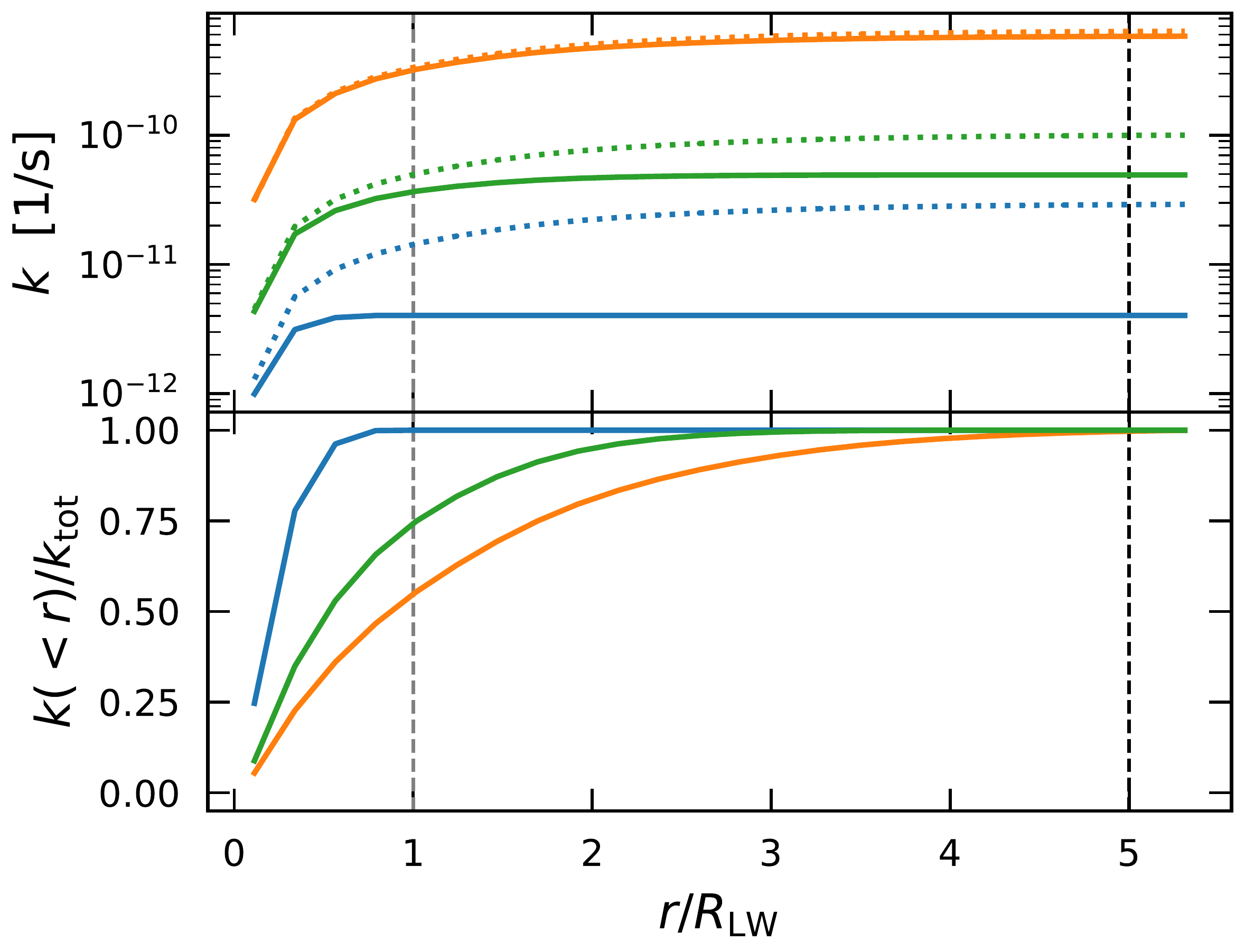}
    \vspace{-0.6cm}
    \caption{\textit{Top panel}. Cumulative contribution to the LW background of the sources inside a sphere of radius $r$ (normalised to $R_\mathrm{LW}$), for the XL simulation at $z=8.9$ and for the \textquotesingle FID\textquotesingle \ choice of SEDs, with (solid lines) and without (dotted lines) considering the IGM absorption. The three rates, \ce{H_2} dissociation, \ce{H^-} detachment and \ce{H^+_2} dissociation, are color coded as in Fig.~\ref{fig:fmod_fit_3reac}. As expected, including the IGM optical depth has a stronger impact on the \ce{H_2} dissociation rate, while the \ce{H^-} detachment rate is almost unaffected.
    \textit{Bottom panel}. Cumulative \textit{fractional} contribution to the LW background, inside a sphere of radius r as in the top panel. Here we account for the IGM absorption. By definition the \ce{H_2} dissociation rate converges within $R_\mathrm{LW}$ (grey dashed vertical line). We consider a maximum distance of $\sim5R_\mathrm{LW}$ (black dashed vertical line) to have a convergence of $k_{\ce{H^+_2}}$ and $k_{\ce{H^-}}$ too.}
    \vspace{-0.2cm}
    \label{fig:horizon_frac}
\end{figure}

In the top panel of Fig.~\ref{fig:horizon_frac} we report the cumulative contribution to the LW background of spheres of increasing radius and centred on random points within the simulation volume. We use here the XL box at $z=8.9$ and the \textquotesingle FID\textquotesingle \ choice of SEDs. The solid and dotted lines indicate whether the IGM optical depth is included (solid) or not (dotted). The largest sphere has a radius of $\sim500\alpha$ cMpc, the same maximum distance considered for the evaluation of the modulation factor (Fig.~\ref{fig:fmod_fit_3reac}). In order to reach a volume that is larger than the simulated box, we stack several copies of the box until the target sphere is reached. As expected, including the IGM optical depth has the strongest impact on the \ce{H_2} dissociation rate, that is reduced by a factor of $5-10$. The \ce{H^+_2} dissociation rate is only moderately reduced, while the \ce{H^-} detachment rate is almost unaffected.

In the bottom panel we demonstrate instead the convergence of the three rates, in terms of cumulative \textit{fractional} contribution to the LW background. By definition, the \ce{H_2} dissociation rate converges within $R_\mathrm{LW}$ (grey dashed vertical line). \ce{H^-} detachment and \ce{H^+_2} dissociation instead converge somewhere between 1 and 5 $R_\mathrm{LW}$, and the exact distance slightly depends on the star formation history and the stellar models. We hence assume that all three photochemical rates converge within a maximum distance of $5R_\mathrm{LW}$. This sets the volume that needs to be considered around a given observer in order to determine the LW background in that point. Only the total \ce{H^-} detachment rate could be underestimated by not more than 5\% at low redshift, due to $f_{\mathrm{mod}}$ being significantly greater than zero at any distance.

The simulation boxes described in Table~\ref{Table:FiBYruns} (with sizes ranging from 4 to 32 cMpc) are smaller than $5R_\mathrm{LW}$. As already done in \cite{Ahn:2009}, we account for this by attaching multiple copies of the simulation box next to the central one, until the maximum distance is reached. This ensures that we include all the sources that contribute to the radiation background as measured in the central box. The drawback of our method is that the conclusions we can draw on the inhomogeneities of the radiation field are certainly limited, as the simulated volume is not able to capture the total variance of the cosmic structures that we would expect to find in a sphere with radius $\sim500\alpha$ cMpc.

\section{Results}
\label{sec:results}

\subsection{Mean LW background}
\label{sec:LWB}

\begin{figure}
    \centering
    \includegraphics[width=\linewidth]{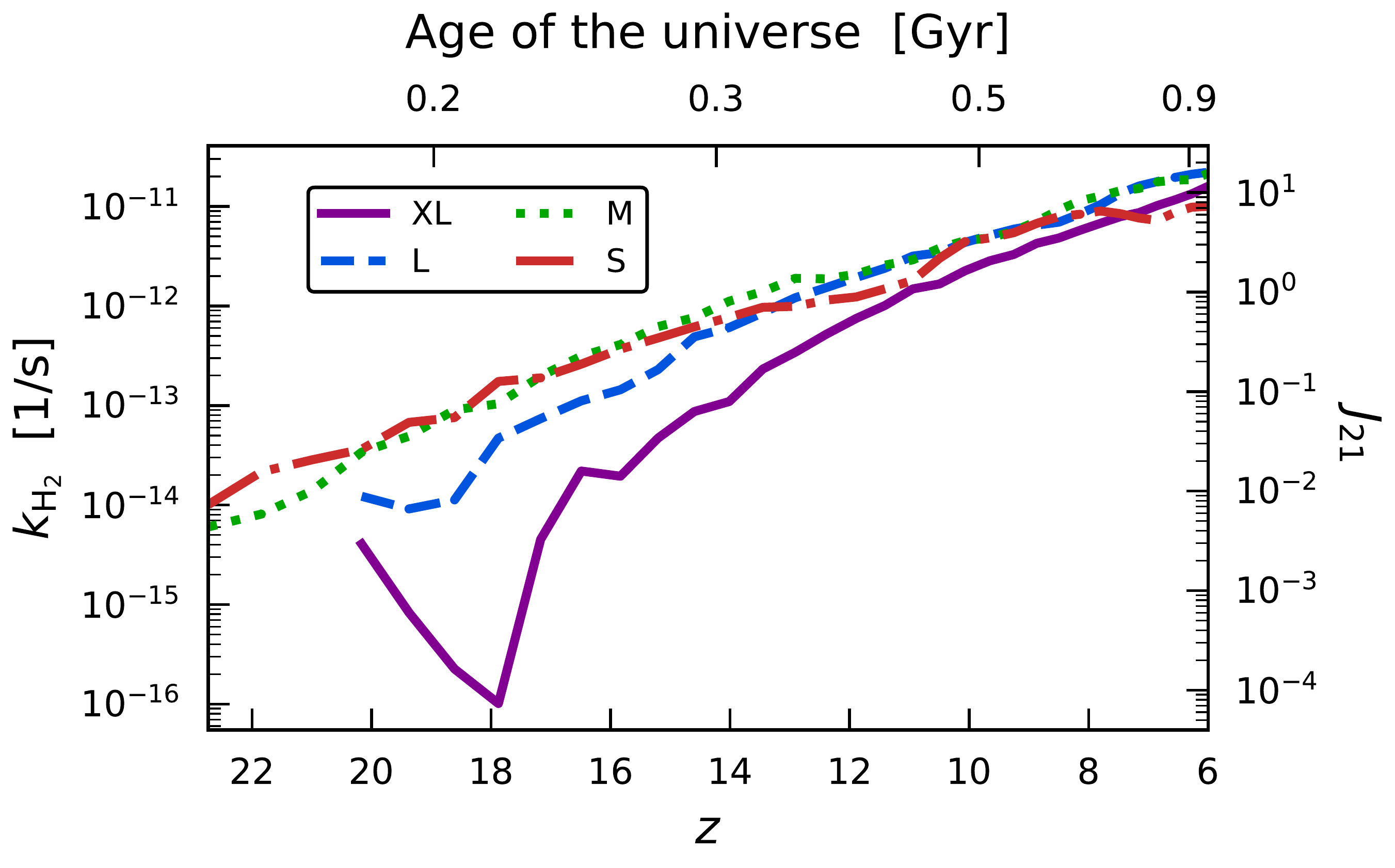}
    \vspace{-0.6cm}
    \caption{The evolution with redshift of the mean \ce{H_2} dissociation rate, given by the LW radiation measured in the XL (solid purple), L (dashed blue), M (dotted green) and S (dash-dotted red) FiBY simulations. The colour and line-style scheme is consistent with Fig.~\ref{fig:SFRD_SFRthresh}. On the secondary y-axis we express the LW radiation intensity at 13.6 eV in units of $J_{21} = 10^{-21} \ \mathrm{erg} \ \mathrm{s}^{-1} \ \mathrm{Hz}^{-1} \ \mathrm{sr}^{-1} \ \mathrm{cm}^{-2}$, where we use the approximate relation $k_{\ce{H_2}}=1.38\times10^{-12} \ J_{21} \ \mathrm{s}^{-1}$ commonly used in the literature.}
    \vspace{-0.2cm}
    \label{fig:LWB_FiBY_only}
\end{figure}

In Fig.~\ref{fig:LWB_FiBY_only} we show the LW background obtained with the post-processing method described in Section~\ref{sec:methods}, using the \textquotesingle FID\textquotesingle \ choice for the stellar models. The four coloured lines represent the mean \ce{H_2} dissociation rate measured in the XL (solid purple), L (dashed blue), M (dotted green) and S (dash-dotted red) FiBY simulations. The secondary y-axis expresses the common conversion between the \ce{H_2} dissociation rate and the LW radiation intensity at the Lyman limit (13.6 eV) $J_{21}$, in units of $10^{-21} \ \mathrm{erg} \ \mathrm{s}^{-1} \ \mathrm{Hz}^{-1} \ \mathrm{sr}^{-1} \ \mathrm{cm}^{-2}$. With this definition, and under the assumption of low-density gas in the optically-thin limit and a flat incident spectrum, the \ce{H_2} dissociation rate is $k_{\ce{H_2}}=1.38\times10^{-12} \ J_{21} \ \mathrm{s}^{-1}$  \citep{Abel:1997}\footnote{Note that this scaling is generally only valid for young stellar populations \citep{Shang:2010,Agarwal:2015,Glover:2015b}.}.

The LWB intensity generally increases with time, as primordial low-mass halos and then proto-galaxies grow in mass and trigger the formation of more and more stars. The LWB reaches mean values well above $J_{21}\sim1$ at $z\lesssim10$ in all simulations. We attribute the small difference in the XL run to the slightly lower $\rho_\mathrm{SFR}$ in massive galaxies (Fig.~\ref{fig:SFRD_SFRthresh}), that in turn can be explained with a systematic degradation of the star formation efficiency in lower-resolution simulations, as already found in IllustrisTNG \citep{Pillepich:2018b}. The S box shows a more irregular evolution at low redshift due to low number of high-mass halos that dominate the photon budget in the small volume. On the other hand, at $z>12$ the different halo mass range that is resolved by each simulation determines when the LWB starts to build up and its intensity. Both M and S simulations resolve star formation in low-mass halos and hence show a good convergence from early times, even if there is a hint of a missing contribution from $M_\mathrm{h}\lesssim10^6 \ \mathrm{M_\odot}$ halos (not resolved in M, see the last column in Table~\ref{Table:FiBYruns}) at $z\geq20$. L and XL, instead, have delayed PopIII star formation, as they resolve only halos with $M_\mathrm{h}\gtrsim10^7 \ \mathrm{M_\odot}$ and $M_\mathrm{h}\gtrsim10^8 \ \mathrm{M_\odot}$ respectively. This is reflected into a delayed build-up of the LW intensity, that is $5-10$ ($100$) times lower in L (XL) than in M at $z\sim20-15$.

Overall, we find that the LWB from the FiBY simulations, when the relevant halo mass range is resolved, is well fitted by the following polynomial, with $6<z<23$:
\begin{equation}
\label{eq:fit_FiBY_LWB}
    \log J_{21} = A + B(1+z) + C(1+z)^2
\end{equation}
with $A=2.119$, $B=-1.117\times10^{-1}$ and $C=-2.782\times10^{-3}$.

\subsubsection{Effective LW spectral shape}
\label{sec:effective_spectral_shape}

\begin{figure}
    \centering
    \includegraphics[width=\linewidth]{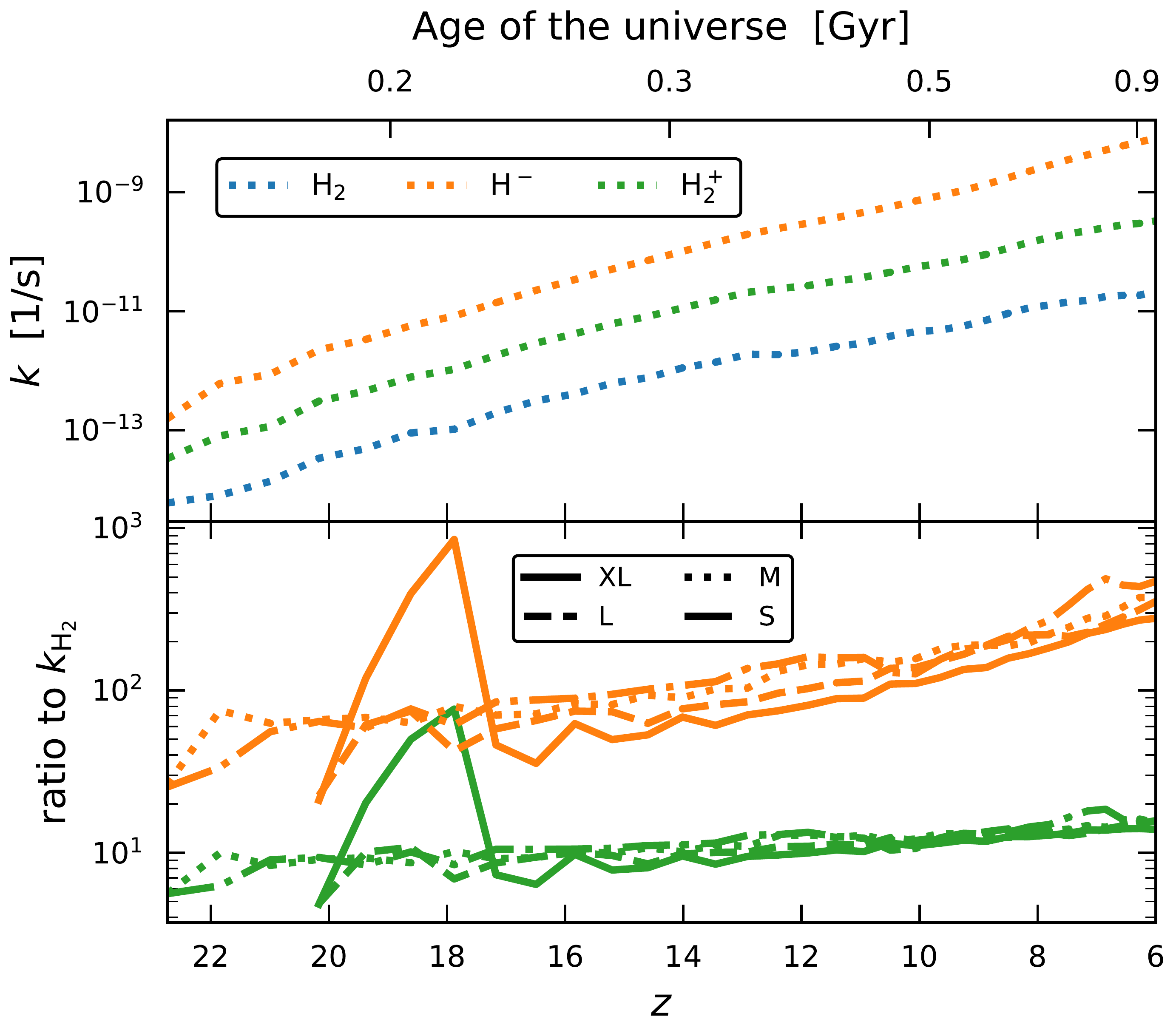}
    \vspace{-0.6cm}
    \caption{\textbf{\textit{Top panel}}: the evolution of the three photochemical rates considered in this work, in the M simulation and for our \textquotesingle FID\textquotesingle \ choice of SEDs. \textbf{\textit{Bottom panel}}: the ratio of $\ce{H^-}$ detachment rate (orange) and $\ce{H^+_2}$ dissociation rate (green) to $\ce{H_2}$ dissociation rate, for the FiBY simulations.}
    \vspace{-0.2cm}
    \label{fig:all_rates}
\end{figure}

As already stated above, in this work we include also the \ce{H^-} detachment and \ce{H^+_2} dissociation rate, that are important to determine the rate at which \ce{H_2} molecules form during the initial phases of gas collapse.
The top panel of Fig.~\ref{fig:all_rates} illustrates the evolution of these rates in the M box, for our \textquotesingle FID\textquotesingle \ choice of SEDs. The blue line is the LWB previously shown in Fig.~\ref{fig:LWB_FiBY_only}, while the orange and the green lines are $k_{\ce{H^-}}$ and $k_{\ce{H^+_2}}$ respectively. The rates concurrently grow in time, as the UV photons emitted by young massive stars are the major contributors to all of them; nevertheless, an increasing additional contribution of IR photons is present in $k_{\ce{H^-}}$ and $k_{\ce{H^+_2}}$.

The bottom panel of the same figure presents the ratio between the latter two rates and $k_{\ce{H_2}}$, for the same four FiBY runs as in Fig.~\ref{fig:LWB_FiBY_only}. At the zero-th order \citep[see e.g.][]{Latif:2015}, the \ce{H^-} detachment rate (in orange) and the \ce{H^+_2} dissociation rate (in green) are approximately two and and one order(s) of magnitude higher than the \ce{H_2} dissociation rate respectively. These differences increase for softer spectra (with lower black-body temperature, in their approximate treatment): the \ce{H^-} detachment rate, in particular, is more sensitive to the spectral shape due to the wider photon energy range of its cross section. This is reflected into the evolution of these ratios with time: they increase with decreasing redshift, as older stellar populations and PopII stars, that are predicted to dominate at later times, have softer spectra with a higher IR-to-UV ratio. In particular, given the star formation history of the FiBY simulations and our fiducial set of SEDs, we can tentatively describe the spectral shape of the LW background by assigning it an effective black-body temperature, based on the ratios in Fig.~\ref{fig:all_rates}: we predict that the LW background spectral shape evolves from $T_{\mathrm{eff}}=6\times10^4 \ \mathrm{K}$ at $z=23$ to $T_{\mathrm{eff}}=2\times10^4 \ \mathrm{K}$ at $z=6$. In Appendix~\ref{appen:SEDs} we show that the choice of the specific set of stellar SEDs does not drastically change these results.

The differences in the star formation history of the different FiBY runs results in a scatter of $0.1-0.2$ dex in $k_{\ce{H^-}}/k_{\ce{H_2}}$ and no appreciable scatter in $k_{\ce{H^+_2}}/k_{\ce{H_2}}$, with the exception of the bump at $z\sim18$ for the XL box, that is caused by the high stochasticity of the first star formation episodes at early times. The scatter in $k_{\ce{H^-}}/k_{\ce{H_2}}$ is due to the dependence of the timing of the transition from PopIII- to PopII-dominated star formation on the spatial resolution, discussed in Section~\ref{sec:age_contrib}. Here we neglect this second order effect and fit the ratios with Eq.~\ref{eq:fit_ratios}:
\begin{equation}
\label{eq:fit_ratios}
    \log\left(\frac{k_{\mathrm{X}}}{k_{\ce{H_2}}}\right) = A + B(1+z) + C(1+z)^2
\end{equation}
where the best fit values for $A$, $B$ and $C$ are reported separately for $k_{\ce{H^-}}$ and $k_{\ce{H^+_2}}$ in Table~\ref{Table:fit_ratios_params}.

\begin{table}
    \centering
    \renewcommand{\arraystretch}{1.5}
    \setlength{\heavyrulewidth}{1.5pt}
    \setlength{\abovetopsep}{4pt}
    \begin{tabular}{ cccc }
        \toprule
        Ratio & $A$ & $B$ & $C$ \\
        \midrule
        $k_{\ce{H^-}}/k_{\ce{H_2}}$ & $3.06$ & $-8.70 \times 10^{-2}$ & $1.03 \times 10^{-3}$ \\
        $k_{\ce{H^+_2}}/k_{\ce{H_2}}$ & $1.37$ & $-2.84 \times 10^{-2}$ & $3.09 \times 10^{-4}$ \\
        \bottomrule
    \end{tabular}
    \caption{Parameters that reproduce the ratio of $k_{\ce{H^-}}$ and $k_{\ce{H^+_2}}$ to $k_{\ce{H_2}}$. They are valid for the \textquotesingle FID\textquotesingle \ case, but in Appendix~\ref{appen:SEDs} we demonstrate that the rates given by other stellar SEDs differ by not more than a factor of $2-3$.}
\label{Table:fit_ratios_params}
\end{table}

\subsubsection{Contribution from PopIII/PopII and young/old stars}
\label{sec:age_contrib}

\begin{figure}
    \centering
    \includegraphics[width=\linewidth]{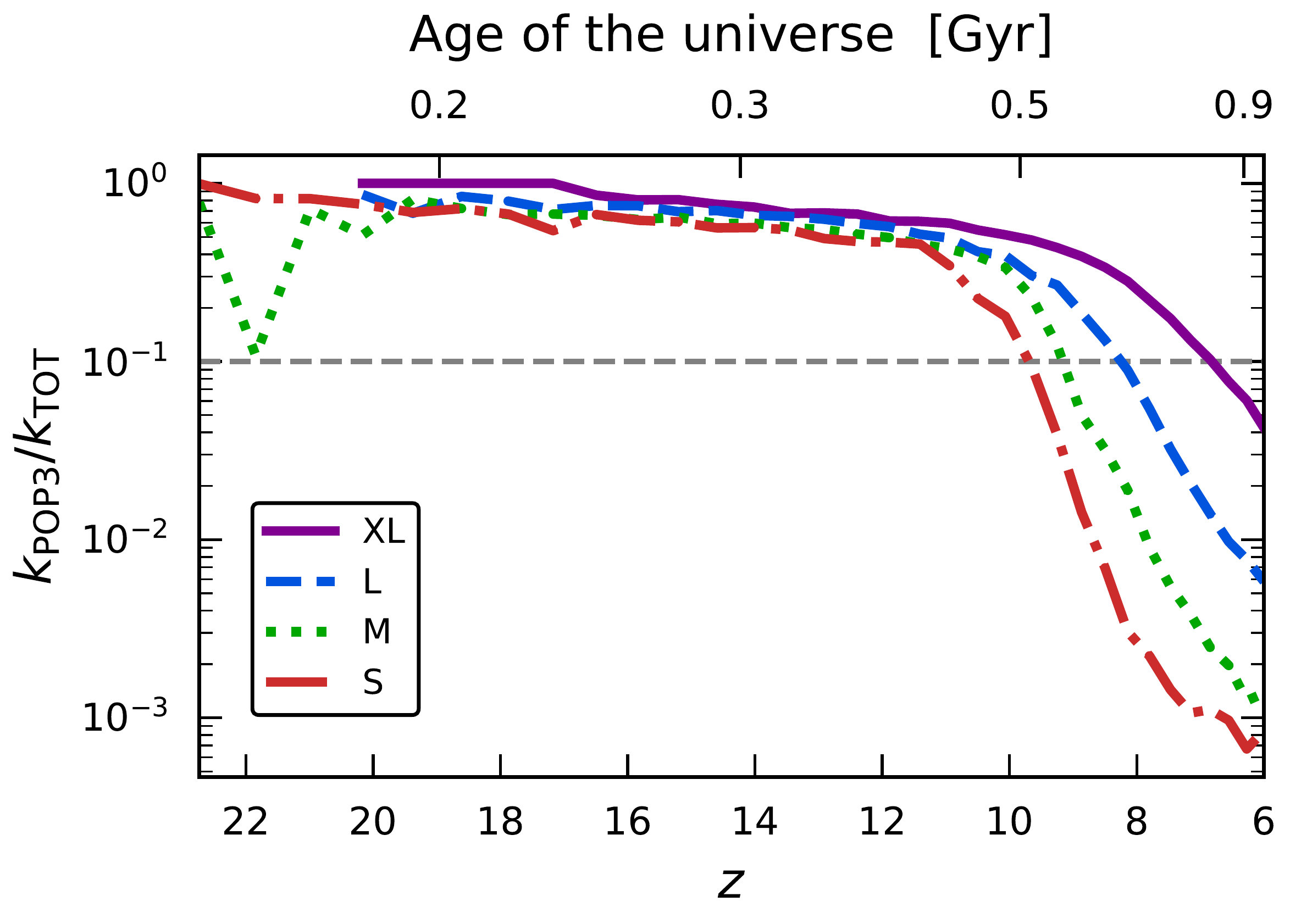}
    \vspace{-0.6cm}
    \caption{Fraction of the LWB due to the radiation emitted by Population III stars, in the same FiBY simulations as in Fig.~\ref{fig:SFRD_SFRthresh} and Fig.~\ref{fig:LWB_FiBY_only}.}
    \vspace{-0.2cm}
    \label{fig:pop3_contrib}
\end{figure}

\begin{table}
    \centering
    \renewcommand{\arraystretch}{1.5}
    \setlength{\heavyrulewidth}{1.5pt}
    \setlength{\abovetopsep}{4pt}
    \begin{tabular}{ ccccccc }
        \toprule
        Simulation & \multicolumn{3}{c}{$z_{50}$} & \multicolumn{3}{c}{$z_{10}$} \\
        \midrule
        {} & $\ce{H_2}$ & $\ce{H^-}$ & $\ce{H^+_2}$ & $\ce{H_2}$ & $\ce{H^-}$ & $\ce{H^+_2}$ \\
        \midrule
        \textbf{XL} & 9.9 & 11.9 & 10.0 & 6.8 & 7.1 & 6.7 \\
        \textbf{L} & 11.1 & 13.0 & 11.2 & 8.2 & 8.2 & 8.0 \\
        \textbf{M} & 12.0 & 15.6 & 12.2 & 9.2 & 8.8 & 8.8 \\
        \textbf{S} & 13.0 & 16.4 & 12.9 & 9.7 & 9.5 & 9.4 \\
        \bottomrule
    \end{tabular}
    \caption{Redshift after which the contribution from PopIII stars to the three rates falls below 50\% (first three columns) and 10\% (second group of three colums), as shown in Fig.~\ref{fig:pop3_contrib} for the $\ce{H_2}$ dissociation rate.}
\label{Table:pop3_contrib}
\end{table}

Fig.~\ref{fig:pop3_contrib} shows the fraction of the LWB that is emitted by PopIII stars, again for the \textquotesingle FID\textquotesingle \ SEDs. Metal-free stars dominate in the early universe, but their contribution is slowly reduced to $\sim50\%$ at $z=11$, before quickly dropping to less than 10\% at $z\sim8-10$, following the fast metal injection from PopIII CCSNe and PISNe that boosts the metallicity above the threshold for PopII star formation.
When the resolution limits the halo masses that can be resolved, the sequence \textquotesingle PopIII formation - metal enrichment - PopII formation\textquotesingle \ is delayed by a few hundreds Myr (as already shown i.e. by \citealt{Maio:2010}) and this is reflected in the shallower and delayed drop in the PopIII contribution in the XL simulation. We summarise these results in Table~\ref{Table:pop3_contrib}, where we report $z_{50}$ and $z_{10}$, the redshifts at which the contribution from PopIII stars falls below 50\% and 10\% respectively. We estimate them for all the three photochemical rates considered in this work, while only the \ce{H_2} dissociation rate is shown in Fig.~\ref{fig:pop3_contrib}. $z_{10}$ doesn't significantly change according to the specific rate considered and the lower IGM optical depth associated with $\ce{H^-}$ and $\ce{H^+_2}$ (see Sec~\ref{sec:fmod}), that marginally increases the contribution from distant sources, does not have any impact. $z_{50}$, on the other hand, is appreciably higher for the \ce{H^-} detachment rate: the contribution from PopII stars shows indeed a more rapid and steady growth with redshift, due to their softer spectrum.

\begin{figure}
    \centering
    \includegraphics[width=\linewidth]{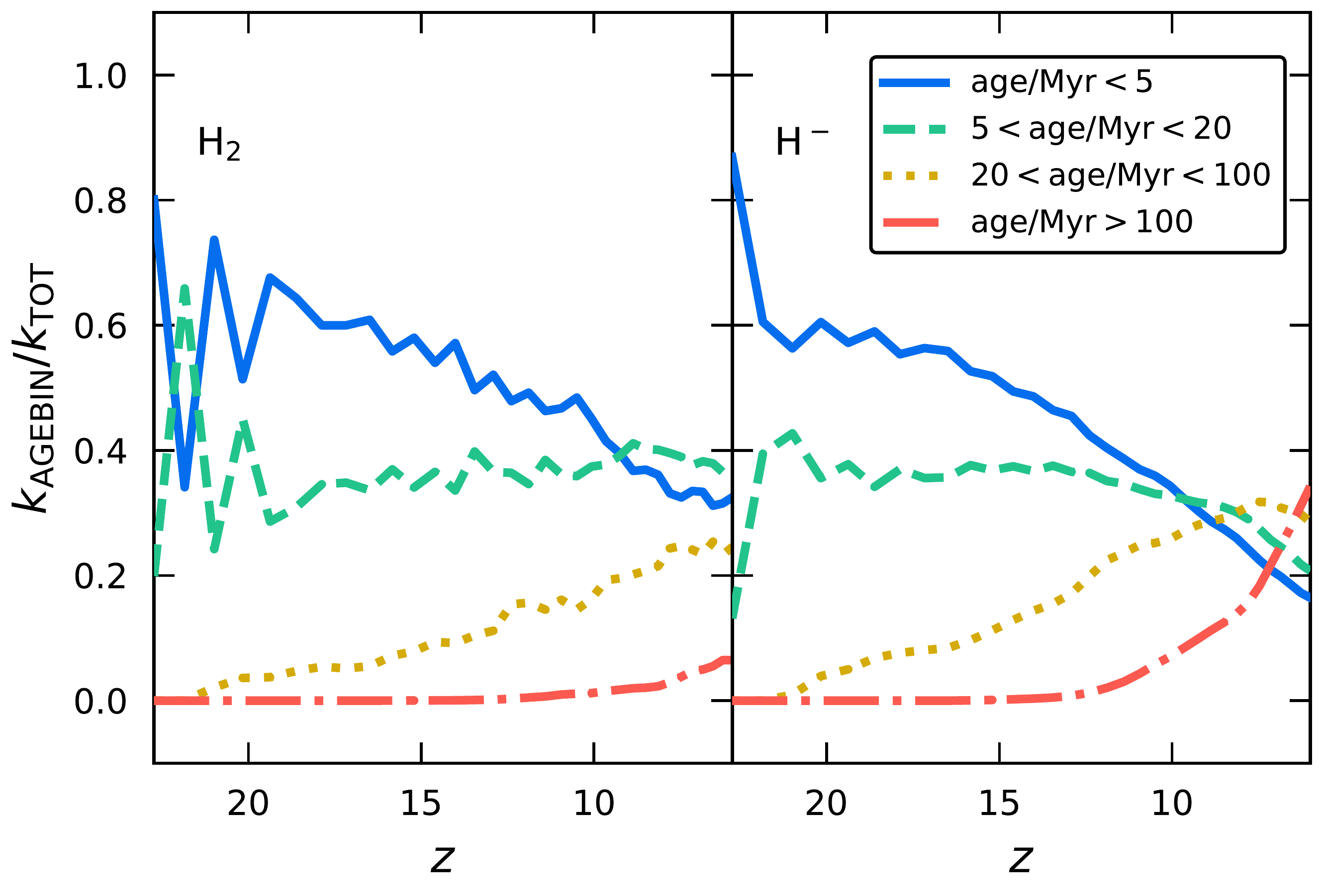}
    \vspace{-0.6cm}
    \caption{Fraction of the $\ce{H_2}$ dissociation rate (left panel) and $\ce{H^-}$ detachment rate (right panel) in the M simulation originated by newly-born stellar particles ($\mathrm{age} < 5 \ \mathrm{Myr}$, solid blue), young ($5 \ \mathrm{Myr} < \mathrm{age} < 20 \ \mathrm{Myr}$, dashed green), intermediate ($20 \ \mathrm{Myr} < \mathrm{age} < 100 \ \mathrm{Myr}$, dotted gold) and old stars ($\mathrm{age} > 100 \ \mathrm{Myr}$, dash-dotted red).}
    \vspace{-0.2cm}
    \label{fig:agebins}
\end{figure}

Beyond the distinction between metal-free PopIII and metal-poor PopII stars, it is commonly accepted in the literature that young stellar populations are the major contributors to the UV radiation field, as the short-lived massive stars dominate over the more abundant low-mass stars by several orders of magnitude, due to their hotter atmospheres and larger luminosities. We quantify this in Fig.~\ref{fig:agebins}, where we show the contribution from stellar populations with different ages, as concerns the $\ce{H_2}$ dissociation rate (left panel) and the $\ce{H^-}$ detachment rate (right panel).
In particular, we split the rates estimated from the M simulation into four bins depending on the stellar age: newly-born stellar particles ($\mathrm{age} < 5 \ \mathrm{Myr}$, solid blue), young ($5 \ \mathrm{Myr} < \mathrm{age} < 20 \ \mathrm{Myr}$, dashed green), intermediate ($20 \ \mathrm{Myr} < \mathrm{age} < 100 \ \mathrm{Myr}$, dotted gold) and old stars ($\mathrm{age} > 100 \ \mathrm{Myr}$, dash-dotted red).

Newly-born stars dominate both rates at early times, when PopIII star formation occurs at sustained rate. The contribution from young stars is approximately constant at all $z$ ($\sim 35-40\%$), while an increasing importance of older populations can be seen at $z<15$ and is $>20\%$ ($>30\%$) at $z<10$ for $\ce{H_2}$ dissociation ($\ce{H^-}$ detachment). The latter is actually dominated by stars older than 100 Myr during the latest stages of the simulation, while they never account for more than 5\%-10\% in the \ce{H_2} dissociation rate. Such a different behaviour is expected, as for an ageing stellar population the \ce{H_2} dissociation rate due to the emitted radiation drops much faster than the corresponding $\ce{H^-}$ detachment rate.\footnote{We will show a more in-depth analysis in a companion paper focused on the detailed calculation of the rates and their dependence on the spectral shape \textcolor{blue}{(Incatasciato et al., in prep.)}.}
We do not show here the dissociation of $\ce{H^+_2}$ as, with regard to this discussion, it qualitatively lies between $\ce{H_2}$ and $\ce{H^-}$: far UV photons at $\sim10$ eV contribute the most, but the energy threshold is well within the VIS and IR range ($\sim 0.5$ eV for the excited molecular states).

In conclusion, we confirm that young stellar populations (with $\mathrm{age}<20 \ \mathrm{Myr}$, in our treatment) are the major contributors to the UV radiation field at $z \geq 10$. The star formation rate history hence needs to be well modeled in order to estimate a realistic LWB. However, at $z\lesssim12$ the role of older stars cannot be neglected and at later times they even dominate over young stars in the $\ce{H^-}$ detachment rate.
The radiation background and its negative feedback on the star formation could then be underestimated if the contribution of older stellar populations is neglected, especially at $z\sim6-10$, when PopIII star formation episodes are mainly restricted to low density regions still marginally affected by metal enrichment (\citealt{Tornatore:2007,Maio:2010}, but see \citealt{Liu:2020}). These results are only mildly dependent on the choice of the IMF and spectra for PopIII and PopII stars. We report further discussions in Appendix~\ref{appen:SEDs}, where in particular we show that a bottom-heavy (top-heavy) IMF increases (decreases) the contribution from old stellar populations to up to 40\% (20\%) and 80\% (40\%) in the $\ce{H_2}$ dissociation and $\ce{H^-}$ detachment rate respectively.

\subsubsection{Connecting stellar mass densities and the LWB}
\label{sec:fits}

\begin{figure}
    \centering
    \includegraphics[width=\linewidth]{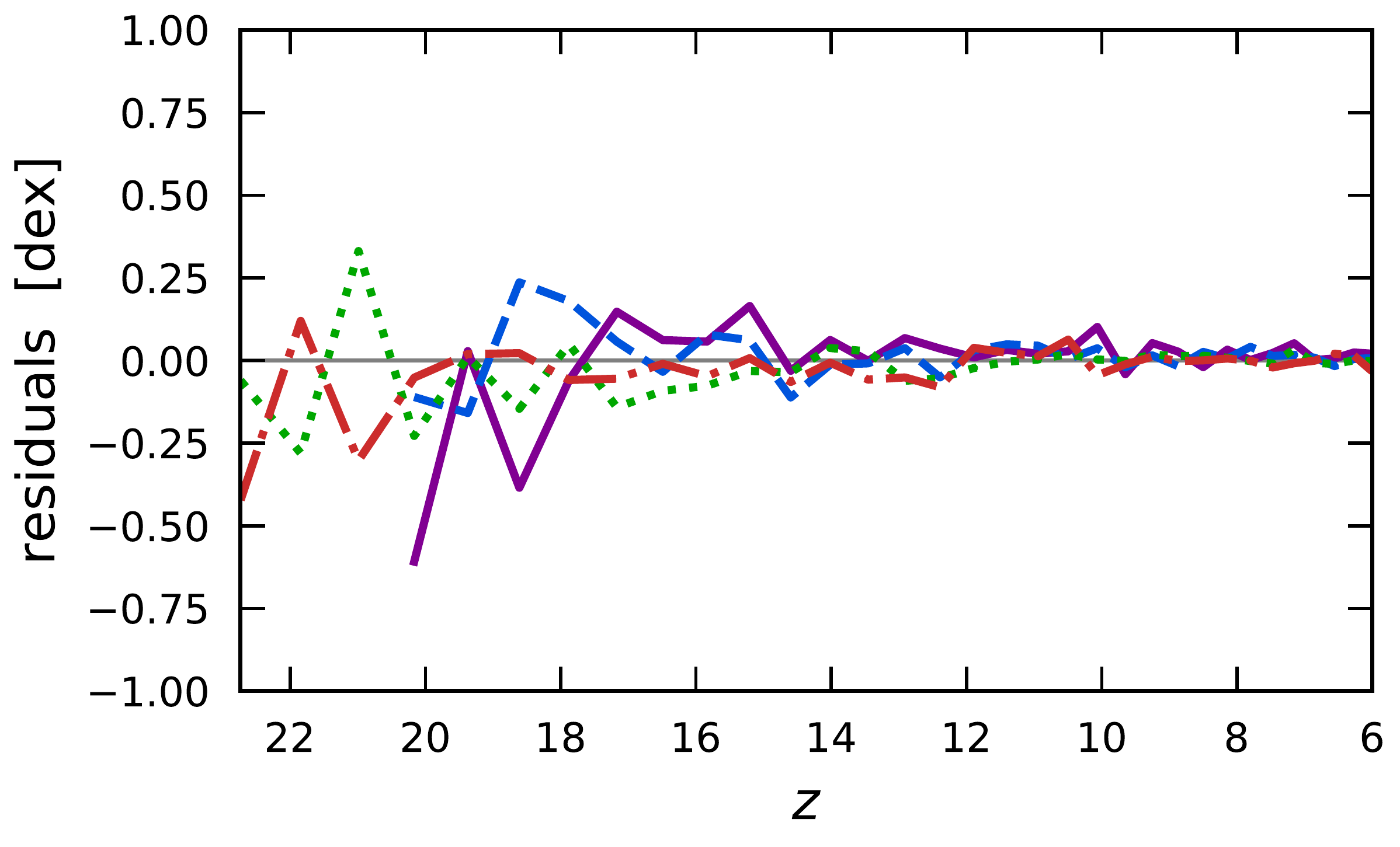}
    \vspace{-0.6cm}
    \caption{Residuals of the reconstructed LWB, based on the fit of the contribution from the different stellar populations modeled as in Eq.~\ref{eq:fit_LWB_agebins}, and the median LW radiation intensity obtained in post-processing from the FiBY simulations. The colour and line-style scheme follows Fig.~\ref{fig:SFRD_SFRthresh}. The reconstructed LWB closely follows the evolution of the mean LWB: it is always within 0.1 dex (25\%) at $z\lesssim17$ and only at higher redshift the residuals are as high as 0.3 dex (a factor of 2).}
    \vspace{-0.2cm}
    \label{fig:fit_LWB}
\end{figure}

In this paper we present a method to accurately determine the simulated radiation field in the Lyman-Werner energy range. In particular we describe its effect on the atomic and molecular gas by explicitly calculating the photochemical rates $k_{\ce{H_2}}$, $k_{\ce{H^-}}$ and $k_{\ce{H^+_2}}$ given the radiation emitted by all the stars formed in our simulations. The high computational cost of our algorithm makes it unfeasible to be used on-the-fly to self-consistently determine the LW background that develops in a cosmological simulation, during the formation of the first minihalos up to the Epoch of Reionisation. The relations in Equations \ref{eq:fit_FiBY_LWB} and \ref{eq:fit_ratios} provide an estimate of the three time-varying photochemical rates associated with the LW radiation; however, they rely on the physical processes included in the FiBY suite of simulations.

Alternatively, we present here a method to quickly reconstruct the LWB from the star formation history of a generic cosmological volume, hence making it independent from the specific predictions of FiBY on the formation and evolution of the early galaxy populations at $z>6$. In Section~\ref{sec:age_contrib} we split the LWB into four bins depending on the age of the stellar populations contributing to it. We proceed here along the same path. For PopIII and PopII stars individually, we consider the contribution from each bin ($k_{\ce{H_2},\mathrm{i}}$) and divide it by the comoving stellar density in that bin ($\rho_{\star,\mathrm{i}}$). We here consider only the density within the central box, thus neglecting the additional copies introduced to reach the LW horizon.\footnote{Despite this not being the most accurate procedure, it is the most straightforward to be applied on-the-fly in a cosmological simulation.}
By doing so, the values from all the FiBY simulations collapse onto the same relation, that can be modeled with Eq.~\ref{eq:fit_LWB_agebins}:
\begin{equation}
\label{eq:fit_LWB_agebins}
    \frac{k_{\ce{H_2},\mathrm{i}}}{\rho_{\star,\mathrm{i}}} = A + B(1+z)
\end{equation}
where the moderate dependence on the redshift mainly includes the impact of the varying IGM modulation factor, while $A$ and $B$ are free parameters in units of $\mathrm{cMpc}^3 \ \mathrm{M}_\odot^{-1} \ \mathrm{s}^{-1}$, evaluated with the MCMC fitting procedure of the \verb|emcee| library. The resulting parameters are listed in Table~\ref{Table:fit_LWB_params} for the \textquotesingle FID\textquotesingle \ choice of stellar SEDs.

\begin{table}
    \centering
    \renewcommand{\arraystretch}{1.5}
    \setlength{\heavyrulewidth}{1.5pt}
    \setlength{\abovetopsep}{4pt}
    \begin{tabular}{ cccc }
        \toprule
        SED & age & A & B \\
        \midrule
        \multirow{4}{*}{PopIII\_Ygg2} & newly-born & -2.782e-17 & 1.241e-17 \\
         & young & -5.425e-18 & 2.223e-18 \\
         & intermediate & -1.437e-18 & 2.718e-19 \\
         & old & -1.602e-19 & 2.151e-20\\
        \midrule
        \multirow{4}{*}{PopII\_BPASS\_Chab} & newly-born & 3.958e-18 & 3.841e-18 \\
         & young & -8.967e-18 & 2.761e-18 \\
         & intermediate & -2.673e-18 & 5.651e-19 \\
         & old & -5.222e-19 & 9.660e-20 \\
        \bottomrule
    \end{tabular}
    \caption{Parameters that reproduce the LWB intensity in the FiBY simulations, following Eq.~\ref{eq:fit_LWB_agebins}. Both parameters are in units of $\mathrm{cMpc}^3 \ \mathrm{M}_\odot^{-1} \ \mathrm{s}^{-1}$. PopIII and PopII stellar populations are split into four bins according to their age and their SEDs are the ones included in our \textquotesingle FID\textquotesingle \ choice. The result of the fit is shown in Fig.~\ref{fig:fit_LWB}.}
\label{Table:fit_LWB_params}
\end{table}

The contributions from all the bins of both PopIII and PopII stars have to be added up to obtain the total LWB. With these parameters we are able to reconstruct the mean LWB with a good precision. In Figure \ref{fig:fit_LWB} we show the residuals between the reconstructed and the mean LWBs, for the same FiBY simulations and the associated color scheme represented in Fig.~\ref{fig:LWB_FiBY_only}-\ref{fig:pop3_contrib}. The reconstructed LWB is consistently within 0.3 dex (a factor of 2) from the mean values and especially at $z\lesssim17$ is extremely close to it, within $20\%-25\%$ (0.1 dex). The poorer performance at high-$z$ can be motivated by the fact that the star formation rate density is still quite stochastic, hence it's harder to establish a strong correlation between the stellar density and the mean LW background intensity.

In conclusion, in Sec.~\ref{sec:LWB}-\ref{sec:age_contrib} we have described the LWB obtained with our postprocessing methods applied to the FiBY simulations. In particular Eqs.~\ref{eq:fit_FiBY_LWB}-\ref{eq:fit_ratios} and the relative parameters provide a simple fit to the three mean photochemical rates needed to determine the $\ce{H_2}$ content of the high-$z$ universe under the influence of a stellar LWB,  estimated directly from the FiBY simulations.
In addition, in this Section we have introduced a new way to approximate the LWB using the stellar density within any given simulated volume, under the only assumption that the stellar SEDs employed in this work are sensible enough to model the radiation emitted by PopIII and PopII stars.
Alternative (and more computationally expensive) methods, such as on-the-fly radiative transfer, exact or approximated such as in this work, can be therefore limited to small portions of the simulated volume in order to calculate the rare peaks of the LWB \citep[e.g.][]{Lupi:2021}.

\subsection{Spatial inhomogeneities}
\label{sec:histogs}

In this Section we explore the spatial inhomogeneities of the LW radiation background beyond the mean value. Despite the very long mean free path of LW photons, in fact, the LW intensity is unavoidably influenced by the spatial distribution of galaxies and of the underlying dark matter field, with correlation lengths of a few Mpc $\mathrm{h}^{-1}$ as studied over a wide redshift range \citep[see e.g.][]{Iliev:2003,Adelberger:2005,Guzzo:2014}.

\begin{figure}
    \centering
    \includegraphics[width=\linewidth]{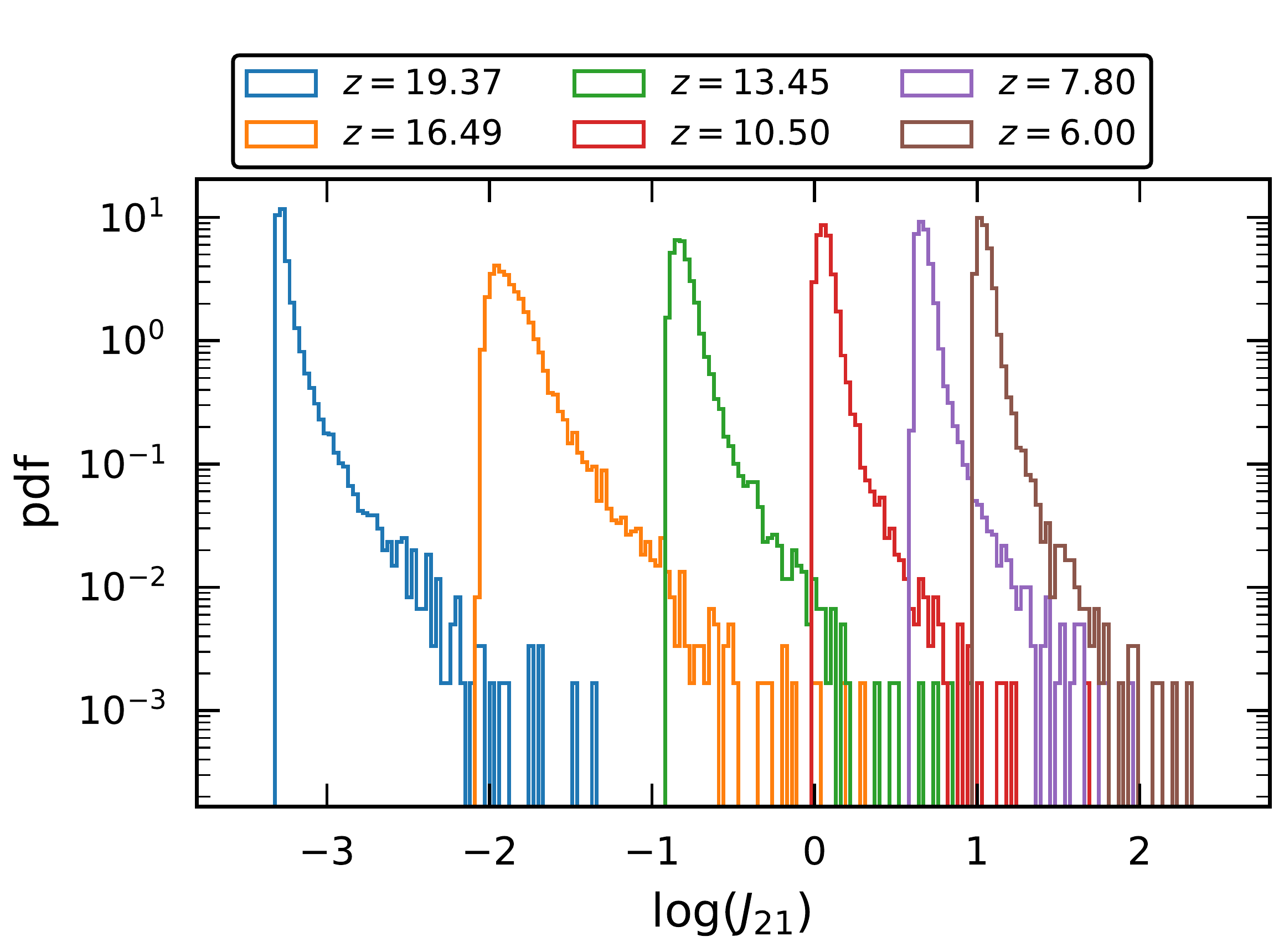}
    \vspace{-0.6cm}
    \caption{Probability density distribution (pdf) of the \ce{H_2} dissociation rate, expressed in terms of LW radiation intensity $J_{21}$, for the XL simulation and for six different redshifts (from $z\sim19$ to $z=6$). The distributions are highly right-skewed, with long tails that extend up to two orders of magnitude above the pdf peaks.}
    \vspace{-0.2cm}
    \label{fig:histogs}
\end{figure}

In Fig.~\ref{fig:histogs} we show the probability distribution function of the $\ce{H_2}$ dissociation rate, expressed in terms of the LW radiation intensity $J_{21}$, as determined with our postprocessing method in randomly selected points within the XL simulation at six different redshifts (from $z\sim19$ to $z=6$). We choose the largest volume available in order to include the largest cosmological structures simulated in the FiBY suite, that instead are less likely to be found in the smaller simulations.

At all redshifts the \ce{H_2} dissociation rate shows a pronounced right-skewed distribution, with a long tail that can extend up to two orders of magnitude above the peak of the distribution, while the minimum is always very close to it. Fig.~\ref{fig:histogs} qualitatively suggests also that the distribution becomes more and more narrow from $z=13$ to $z=6$. The redshifts shown in Fig.~\ref{fig:histogs} are approximately the same as in \citet[their Figure 11, left panel]{Ahn:2009}, with the exception of $z=6$ (their simulation stopped at $z=7.8$). As discussed in Section~\ref{sec:conclusion}, our mean LWB at low-$z$ is systematically lower than what they find, but the distribution at each $z$ is consistent with theirs and with the one shown in \cite{Dijkstra:2008}.

The LW radiation field extracted from the other FiBY simulations shows a similar distribution, but with a smaller scatter. This is not surprising, given that the smaller volumes can resolve the ubiquitous low-mass halos contributing to the overall LWB, but do not contain enough dense regions where we expect to find the intensity peaks. Finally, we find similar distributions for $k_{\ce{H^-}}$ and $k_{\ce{H^+_2}}$ as well, but with smaller spatial variations with respect to the \ce{H_2} dissociation rate. We motivate this with the lower IGM absorption associated with these rates (Fig.~\ref{fig:fmod_fit_3reac}), that enhances the contribution from sources further away and decreases the importance of the inhomogeneous distribution of galaxies at small scales.

\begin{figure}
    \centering
    \includegraphics[width=\linewidth]{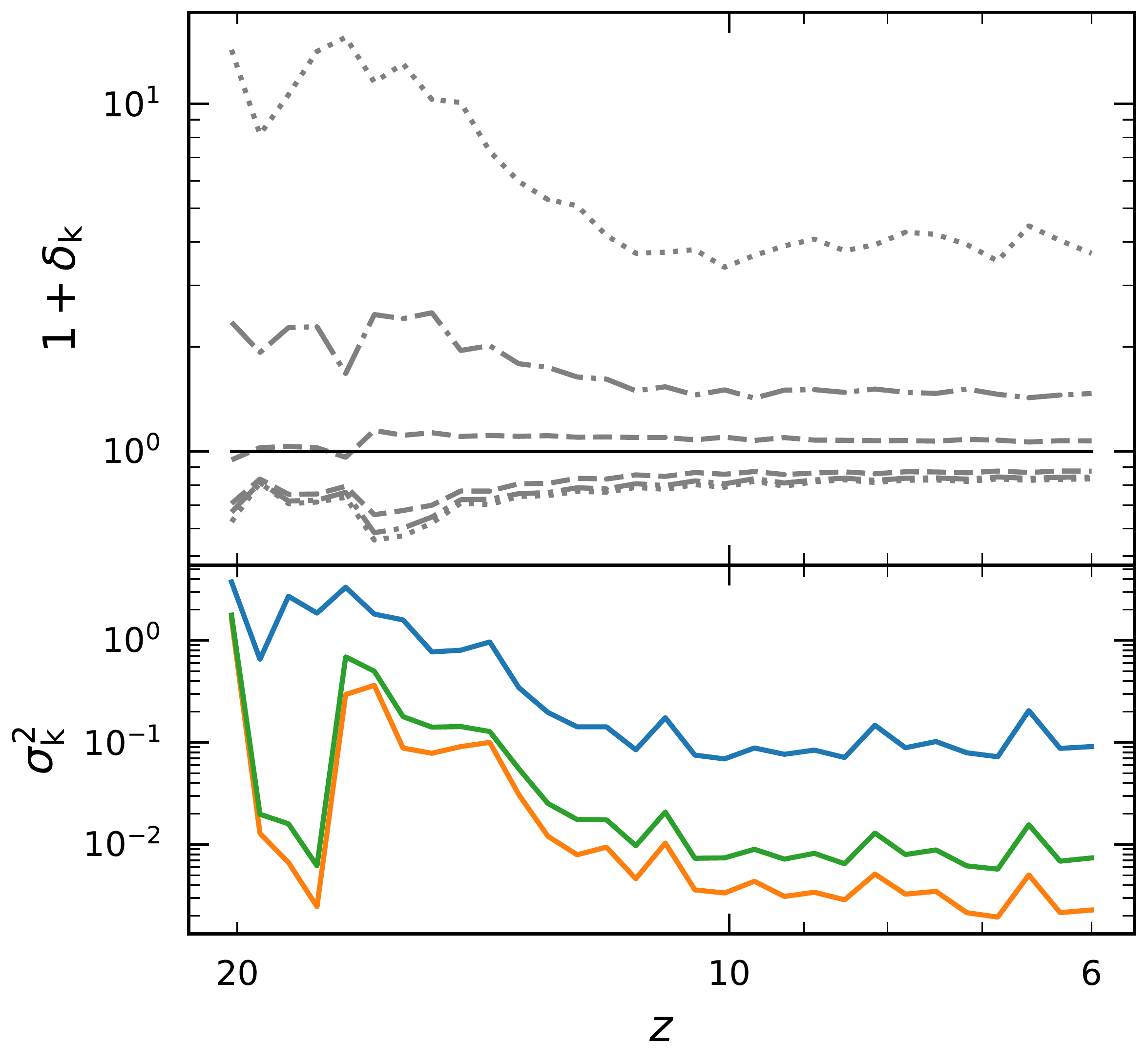}
    \vspace{-0.6cm}
    \caption{Quantitative analysis of the \ce{H_2} dissociation rate pdf shown in Fig.~\ref{fig:histogs}. \textbf{\textit{Top panel}}: the evolution of the deviation from the mean, expressed as $1+\delta_{\mathrm{k}}$, $\delta_{\mathrm{k}}$ being $(k - \langle k \rangle) / \langle k \rangle$. The lines represent the 68.3\% (dashed), 95.9\% (dash-dotted) and 99.7\% (dotted) contours in the XL simulation. \textbf{\textit{Bottom panel}}: the variance $\sigma^2_{\mathrm{k}} = \langle \delta^2_{\mathrm{k}} \rangle$ of the three photochemical rates ($k_\mathrm{\ce{H_2}}$, blue, $k_\mathrm{\ce{H^-}}$, orange, and $k_\mathrm{\ce{H^+_2}}$, green): the latter two have a lower variance, mainly due to the lower IGM absorption in the relevant energy range.}
    \vspace{-0.2cm}
    \label{fig:deltaJ}
\end{figure}

We continue this analysis in Fig.~\ref{fig:deltaJ}, where the deviation from the mean is expressed as $1 + \delta_\mathrm{k}$, with $\delta_\mathrm{k}=(k - \langle k \rangle) / \langle k \rangle$. In the top panel we show the evolution of the 68.3\% (dashed line), 95.9\% (dash-dotted) and 99.7\% (dotted) contours in the XL simulation for the $\ce{H_2}$ dissociation rate. All the lines approach the mean at lower redshift, reflecting the fact that the radiation field becomes more and more homogeneous at later times, when even the most remote and underdense regions receive the photons emitted in the large volume comprised in the LW horizon. The LWB has a minimum (lower dotted line) that is always within a factor of two from the mean ($\delta\sim-0.5$ at $z\sim17$, but $\delta>-0.8$ at $z<12$), while the maximum (upper dotted line) is $\delta\sim10$ at early times and decreases to $\delta\sim2-3$ at $z<12$. Our estimates are in good agreement with the right panel of Figure 11 of \cite{Ahn:2009}, despite our simulated volume being $\sim3.5$ times smaller.

In the bottom panel of Fig.~\ref{fig:deltaJ}, instead, we include all three photochemical rates ($\ce{H_2}$ in blue, $\ce{H^-}$ in orange and $\ce{H^+_2}$ in green) to show how their variance, defined as $\sigma^2_\mathrm{k}=\langle\delta^2_\mathrm{k}\rangle$, evolves with $z$. All the rates show a similar decreasing trend with decreasing redshift, but $\ce{H^+_2}$ and $\ce{H^-}$ have a variance that is $1-2$ orders of magnitude lower than $\ce{H_2}$.

\subsubsection{Local contribution}
\label{sec:local_contrib}

The tail of the LW intensity distribution shown in Fig.~\ref{fig:histogs}  requires more attention, due to importance of the highest peaks of the LW radiation field in the most common theoretical models of formation of massive black hole seeds \citep{Agarwal:2012,Fernandez:2014,Lupi:2021,Sassano:2021}. These are regions where the contribution from one or few galaxies dominates over the homogeneous background \citep[e.g.][]{Agarwal:2014,Wise:2019,Spinoso:2023}.

\begin{figure}
    \centering
    \includegraphics[width=\linewidth]{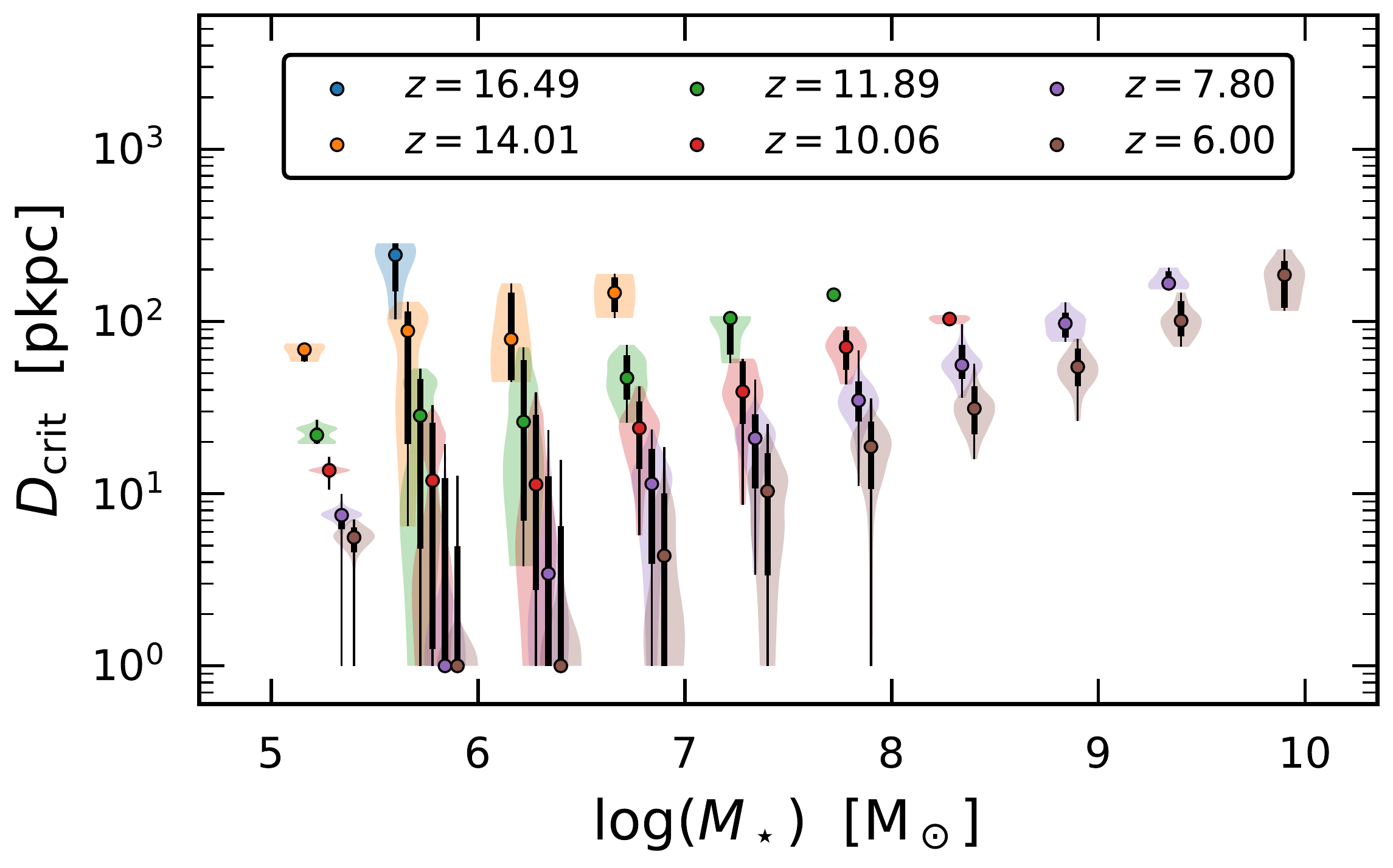}
    \vspace{-0.6cm}
    \caption{\textquotesingle Critical distance\textquotesingle \ within which the radiation coming from a galaxy is higher than the mean LWB, for the XL simulation and the \textquotesingle FID\textquotesingle \ choice of SEDs. Galaxies are split into 0.5-dex-wide bins according to their stellar mass. The dot at the centre of the violin plot shows the median value, the thick black line the $10\%-90\%$ percentiles and the thin black line the minimum and maximum value of $D_\mathrm{{crit}}$ for each bin. For a given stellar mass bin, the violin plots are horizontally displaced for visualisation purposes only. Note that $D_\mathrm{{crit}}$ is expressed in \textit{proper} kpc.}
    \vspace{-0.2cm}
    \label{fig:local_contrib_SM}
\end{figure}

We quantify the size of these regions in Fig.~\ref{fig:local_contrib_SM}: for each galaxy in the XL simulation we calculate the distance at which the \ce{H_2} dissociation rate due to the radiation emitted by the galaxy itself is equal to the mean LW background. The galaxies are then grouped according to their stellar mass, with each bin being 0.5 dex wide. The resulting violin plot shows this \textquotesingle critical distance\textquotesingle \ ($D_\mathrm{crit}$) as a function of the stellar mass, where the shaded region represents the distribution of $D_\mathrm{crit}$ in each bin and the median, $10\%-90\%$ and min-max values are shown by the circles, thick and thin black lines respectively. The minimum distance considered is 1 \textit{physical} kpc, to include only the region outside the virial radius of the galaxies.

At fixed redshift, the median $D_\mathrm{crit}$ increases with the galaxy stellar mass. Massive galaxies dominate over the LW background at distances as large as 100 pkpc even at $z=6$, when the LWB has reached $J_{21}\sim10$. Low-mass galaxies, instead, show large variations at any redshift, shown by the shaded region of the violins, as the emitted UV radiation strongly depends on the particular star formation history of each galaxy \citep{Lee:2009}. For a fixed stellar mass, the decreasing trend with decreasing $z$ (e.g., from 100 kpc at $z=14$ to 4 kpc at $z=6$ for galaxies with $6.5 < \log(M_\star/M_\odot) < 7$) is in first order explained with the evolution of the LW mean intensity (from $J_{21}=0.1$ to $J_{21}=10$). Once the latter is taken into account, $D_\mathrm{crit}(M_\star,z)$ collapses into a single $\mathcal{D}(M_\star)$, defined as in Eq.~\ref{eq:normalise_Dcrit}, and fitted by the relation in Eq.~\ref{eq:fit_Dcrit}:

\begin{gather}
\label{eq:normalise_Dcrit}
    D_\mathrm{crit}(M_\star,z) = \mathcal{D}(M_\star) \times J_{21}(z)^{-1/2}\\
\label{eq:fit_Dcrit}
    \log(\mathcal{D}(M_\star)) = A + B\log(M_\star) + C\log^2(M_\star)
\end{gather}

where $M_\star$ is in $\mathrm{M}_\odot$, $\mathcal{D}$ is in pkpc, $A=1.008$, $B=1.890\times10^{-1}$ and $C=1.519\times10^{-2}$. Our aim here is to complement the LW modelling discussed in Sec.~\ref{sec:LWB} (Eqs.~\ref{eq:fit_FiBY_LWB}-\ref{eq:fit_ratios}-\ref{eq:fit_LWB_agebins}) to include the spatial fluctuations beyond the homogeneous approximation. Our results (see e.g. Fig.~\ref{fig:histogs}, consistent with comparable works in the literature such as \citealt{Dijkstra:2008} and \citealt{Ahn:2009}), indicate that the long tail of high $J_{21}$, well above the mean LWB, is the effect of the radiation emitted by close luminous galaxies, that dominate over the homogeneous radiation field within radii of the order of $D_\mathrm{crit}$, described with Eq.~\ref{eq:normalise_Dcrit} and Eq.~\ref{eq:fit_Dcrit}. This represents an easy-to-use recipe to include spatial inhomogeneities in the LW radiation on-the-fly, while a simulation is performed, by focusing such calculations only to radii smaller than $D_\mathrm{crit}$. Its only limitation is that we observe a slight evolution with redshift (see Fig.~\ref{fig:Dcrit_normalised}): $\mathcal{D}(M_\star)$ decreases by $\sim0.3$ dex with decreasing $z$, as the UV emissivity per stellar mass changes due to the progressive shift of the dominant stellar population from PopIII to PopII.

Here we have not been considering $\ce{H^+_2}$ and $\ce{H^-}$ rates. Given the results presented in Section~\ref{sec:histogs}, the small spatial variations of these two rates do not require any further analysis. Fig.~\ref{fig:local_contrib_SM} can be considered as a very safe upper limit for them as well.



\subsection{Negative feedback of the LW radiation}
\label{sec:minhmass}

\subsubsection{Minimum halo mass for PopIII star formation}
\label{sec:minhmass_literature}

\begin{figure}
    \centering
    \includegraphics[width=\linewidth]{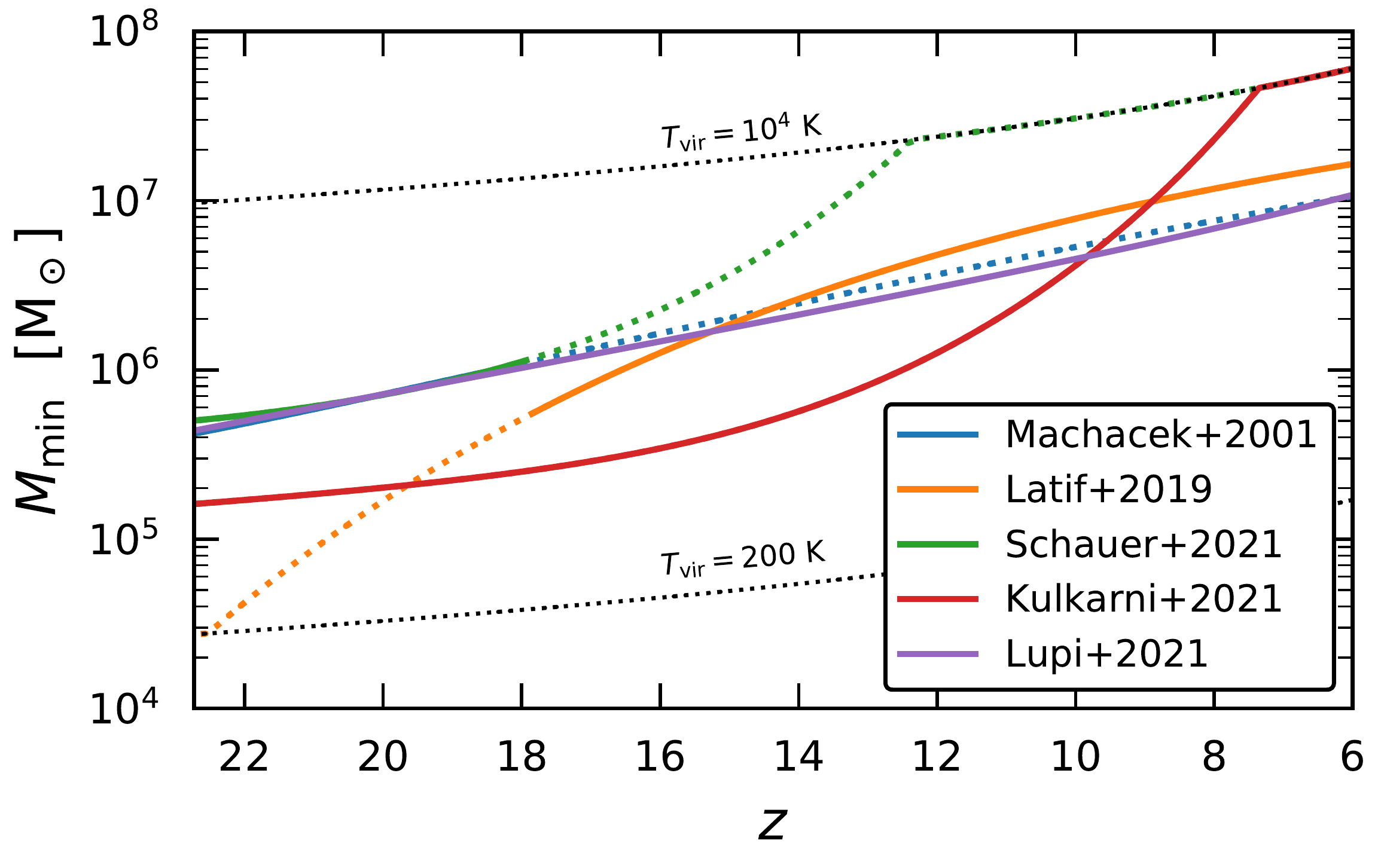}
    \vspace{-0.6cm}
    \caption{Compilation of estimates of the minimum halo mass for PopIII star formation from the literature: \protect\cite{Machacek:2001,Latif:2019,Schauer:2021,Kulkarni:2021,Lupi:2021}. $M_\mathrm{min}$ in general depends on the LWB intensity at a given $z$: in the respective analytical formulae we have been using the LWB obtained in this work, as described in Eq.~\ref{eq:fit_FiBY_LWB}, in the \textquotesingle FID\textquotesingle \ case. The coloured dotted lines indicate the extrapolation of $M_\mathrm{min}$ outside the corresponding range of $J_{21}$ investigated by the authors. The black dotted lines show $M_\mathrm{h}$ for $T_\mathrm{vir}=10^4 \ \mathrm{K}$, at the top, and $T_\mathrm{vir}=200 \ \mathrm{K}$, at the bottom \protect\citep{Bromm:2011}.}
    \vspace{-0.2cm}
    \label{fig:minMh_SFR_literature_M_FID}
\end{figure}

As already mentioned in Sec.~\ref{sec:introduction}, even a moderate intensity of LW radiation (as low as $J_{21}\sim10^{-2}$, \citealt{Haiman:1997}) can delay or even prevent star formation in low-mass molecular-cooling halos with virial temperature between 200 K and 10$^4$ K. For a given LW intensity, the minimum halo mass required to overcome this negative feedback, allowing the gas to increase its \ce{H_2} abundance and subsequently form stars, can be estimated with both analytical arguments and high-resolution simulations.

Using the LWB derived from the FiBY simulations, we predict the associated minimum halo mass as suggested by \cite{Machacek:2001,Latif:2019,Schauer:2021,Kulkarni:2021,Lupi:2021} in Fig.~\ref{fig:minMh_SFR_literature_M_FID}. The latter provides an analytical estimate where halos with $M_\mathrm{h} = M_\mathrm{min}$ experience enough \ce{H_2} cooling rate to ensure a cooling time comparable to the Hubble time, and the \ce{H_2} abundance is given by the equilibrium between formation through the \ce{H^-} channel and the LW dissociation. The other references, instead, investigate how $M_\mathrm{min}$ depends on the LW intensity (and other relevant factors, such as baryonic streaming \citealt{Tseliakhovich:2010,Schauer:2021}) by exploring the parameter space with a large number of high-resolution cosmological simulations. This also sets the range of validity for their relations with respect to $J_{21}$. We here employ the homogeneous LWB obtained in Sec.~\ref{sec:LWB}, as described in Eq.~\ref{eq:fit_FiBY_LWB}. The dotted lines show the extrapolation needed when the LWB is below or above the range of validity of each reference.

If we restrict ourselves to the solid lines, we can observe a general concordance among the authors, that set the minimum halo mass for PopIII star formation at a few times $10^5 \ \mathrm{M}_\odot$ at $z\sim20$ ($J_{21}\sim10^{-2}$), increased to up to $2\times10^7 \ \mathrm{M}_\odot$ at $z\lesssim10$ ($J_{21}\gtrsim1$). Only \cite{Kulkarni:2021} shows a different normalisation at $z\geq15$ and a different evolution at $z\leq15$ (red solid line). This can be explained with the fact that, in addition to the direct dependence on $J_{21}(z)$, they find a stronger explicit dependence on the redshift - $M_\mathrm{min}(J_{21}=\mathrm{const}) \propto(1+z)^{1.64(1+J_{21})^{0.36}}$ - that is not found in other works.

One important limitation of these studies is that they do not include \ce{H^+_2} dissociation and \ce{H^-} detachment rates in their chemical networks, with the exception of \cite{Latif:2019}, who consider only the latter. Another caveat is that in \cite{Machacek:2001,Latif:2019,Schauer:2021,Kulkarni:2021} the homogeneous LW intensity is kept constant throughout the simulations, while as we show in Fig.~\ref{fig:LWB_FiBY_only} the LWB grows by 3-4 orders of magnitude during the first billion years after the Big Bang. We plan to address these limitations in a future work, that will estimate the minimum halo mass for PopIII star formation under the influence of the LWB obtained in this work.

\subsubsection{Effect on molecular-cooling halos in FiBY}
\label{sec:minhmass_fiby}

\begin{figure}
    \centering
    \includegraphics[width=\linewidth]{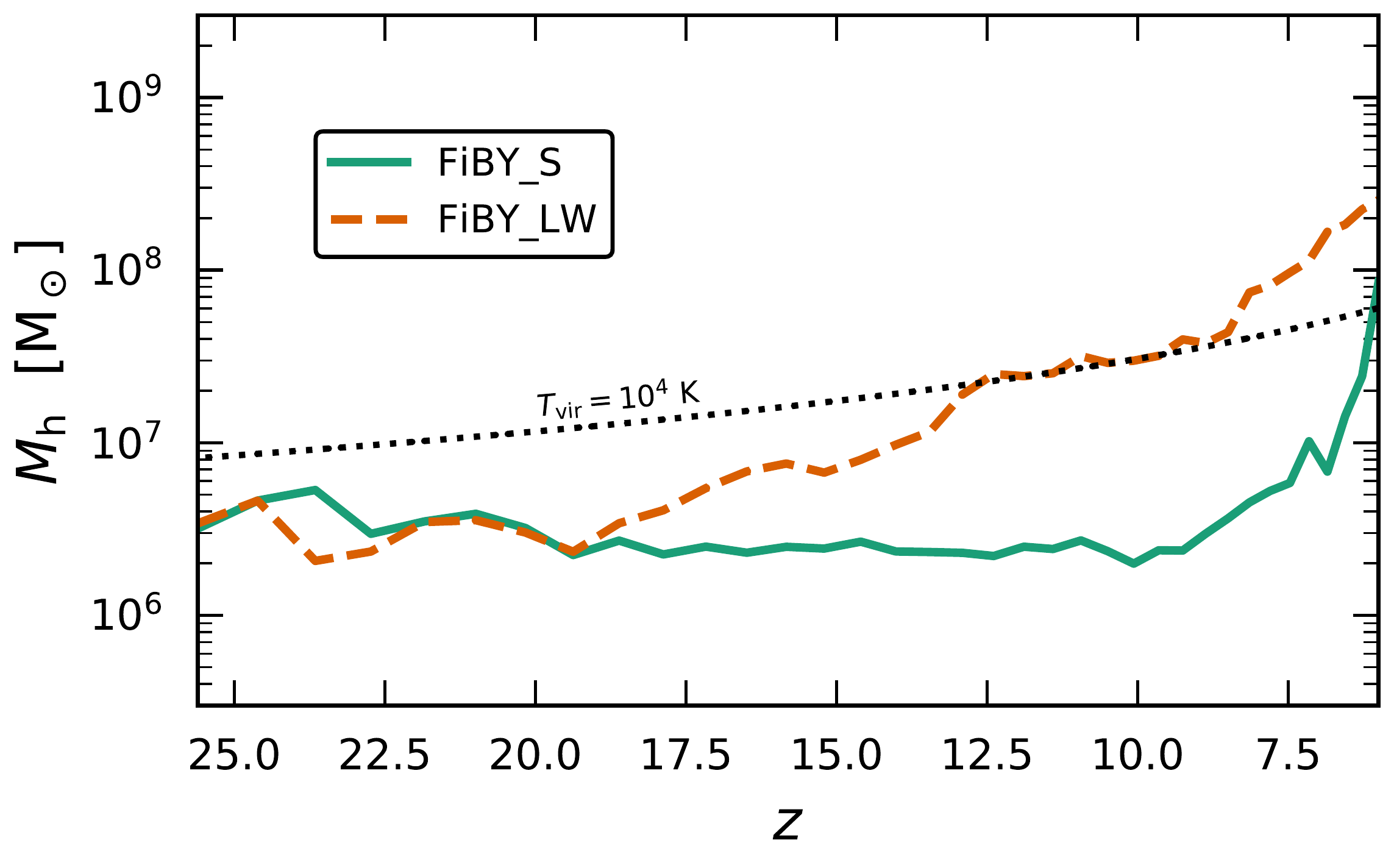}
    \vspace{-0.6cm}
    \caption{Minimum mass of star-forming halos in the simulations with the highest resolution: FiBY\_S (solid green) and FiBY\_LW (dashed orange). The halos considered are the ones with at least one gas particle tagged as star forming in the halo catalog. The black dotted line shows $M_\mathrm{h}$ for $T_\mathrm{vir}=10^4 \ \mathrm{K}$ \citep{Bromm:2011}.}
    \vspace{-0.2cm}
    \label{fig:minMh_SFR_Sboxes}
\end{figure}

The FiBY suite of simulations offers an optimal setup to study PopIII star formation in molecular cooling halos at $z\geq10$, as it includes the essential physical processes needed to simulate their dynamical evolution and, in particular, it employs a basic chemical network to track the formation of \ce{H_2} through \ce{H^-} at moderate densities. The simulation with the highest level of resolution (the S box) resolves low-mass halos with $\sim3\times10^5-10^6 \ \mathrm{M}_\odot$, significantly below the atomic-cooling limit at $10^7-10^8 \ \mathrm{M}_\odot$ \citep{Bromm:2011}. Additionally, FiBY\_LW (see the last row in Table~\ref{Table:FiBYruns}) couples the chemical network with the \ce{H_2} dissociation rate due to LW radiation, calculated on-the-fly with a homogeneous component, proportional to the global star formation rate, and the spatial fluctuations due to the local distribution of young stellar populations.

In Fig.~\ref{fig:minMh_SFR_Sboxes} we show the minimum mass of the halos that are experiencing star formation (i.e., where at least one gas particle is tagged as star-forming) in each snapshot, as measured in FiBY\_S (solid green) and FiBY\_LW (dashed orange).

FiBY\_S does not include any LW radiation, hence the minimum mass $\sim10^6 \ \mathrm{M}_\odot$ is approximately constant at $z\geq10$. This value is $\sim5$ times higher than the mass of the smallest halos considered in the creation of the halo catalogs and substantially higher than the halo mass corresponding to a virial temperature of 200 K (black dotted line in the lower part of Fig.~\ref{fig:minMh_SFR_Sboxes}), the lowest temperature at which $\ce{H_2}$-cooling is efficient. Besides the limitations imposed by the resolution, hydrodynamical effects such as pressure support from turbulence \citep{Latif:2022b} and dynamical heating due to intense accretion flows \citep{Fernandez:2014} can delay star formation even when the LW radiation is not included \citep{Regan:2022}. On the other hand, FiBY\_LW shows a very clear evolution with redshift, differentiating from FiBY\_S at $z\lesssim20$ due to the impact of the LW radiation on the dynamical evolution of molecular-cooling halos. $M_\mathrm{min}$ grows from $2\times10^6 \ \mathrm{M}_\odot$ at $z=20$ ($J_{21}\sim0.05$, if we consider the \textquotesingle BB\textquotesingle \ case that reproduces very closely the original LW calculation in FiBY) to $3\times10^7 \ \mathrm{M}_\odot$ at $z=13$ ($J_{21}\sim0.2$). Afterwards, stars form only in atomic-cooling halos, indicated with the upper black dotted line corresponding to $T_\mathrm{vir}=10^4 \ \mathrm{K}$.

At $z\leq10$, $M_\mathrm{min}$ in FiBY\_S rapidly increases up to (and eventually above, at $z<6$) the atomic-cooling limit, even in absence of a LWB, due to the ionising UV background. An increase can also be seen in FiBY\_LW, despite it being already at $10^4 \ \mathrm{K}$: the reason can be traced back to the large amount of stellar feedback that follows the sudden increase of the global star formation rate at $z\sim11$ (see Figure 1 of \citealt{Johnson:2013}, and a very similar trend has been found in \textsc{ramses-rt} simulations by \citealt{Sarmento:2022}).

Only stars younger than 5 Myr are considered for the on-the-fly calculation of the LWB in the FiBY\_LW simulation. We have shown in Sec.~\ref{sec:age_contrib} that these stars give the largest contribution to the \ce{H_2} dissociation rate, but never account for more than 60\%-70\% of the total rate (and this number is even lower for \ce{H^-} detachment). Thus, the effect of the LWB in \cite{Johnson:2013} and here in Fig.~\ref{fig:minMh_SFR_Sboxes} can be considered as a conservative estimate of the impact of the LW radiation in delaying star formation in low-mass halos.

\section{Summary and discussion}
\label{sec:conclusion}

This work is aimed at estimating the evolution of the LW radiation field at $6<z<25$, with the use of the FiBY suite of high-resolution and physics-rich cosmological simulations described in Sec.~\ref{sec:fiby}. To do so, we accurately calculate the three photochemical rates needed to model the abundance of \ce{H_2} molecules, that represent the primary cooling channel of gas in the high-$z$ universe: the \ce{H_2} and \ce{H^+_2} dissociation and the \ce{H^-} detachment (Sec.~\ref{sec:rates}). The radiation is emitted by all the stellar sources in the simulated volumes (Sec.~\ref{sec:SEDs}) and we also account for the IGM optical depth beyond the approximate treatment of \cite{Haiman:2000} and \cite{Ahn:2009}, as reported in Sec.~\ref{sec:fmod}.
We here present a summary of our findings and discuss them in the broader context of cosmological structures formation during the first billion years after the Big Bang.

\begin{figure}
    \centering
    \includegraphics[width=\linewidth]{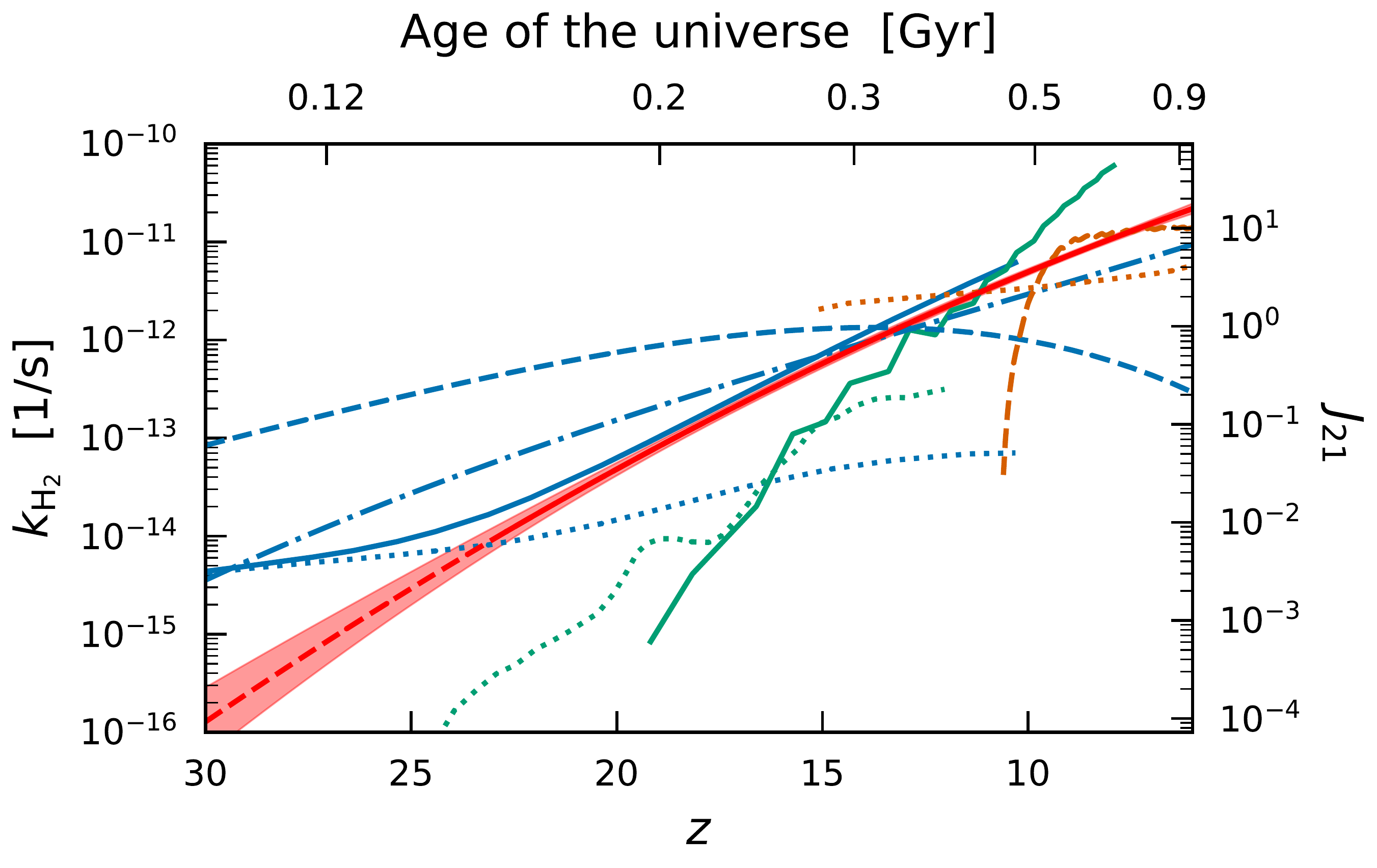}
    \vspace{-0.6cm}
    \caption{The LWB derived from the FiBY simulations, described with the fit in Eq.~\ref{eq:fit_FiBY_LWB}, is shown with the red solid line in the redshift range of the simulations ($6<z<23$) and extrapolated to $z=30$ with the red dashed line. Alongside our estimates, we report a number of LWB models available in the literature: \protect\citet[dotted blue, fiducial, and solid blue, with an external radiative field]{Trenti:2009}, \protect\citet[dashed blue]{Wise:2005}, \protect\citet[dash-dotted blue]{Qin:2020}, \protect\citet[solid green]{Ahn:2009}, \protect\citet[dotted green, Renaissance \textquotesingle Normal\textquotesingle \ box]{Xu:2016}. The dashed and dotted orange lines represent the \ce{H_2} dissociation rate derived by \protect\cite{Smith:2015} from the UV background of \protect\cite{Faucher-Giguere:2009} and \protect\cite{Haardt:2012} respectively.
    The y-axis refers to the \ce{H_2} dissociation rate on the left and the corresponding $J_{21}$ LW intensity on the right, as already in Fig.~\ref{fig:LWB_FiBY_only}.}
    \vspace{-0.2cm}
    \label{fig:LWB_with_lit}
\end{figure}

\begin{itemize}

    \item The mean LW intensity (Fig.~\ref{fig:LWB_FiBY_only}) grows from $J_{21}\sim10^{-2}$ at $z\sim23$ to $J_{21}\sim10$ at $z\sim6$ in the FiBY simulations that have enough resolution to resolve star formation in $\sim10^6-10^7 \ \mathrm{M}_\odot$ halos (M and S). Our predicted LWB is strong enough to delay PopIII star formation in low-mass \ce{H_2}-cooling halos \citep{Machacek:2001,Wise:2007,O'Shea:2008,Latif:2019,Schauer:2021,Kulkarni:2021,Lupi:2021,Park:2021}, but on average is a few orders of magnitude below the intensity needed for the formation of massive black hole seeds at $z\gtrsim10$, broadly located between $J_{21}\sim10$ and $J_{21}\sim10^4$, as shown by \citealt{Sugimura:2014,Agarwal:2016,Wolcott-Green:2017}, with large uncertainties due to different treatments of the gas chemistry \citep{Glover:2015a,Glover:2015b}, $\ce{H_2}$ self-shielding \citep{Wolcott-Green:2011,Hartwig:2015a,Wolcott-Green:2019} and the radiation spectral shape \citep{Latif:2015}. This suggests that close proximity to an intense LW source is needed for haloes to yield direct collapse (see also e.g. \citealt{Agarwal:2019}).
    
    \item In Fig.~\ref{fig:LWB_with_lit} we complement our results with a number of LW background models available in the literature: blue lines indicate semi-analytical models, such as \citet[dotted, fiducial, and solid, with an external radiative field]{Trenti:2009}, \citet[dashed]{Wise:2005} and \citet[dash-dotted]{Qin:2020}, while green lines are models obtained from cosmological simulations, either in post-processing on top of a dark-matter only simulation, as \citet[solid]{Ahn:2009} or on the fly, as the Renaissance \textquotesingle Normal\textquotesingle \ box in \citet[dotted]{Xu:2016}. The orange lines represent the LWB derived from the UV backgrounds of \citet[dashed]{Faucher-Giguere:2009} and \citet[dotted]{Haardt:2012}, as tabulated in the Grackle astrochemistry library \citep{Smith:2015}. Our mean LWB can be expressed analytically with the simple second-order polynomial in Eq.~\ref{eq:fit_FiBY_LWB}, shown with the red solid line (and extrapolated to $z=30$ with the dashed line).
    Studies in the literature have conflicting predictions for the evolution of the LWB, depending on the specific methods and parameters employed. \cite{Ahn:2009} predicts a late and steep build-up and generally shows a similar evolution to the FiBY XL simulation (the purple line in Fig.~\ref{fig:LWB_FiBY_only}), as expected since they resolve only atomic-cooling halos with $M_\mathrm{h}\gtrsim10^8 \ \mathrm{M_\odot}$ in their treatment. Their LWB, however, is systematically a factor of $3-10$ higher than ours and increases rapidly at $z\lesssim10$, while all the other models suggest a milder evolution. We have verified that this mismatch can be explained with the higher star formation rate predicted by their analytical model painted on top of the dark matter halos (that can be estimated from the emission coefficient in their Figure 7), while their assumptions for the stellar emission in terms of LW photons per stellar baryon are consistent with our \textquotesingle FID\textquotesingle \ SEDs choice (Fig.~\ref{fig:SEDs_LW_emission}). The LWB from the Renaissance simulation \citep[][green dotted line]{Xu:2016}, instead, is constantly at least one order of magnitude below our estimates, despite a comparable mass resolution. Their treatment of the sources outside the simulation box is consistent with ours, while the different assumptions in terms of PopIII stellar emissions and more importantly our updated treatment of the IGM optical depth (see Sec.~\ref{sec:fmod}) can partially explain the large difference. We have also verified that the stellar mass functions in the two simulations differ quite considerably in the low-mass end ($M_\star\sim10^3-10^6 \ \mathrm{M}_\odot$), as in the Renaissance \textquotesingle Normal\textquotesingle \ simulation the LWB is calculated on-the-fly and included in the evolution of the \ce{H_2} abundance. Our LWB is reasonably consistent with semi-analytical models of \cite{Trenti:2009} (blue solid line, the model where an external radiation field due to PopII stars is added to artificially match a realistic reionisation history) and \cite{Qin:2020}, both in terms of normalisation and evolution with redshift, confirming the robust results of the FiBY model for PopIII star formation in low-mass halos and our sensible choice of stellar emission. \cite{Wise:2005}, on the other hand, predicts a LWB that is more than one order of magnitude higher than the FiBY at $z\sim20$, possibly due to the different choices of PopIII IMF in their model. The decreasing evolution at later times disagrees with our results and all the other models in the literature and is subject to large uncertainties due to the choices on the star-formation efficiency and ionising photon escape fraction \citep{Xu:2016}. Finally, the UV background of \cite{Faucher-Giguere:2009} provides a steeply-increasing LWB that is only consistent with ours at $z\lesssim10$, and \cite{Haardt:2012}-derived LWB only increases by a factor of 2 between $z=15$ and $z=6$.
    
    \item For the first time we show the mean \ce{H^-} detachment and \ce{H^+_2} dissociation rates (Fig.~\ref{fig:all_rates}), necessary to properly model \ce{H_2} formation \citep{Glover:2015a,Sugimura:2016}. Based on how these rates evolve with $z$ with respect to the \ce{H_2} dissociation rate (Eq.~\ref{eq:fit_ratios}), we find that the resulting spectral shape of the LWB can be approximated with a black-body spectrum with an effective temperature that evolves from $6\times10^4$ K at $z\sim23$ to $2\times10^4$ K at $z\sim6$. The same analysis performed on the rates derived from the UV backgrounds of \cite{Haardt:2012} and \cite{Faucher-Giguere:2009} gives a consistent evolution for the former (from $3\times10^4$ K at $z\sim15$ to $1.5\times10^4$ K at $z\sim6$), and a very hard spectrum with $T_\mathrm{eff}\sim10^5 \ \mathrm{K}$ for the latter, that would require massive PopIII stars to dominate the UV radiation field even at $z\sim10$.
    
    \item The high-resolution FiBY simulations suggest that the contribution from PopIII stars is dominant at $z>12$ (Fig.~\ref{fig:pop3_contrib}). However, it is worth noting that the exact transition time somewhat depends on the resolution of the simulations and the associated metal enrichment: lower mass and spatial resolution delays the transition from a PopIII- to a PopII-dominated star formation to as late as $z\sim10$ \citep{Maio:2010}.
    
    \item Young stellar populations undoubtedly provide the largest contribution to the LWB (Fig.~\ref{fig:agebins}), thanks to the presence of short-lived hot massive stars \cite[][and references therein]{Eldridge:2022}. Nonetheless, since we follow the evolution of the stellar spectra during their entire lifetime (up to 1 Gyr, more than the age of the universe at $z=6$), we are able to determine also the contribution from old stars. We find that stars older than 20 Myr account for $20\%-30\%$ of the \ce{H_2} dissociation rate at $z\lesssim10$, and up to 60\% in the \ce{H^-} detachment rate. This shows that the LW radiation intensity is often underestimated in simulations in the literature. FiBY\_LW, for example, considers only the LW radiation emitted by stars younger than 5 Myr \citep{Johnson:2013}: however, those stellar populations never contribute to more than 60\% of the total LW intensity and can reach as low as 20\% of the \ce{H^-} detachment rate at $z\lesssim10$. The Renaissance suite of simulations \citep{O'Shea:2015}, as well as its progenitor \citep[][]{Wise:2012a} and descendant \citep[Phoenix,][]{Wells:2022}, adopts a similar approach, where PopII stars contribute to the UV radiation field only if younger than 20 Myr, with a constant luminosity equal to their lifetime-averaged luminosity. In their case, however, PopIII stellar evolution is followed, under the assumption of a delta function as IMF, centred on 40 M$_\odot$, 100 M$_\odot$ and 20 M$_\odot$ respectively. Similarly, \cite{Ahn:2009} in their semi-analytical model consider a constant stellar emissivity that approximates the spectrum of a young ($<20 \ \mathrm{Myr}$) stellar population.
    
    \item Our fiducial choice for stellar models and IMFs reflects the fact that PopIII stars have generally a higher characteristic mass and hotter atmospheres compared to solar-metallicity models with standard \cite{Salpeter:1955} or \cite{Chabrier:2003} IMFs. Nevertheless, the ongoing debate on the IMF of metal-free and metal-poor stars \citep[e.g.][]{Abel:2002,Frebel:2007,Hirano:2015,Stacy:2016,Rossi:2021} leads us to relax the initial hypothesis and consider multiple sets of stellar models (see Tables~\ref{Table:SEDpop3}-\ref{Table:SEDpop2}). The mean LW intensity is increased (decreased) by a factor of 2-3 with more top-heavy (bottom-heavy) IMFs (Fig.~\ref{fig:rates_combs_Mbox}); interestingly, a black-body with $T=10^4 \ \mathrm{K}$, as often assumed in the literature \citep{Shang:2010,Johnson:2013,Glover:2015a}, gives a factor of 400 and 30 higher \ce{H^-} detachment and \ce{H^+_2} dissociation rate respectively, due to the extremely different spectral shape that does not resemble the common SED models of young PopII stellar populations (Fig.~\ref{fig:example_spectra}). Present and future observations will be key to complement high-resolution simulations of PopIII star formation \citep{Hirano:2015,Stacy:2016,Park:2021} and specific models for PopIII stars \citep{Schaerer:2002,Raiter:2010,Gessey-Jones:2022,Larkin:2023}, to put tighter constraints on their IMF. JWST and ALMA will investigate spectral signatures of PopIII-dominated galaxies \citep{Yajima:2017,Woods:2021,Nakajima:2022,Latif:2022a}, while precise metal abundances measurements in Damped Lyman-$\alpha$ systems \citep{Welsh:2019,Welsh:2022} model the chemical enrichment from PopIII supernovae and stellar archaeology is already constraining the IMF low-mass end with local observations of extremely-metal-poor stars \citep{Frebel:2007,Rossi:2021,Hartwig:2015b,Hartwig:2022}. This will help reducing the uncertainties in the LWB modelling presented here.
    
    \item Eq.~\ref{eq:fit_FiBY_LWB} models the LWB determined with the methods presented in this work and can be safely used as a realistic homogeneous background in simulations that do not resolve low-mass halos or do not have enough volume to reach the LW horizon. However, it is tightly connected with the modelling of star formation in FiBY. Explicitly taking into account the amount of young and old stars allows to evaluate on-the-fly a LWB that is more general and can be applied to any simulation. For this reason we include Eq.~\ref{eq:fit_LWB_agebins} and the fitting parameters in Table~\ref{Table:fit_LWB_params}, that if applied to the stellar densities in FiBY reconstruct the LWB presented in this work (Fig.~\ref{fig:fit_LWB}).
    
    \item We also study the spatial fluctuations of the LW radiation field: the \ce{H_2} dissociation rate presents a right-skewed distribution (Fig.~\ref{fig:histogs}), with a minimum that is very close to the mean value and a long tail extending $1-2$ orders of magnitude above the mean. Such tail progressively shrinks at $z\lesssim12$, while the \ce{H^-} detachment and \ce{H^+_2} dissociation rates show lower fluctuations due to the IGM being more transparent in their corresponding energy range (Fig.~\ref{fig:deltaJ}).
    
    \item The highest peaks in the LWB, usually in close proximity to massive star-forming galaxies, have been proposed as the birthplaces of massive black hole seeds at $z\gtrsim10$ \citep[see e.g.][]{Dijkstra:2008,Agarwal:2014,Agarwal:2019,Wise:2019,Lupi:2021}. We model the critical distance at which a single galaxy dominates over the homogeneous background (Fig.~\ref{fig:local_contrib_SM}), that depends on the stellar mass and can be easily rescaled for a LWB intensity that in general can vary with $z$ (Fig.~\ref{fig:Dcrit_normalised}). Our recommendation to simulators is to go beyond the time-varying homogeneous LWB and to include spatial fluctuations due to local sources. Eq.~\ref{eq:normalise_Dcrit}-\ref{eq:fit_Dcrit} can be easily incorporated in a simulation, limiting at the same time the computational domain where the UV radiation needs to be accounted for on-the-fly with computationally-expensive radiative transfer methods.
    
    \item We use the homogeneous background found in this work to estimate the minimum halo mass for PopIII star formation under the influence of the LW radiation, with the use of a number of analytical and numerical studies available in the literature \citep{Machacek:2001,Latif:2019,Schauer:2021,Kulkarni:2021,Lupi:2021}. Using the LWB in Eq.~\ref{eq:fit_FiBY_LWB}, we obtain a minimum mass that approximately evolves from $\sim3\times10^5 \ \mathrm{M}_\odot$ at $z=20$ to $\sim10^7 \ \mathrm{M}_\odot$ at $z\lesssim10$ (Fig.~\ref{fig:minMh_SFR_literature_M_FID}). The numerical experiments citied here unfortunately suffer from important limitations, such as: \textit{(i)} \ce{H^-} detachment and \ce{H^+_2} dissociation are often neglected, and \textit{(ii)} the LW intensity is assumed as constant throughout simulations that run from $z=30$ to $z=15$, a wide time window during which the LWB can grow by $\sim3.5$ orders of magnitude (if Eq.~\ref{eq:fit_FiBY_LWB} is extrapolated to $z=30$). We then show in Fig.~\ref{fig:minMh_SFR_Sboxes} our preliminary results on the effect of LW radiation in delaying PopIII star formation in molecular-cooling halos. In particular, when $J_{21}\sim0.05$ at $z=20$ the minimum mass of star-forming halos in FiBY\_LW starts diverging from the case where the LW radiation is neglected. By $z=13$ ($J_{21}\sim0.2$), only atomic-cooling halos above $T_\mathrm{vir}=10^4 \ \mathrm{K}$ can form stars. Such results should be treated with caution, though, as further analysis is required to distinguish inefficient \ce{H_2}-cooling due to LW radiation from other hydrodynamical processes that might have a similar effect, such as dynamical heating \citep{Fernandez:2014} or pressure support by turbulence \citep{Latif:2022b}.

\end{itemize}

We highlight here three important caveats of this work. \textit{First}, the calculation of the LWB is performed in postprocessing. Compared to FiBY\_S, the star formation history and the balance between PopIII and PopII stars change in FiBY\_LW \citep{Johnson:2013}, not included in the analysis of Sec.~\ref{sec:LWB}. Similar trends are also observed in recent high-$z$ simulations \citep{Sarmento:2022,Wells:2022}, where PopIII star formation is significantly reduced at $z\gtrsim10$ and does not abruptly decrease afterwards, while PopII form at a slower pace because of the delayed metal enrichment, up until $z\sim10$ when star formation quickly grows and overcomes the one in the case where LW radiation is neglected.
\textit{Secondly}, none of the FiBY simulations encompasses enough volume to reach the LW horizon at $\sim100 \ \mathrm{cMpc}$. Following \cite{Ahn:2009}'s approach, we stack the necessary number of copies of the central box until the LW horizon is reached, but in doing so the cosmic variance is certainly lower than the one expected in a full $\sim(100 \ \mathrm{cMpc})^3$ box.
\textit{Finally}, several studies have shown that X-rays with energies of the order $\sim1 \ \mathrm{keV}$ generally favour \ce{H_2} formation, driving H and He ionisation even at the centre of dense gas clouds and thus increasing the abundance of free electrons, that in turn catalyse the formation of \ce{H^-} and \ce{H_2} molecules \citep{Inayoshi:2011,Inayoshi:2015}. X-rays ultimately counterbalance the effect of the LW radiation, enabling star formation \citep{Haiman:2000} and increasing the $J_\mathrm{crit}$ needed for the DCBH scenario \citep{Glover:2016}. The FiBY simulations, however, do not include any high-$z$ X-ray background due to sources such as massive X-ray binaries and accreting light black holes \citep[see e.g. emission models in][]{Tanaka:2012}. 

In conclusion, with this work we hope to provide a useful contribution to the discussion around the LW radiation in the high-$z$ universe and in particular we aim at assisting theoretical astrophysics to include a realistic LWB into their cosmological simulations. The LW radiation is a key ingredient of our current model of galaxy formation at $z\gtrsim10$, but still nowadays most of the numerical efforts in the literature do not include it (see e.g. BlueTides \citealt{Feng:2016}, SPHINX \citealt{Rosdahl:2018}, OBELISK \citealt{Trebitsch:2021}, FLARES \citealt{Lovell:2021,Wilkins:2022}, THESAN \citealt{Garaldi:2022,Kannan:2022,Smith:2022}, ASTRID \citealt{Bird:2022}). At the same time, the first recent results from JWST \citep{Donnan:2023,Harikane:2023} suggest the existence of massive galaxies even at $z\gtrsim15$, when the LW background is rapidly growing ($J_{21}\sim0.1-1$), is still dominated by PopIII stars and greatly affects star formation in low-mass halos. As for ourselves, we plan to continue studying the build-up of the LW radiation and its interplay with PopIII and PopII star formation in low-mass halos in a future set of cosmological simulations, tailored to represent the best trade-off between a high spatial and mass resolution and a large volume and to address all the limitations reported in this work.

\section*{Acknowledgements}

AI acknowledges support from a STFC-ScotDIST studentship. This work made extensive use of \verb|python3| \citep{vanRossum:2009} and of the following open-source libraries: \verb|IPython| \citep{Perez:2007}, \verb|matplotlib| \citep{Hunter:2007}, \verb|numpy| \citep{Harris:2020}, \verb|astropy| \citep{AstropyCollaboration:2013,AstropyCollaboration:2018}, \verb|scipy| \citep{Virtanen:2020}, \verb|h5py| \citep{Collette:2021}, \verb|emcee| \citep{Foreman-Mackey:2013}; we are grateful to the respective communities of developers. AI is grateful to Dr. Eric Tittley for keeping the ROE Cuillin HPC cluster healthy and running. This work made extensive use of the \href{https://ui.adsabs.harvard.edu/#}{NASA Astrophysics DataSystem}, the \href{https://arxiv.org/archive/astro-ph}{astro-ph} pre-print archive, the \href{https://feedly.com/}{Feedly} aggregator and the \href{https://www.mendeley.com/}{Mendeley} reference manager. For the purpose of open access, the authors have applied a Creative Commons Attribution (CC BY) licence to any Author Accepted Manuscript version arising from this submission.

\section*{Data availability}
The data used in this work is available upon reasonable request.

\bibliographystyle{mnras}
\bibliography{mybibliography}

\begin{thebibliography}{}
\makeatletter
\relax
\def\mn@urlcharsother{\let\do\@makeother \do\$\do\&\do\#\do\^\do\_\do\%\do\~}
\def\mn@doi{\begingroup\mn@urlcharsother \@ifnextchar [ {\mn@doi@}
  {\mn@doi@[]}}
\def\mn@doi@[#1]#2{\def\@tempa{#1}\ifx\@tempa\@empty \href
  {http://dx.doi.org/#2} {doi:#2}\else \href {http://dx.doi.org/#2} {#1}\fi
  \endgroup}
\def\mn@eprint#1#2{\mn@eprint@#1:#2::\@nil}
\def\mn@eprint@arXiv#1{\href {http://arxiv.org/abs/#1} {{\tt arXiv:#1}}}
\def\mn@eprint@dblp#1{\href {http://dblp.uni-trier.de/rec/bibtex/#1.xml}
  {dblp:#1}}
\def\mn@eprint@#1:#2:#3:#4\@nil{\def\@tempa {#1}\def\@tempb {#2}\def\@tempc
  {#3}\ifx \@tempc \@empty \let \@tempc \@tempb \let \@tempb \@tempa \fi \ifx
  \@tempb \@empty \def\@tempb {arXiv}\fi \@ifundefined
  {mn@eprint@\@tempb}{\@tempb:\@tempc}{\expandafter \expandafter \csname
  mn@eprint@\@tempb\endcsname \expandafter{\@tempc}}}

\bibitem[\protect\citeauthoryear{{Abel}, {Anninos}, {Zhang}  \&
  {Norman}}{{Abel} et~al.}{1997}]{Abel:1997}
{Abel} T.,  {Anninos} P.,  {Zhang} Y.,   {Norman} M.~L.,  1997, \mn@doi [\na]
  {10.1016/S1384-1076(97)00010-9}, \href
  {https://ui.adsabs.harvard.edu/abs/1997NewA....2..181A} {2, 181}

\bibitem[\protect\citeauthoryear{{Abel}, {Bryan}  \& {Norman}}{{Abel}
  et~al.}{2000}]{Abel:2000}
{Abel} T.,  {Bryan} G.~L.,   {Norman} M.~L.,  2000, \mn@doi [\apj]
  {10.1086/309295}, \href
  {https://ui.adsabs.harvard.edu/abs/2000ApJ...540...39A} {540, 39}

\bibitem[\protect\citeauthoryear{{Abel}, {Bryan}  \& {Norman}}{{Abel}
  et~al.}{2002}]{Abel:2002}
{Abel} T.,  {Bryan} G.~L.,   {Norman} M.~L.,  2002, \mn@doi [Science]
  {10.1126/science.295.5552.93}, \href
  {https://ui.adsabs.harvard.edu/abs/2002Sci...295...93A} {295, 93}

\bibitem[\protect\citeauthoryear{{Abgrall}, {Roueff}, {Launay}, {Roncin}  \&
  {Subtil}}{{Abgrall} et~al.}{1993a}]{Abgrall:1993a}
{Abgrall} H.,  {Roueff} E.,  {Launay} F.,  {Roncin} J.~Y.,   {Subtil} J.~L.,
  1993a, \aaps, \href {https://ui.adsabs.harvard.edu/abs/1993A&AS..101..273A}
  {101, 273}

\bibitem[\protect\citeauthoryear{{Abgrall}, {Roueff}, {Launay}, {Roncin}  \&
  {Subtil}}{{Abgrall} et~al.}{1993b}]{Abgrall:1993b}
{Abgrall} H.,  {Roueff} E.,  {Launay} F.,  {Roncin} J.~Y.,   {Subtil} J.~L.,
  1993b, \aaps, \href {https://ui.adsabs.harvard.edu/abs/1993A&AS..101..323A}
  {101, 323}

\bibitem[\protect\citeauthoryear{{Abgrall}, {Roueff}, {Launay}, {Roncin}  \&
  {Subtil}}{{Abgrall} et~al.}{1993c}]{Abgrall:1993c}
{Abgrall} H.,  {Roueff} E.,  {Launay} F.,  {Roncin} J.~Y.,   {Subtil} J.~L.,
  1993c, \mn@doi [Journal of Molecular Spectroscopy] {10.1006/jmsp.1993.1040},
  \href {https://ui.adsabs.harvard.edu/abs/1993JMoSp.157..512A} {157, 512}

\bibitem[\protect\citeauthoryear{{Abgrall}, {Roueff}, {Liu}  \&
  {Shemansky}}{{Abgrall} et~al.}{1997}]{Abgrall:1997}
{Abgrall} H.,  {Roueff} E.,  {Liu} X.,   {Shemansky} D.~E.,  1997, \mn@doi
  [\apj] {10.1086/304017}, \href
  {https://ui.adsabs.harvard.edu/abs/1997ApJ...481..557A} {481, 557}

\bibitem[\protect\citeauthoryear{{Abgrall}, {Roueff}  \& {Drira}}{{Abgrall}
  et~al.}{2000}]{Abgrall:2000}
{Abgrall} H.,  {Roueff} E.,   {Drira} I.,  2000, \mn@doi [\aaps]
  {10.1051/aas:2000121}, \href
  {https://ui.adsabs.harvard.edu/abs/2000A&AS..141..297A} {141, 297}

\bibitem[\protect\citeauthoryear{{Adelberger}, {Steidel}, {Pettini}, {Shapley},
  {Reddy}  \& {Erb}}{{Adelberger} et~al.}{2005}]{Adelberger:2005}
{Adelberger} K.~L.,  {Steidel} C.~C.,  {Pettini} M.,  {Shapley} A.~E.,  {Reddy}
  N.~A.,   {Erb} D.~K.,  2005, \mn@doi [\apj] {10.1086/426580}, \href
  {https://ui.adsabs.harvard.edu/abs/2005ApJ...619..697A} {619, 697}

\bibitem[\protect\citeauthoryear{{Agarwal} \& {Khochfar}}{{Agarwal} \&
  {Khochfar}}{2015}]{Agarwal:2015}
{Agarwal} B.,  {Khochfar} S.,  2015, \mn@doi [\mnras] {10.1093/mnras/stu1973},
  \href {https://ui.adsabs.harvard.edu/abs/2015MNRAS.446..160A} {446, 160}

\bibitem[\protect\citeauthoryear{{Agarwal}, {Khochfar}, {Johnson}, {Neistein},
  {Dalla Vecchia}  \& {Livio}}{{Agarwal} et~al.}{2012}]{Agarwal:2012}
{Agarwal} B.,  {Khochfar} S.,  {Johnson} J.~L.,  {Neistein} E.,  {Dalla
  Vecchia} C.,   {Livio} M.,  2012, \mn@doi [\mnras]
  {10.1111/j.1365-2966.2012.21651.x}, \href
  {https://ui.adsabs.harvard.edu/abs/2012MNRAS.425.2854A} {425, 2854}

\bibitem[\protect\citeauthoryear{{Agarwal}, {Dalla Vecchia}, {Johnson},
  {Khochfar}  \& {Paardekooper}}{{Agarwal} et~al.}{2014}]{Agarwal:2014}
{Agarwal} B.,  {Dalla Vecchia} C.,  {Johnson} J.~L.,  {Khochfar} S.,
  {Paardekooper} J.-P.,  2014, \mn@doi [\mnras] {10.1093/mnras/stu1112}, \href
  {https://ui.adsabs.harvard.edu/abs/2014MNRAS.443..648A} {443, 648}

\bibitem[\protect\citeauthoryear{{Agarwal}, {Smith}, {Glover}, {Natarajan}  \&
  {Khochfar}}{{Agarwal} et~al.}{2016}]{Agarwal:2016}
{Agarwal} B.,  {Smith} B.,  {Glover} S.,  {Natarajan} P.,   {Khochfar} S.,
  2016, \mn@doi [\mnras] {10.1093/mnras/stw929}, \href
  {https://ui.adsabs.harvard.edu/abs/2016MNRAS.459.4209A} {459, 4209}

\bibitem[\protect\citeauthoryear{{Agarwal}, {Cullen}, {Khochfar}, {Ceverino}
  \& {Klessen}}{{Agarwal} et~al.}{2019}]{Agarwal:2019}
{Agarwal} B.,  {Cullen} F.,  {Khochfar} S.,  {Ceverino} D.,   {Klessen} R.~S.,
  2019, \mn@doi [\mnras] {10.1093/mnras/stz1347}, \href
  {https://ui.adsabs.harvard.edu/abs/2019MNRAS.488.3268A} {488, 3268}

\bibitem[\protect\citeauthoryear{{Ahn}, {Shapiro}, {Iliev}, {Mellema}  \&
  {Pen}}{{Ahn} et~al.}{2009}]{Ahn:2009}
{Ahn} K.,  {Shapiro} P.~R.,  {Iliev} I.~T.,  {Mellema} G.,   {Pen} U.-L.,
  2009, \mn@doi [\apj] {10.1088/0004-637X/695/2/1430}, \href
  {https://ui.adsabs.harvard.edu/abs/2009ApJ...695.1430A} {695, 1430}

\bibitem[\protect\citeauthoryear{{Astropy Collaboration} et~al.,}{{Astropy
  Collaboration} et~al.}{2013}]{AstropyCollaboration:2013}
{Astropy Collaboration} et~al., 2013, \mn@doi [\aap]
  {10.1051/0004-6361/201322068}, \href
  {https://ui.adsabs.harvard.edu/abs/2013A&A...558A..33A} {558, A33}

\bibitem[\protect\citeauthoryear{{Astropy Collaboration} et~al.,}{{Astropy
  Collaboration} et~al.}{2018}]{AstropyCollaboration:2018}
{Astropy Collaboration} et~al., 2018, \mn@doi [\aj] {10.3847/1538-3881/aabc4f},
  \href {https://ui.adsabs.harvard.edu/abs/2018AJ....156..123A} {156, 123}

\bibitem[\protect\citeauthoryear{{Ba{\~n}ados} et~al.,}{{Ba{\~n}ados}
  et~al.}{2018}]{Banados:2018}
{Ba{\~n}ados} E.,  et~al., 2018, \mn@doi [\nat] {10.1038/nature25180}, \href
  {https://ui.adsabs.harvard.edu/abs/2018Natur.553..473B} {553, 473}

\bibitem[\protect\citeauthoryear{{Babb}}{{Babb}}{2015}]{Babb:2015}
{Babb} J.~F.,  2015, \mn@doi [\apjs] {10.1088/0067-0049/216/1/21}, \href
  {https://ui.adsabs.harvard.edu/abs/2015ApJS..216...21B} {216, 21}

\bibitem[\protect\citeauthoryear{{Begelman}, {Volonteri}  \& {Rees}}{{Begelman}
  et~al.}{2006}]{Begelman:2006}
{Begelman} M.~C.,  {Volonteri} M.,   {Rees} M.~J.,  2006, \mn@doi [\mnras]
  {10.1111/j.1365-2966.2006.10467.x}, \href
  {https://ui.adsabs.harvard.edu/abs/2006MNRAS.370..289B} {370, 289}

\bibitem[\protect\citeauthoryear{{Bhowmick}, {Blecha}, {Torrey}, {Kelley},
  {Vogelsberger}, {Nelson}, {Weinberger}  \& {Hernquist}}{{Bhowmick}
  et~al.}{2022}]{Bhowmick:2022}
{Bhowmick} A.~K.,  {Blecha} L.,  {Torrey} P.,  {Kelley} L.~Z.,  {Vogelsberger}
  M.,  {Nelson} D.,  {Weinberger} R.,   {Hernquist} L.,  2022, \mn@doi [\mnras]
  {10.1093/mnras/stab3439}, \href
  {https://ui.adsabs.harvard.edu/abs/2022MNRAS.510..177B} {510, 177}

\bibitem[\protect\citeauthoryear{{Bird}, {Ni}, {Di Matteo}, {Croft}, {Feng}  \&
  {Chen}}{{Bird} et~al.}{2022}]{Bird:2022}
{Bird} S.,  {Ni} Y.,  {Di Matteo} T.,  {Croft} R.,  {Feng} Y.,   {Chen} N.,
  2022, \mn@doi [\mnras] {10.1093/mnras/stac648}, \href
  {https://ui.adsabs.harvard.edu/abs/2022MNRAS.512.3703B} {512, 3703}

\bibitem[\protect\citeauthoryear{{Black} \& {Dalgarno}}{{Black} \&
  {Dalgarno}}{1977}]{Black:1977}
{Black} J.~H.,  {Dalgarno} A.,  1977, \mn@doi [\apjs] {10.1086/190455}, \href
  {https://ui.adsabs.harvard.edu/abs/1977ApJS...34..405B} {34, 405}

\bibitem[\protect\citeauthoryear{{Bonoli}, {Mayer}  \& {Callegari}}{{Bonoli}
  et~al.}{2014}]{Bonoli:2014}
{Bonoli} S.,  {Mayer} L.,   {Callegari} S.,  2014, \mn@doi [\mnras]
  {10.1093/mnras/stt1990}, \href
  {https://ui.adsabs.harvard.edu/abs/2014MNRAS.437.1576B} {437, 1576}

\bibitem[\protect\citeauthoryear{{Bouwens} et~al.,}{{Bouwens}
  et~al.}{2016}]{Bouwens:2016}
{Bouwens} R.~J.,  et~al., 2016, \mn@doi [\apj] {10.3847/1538-4357/833/1/72},
  \href {https://ui.adsabs.harvard.edu/abs/2016ApJ...833...72B} {833, 72}

\bibitem[\protect\citeauthoryear{{Bowler} et~al.,}{{Bowler}
  et~al.}{2015}]{Bowler:2015}
{Bowler} R.~A.~A.,  et~al., 2015, \mn@doi [\mnras] {10.1093/mnras/stv1403},
  \href {https://ui.adsabs.harvard.edu/abs/2015MNRAS.452.1817B} {452, 1817}

\bibitem[\protect\citeauthoryear{{Bressan}, {Fagotto}, {Bertelli}  \&
  {Chiosi}}{{Bressan} et~al.}{1993}]{Bressan:1993}
{Bressan} A.,  {Fagotto} F.,  {Bertelli} G.,   {Chiosi} C.,  1993, \aaps, \href
  {https://ui.adsabs.harvard.edu/abs/1993A&AS..100..647B} {100, 647}

\bibitem[\protect\citeauthoryear{{Bromm} \& {Larson}}{{Bromm} \&
  {Larson}}{2004}]{Bromm:2004}
{Bromm} V.,  {Larson} R.~B.,  2004, \mn@doi [\araa]
  {10.1146/annurev.astro.42.053102.134034}, \href
  {https://ui.adsabs.harvard.edu/abs/2004ARA&A..42...79B} {42, 79}

\bibitem[\protect\citeauthoryear{{Bromm} \& {Loeb}}{{Bromm} \&
  {Loeb}}{2003}]{Bromm:2003}
{Bromm} V.,  {Loeb} A.,  2003, \mn@doi [\nat] {10.1038/nature02071}, \href
  {https://ui.adsabs.harvard.edu/abs/2003Natur.425..812B} {425, 812}

\bibitem[\protect\citeauthoryear{{Bromm} \& {Yoshida}}{{Bromm} \&
  {Yoshida}}{2011}]{Bromm:2011}
{Bromm} V.,  {Yoshida} N.,  2011, \mn@doi [\araa]
  {10.1146/annurev-astro-081710-102608}, \href
  {https://ui.adsabs.harvard.edu/abs/2011ARA&A..49..373B} {49, 373}

\bibitem[\protect\citeauthoryear{{Bromm}, {Coppi}  \& {Larson}}{{Bromm}
  et~al.}{1999}]{Bromm:1999}
{Bromm} V.,  {Coppi} P.~S.,   {Larson} R.~B.,  1999, \mn@doi [\apjl]
  {10.1086/312385}, \href
  {https://ui.adsabs.harvard.edu/abs/1999ApJ...527L...5B} {527, L5}

\bibitem[\protect\citeauthoryear{{Chabrier}}{{Chabrier}}{2003}]{Chabrier:2003}
{Chabrier} G.,  2003, \mn@doi [\pasp] {10.1086/376392}, \href
  {https://ui.adsabs.harvard.edu/abs/2003PASP..115..763C} {115, 763}

\bibitem[\protect\citeauthoryear{{Chiaki} \& {Yoshida}}{{Chiaki} \&
  {Yoshida}}{2022}]{Chiaki:2022}
{Chiaki} G.,  {Yoshida} N.,  2022, \mn@doi [\mnras] {10.1093/mnras/stab2799},
  \href {https://ui.adsabs.harvard.edu/abs/2022MNRAS.510.5199C} {510, 5199}

\bibitem[\protect\citeauthoryear{{Choi}, {Dotter}, {Conroy}, {Cantiello},
  {Paxton}  \& {Johnson}}{{Choi} et~al.}{2016}]{Choi:2016}
{Choi} J.,  {Dotter} A.,  {Conroy} C.,  {Cantiello} M.,  {Paxton} B.,
  {Johnson} B.~D.,  2016, \mn@doi [\apj] {10.3847/0004-637X/823/2/102}, \href
  {https://ui.adsabs.harvard.edu/abs/2016ApJ...823..102C} {823, 102}

\bibitem[\protect\citeauthoryear{{Chuzhoy}, {Kuhlen}  \& {Shapiro}}{{Chuzhoy}
  et~al.}{2007}]{Chuzhoy:2007}
{Chuzhoy} L.,  {Kuhlen} M.,   {Shapiro} P.~R.,  2007, \mn@doi [\apjl]
  {10.1086/521438}, \href
  {https://ui.adsabs.harvard.edu/abs/2007ApJ...665L..85C} {665, L85}

\bibitem[\protect\citeauthoryear{{Collette} et~al.,}{{Collette}
  et~al.}{2021}]{Collette:2021}
{Collette} A.,  et~al., 2021, {h5py/h5py: 3.3.0},
  \mn@doi{10.5281/zenodo.594310}

\bibitem[\protect\citeauthoryear{{Corney}}{{Corney}}{1977}]{Corney:1977}
{Corney} A.,  1977, {Atomic and laser spectroscopy}

\bibitem[\protect\citeauthoryear{{Cullen}, {McLure}, {Khochfar}, {Dunlop}  \&
  {Dalla Vecchia}}{{Cullen} et~al.}{2017}]{Cullen:2017}
{Cullen} F.,  {McLure} R.~J.,  {Khochfar} S.,  {Dunlop} J.~S.,   {Dalla
  Vecchia} C.,  2017, \mn@doi [\mnras] {10.1093/mnras/stx1451}, \href
  {https://ui.adsabs.harvard.edu/abs/2017MNRAS.470.3006C} {470, 3006}

\bibitem[\protect\citeauthoryear{{Dalla Vecchia} \& {Schaye}}{{Dalla Vecchia}
  \& {Schaye}}{2012}]{DallaVecchia:2012}
{Dalla Vecchia} C.,  {Schaye} J.,  2012, \mn@doi [\mnras]
  {10.1111/j.1365-2966.2012.21704.x}, \href
  {https://ui.adsabs.harvard.edu/abs/2012MNRAS.426..140D} {426, 140}

\bibitem[\protect\citeauthoryear{{Dijkstra}, {Haiman}, {Mesinger}  \&
  {Wyithe}}{{Dijkstra} et~al.}{2008}]{Dijkstra:2008}
{Dijkstra} M.,  {Haiman} Z.,  {Mesinger} A.,   {Wyithe} J. S.~B.,  2008,
  \mn@doi [\mnras] {10.1111/j.1365-2966.2008.14031.x}, \href
  {https://ui.adsabs.harvard.edu/abs/2008MNRAS.391.1961D} {391, 1961}

\bibitem[\protect\citeauthoryear{{Dijkstra}, {Ferrara}  \&
  {Mesinger}}{{Dijkstra} et~al.}{2014}]{Dijkstra:2014}
{Dijkstra} M.,  {Ferrara} A.,   {Mesinger} A.,  2014, \mn@doi [\mnras]
  {10.1093/mnras/stu1007}, \href
  {https://ui.adsabs.harvard.edu/abs/2014MNRAS.442.2036D} {442, 2036}

\bibitem[\protect\citeauthoryear{{Donnan} et~al.,}{{Donnan}
  et~al.}{2023}]{Donnan:2023}
{Donnan} C.~T.,  et~al., 2023, \mn@doi [\mnras] {10.1093/mnras/stac3472}, \href
  {https://ui.adsabs.harvard.edu/abs/2023MNRAS.518.6011D} {518, 6011}

\bibitem[\protect\citeauthoryear{{Dotter}}{{Dotter}}{2016}]{Dotter:2016}
{Dotter} A.,  2016, \mn@doi [\apjs] {10.3847/0067-0049/222/1/8}, \href
  {https://ui.adsabs.harvard.edu/abs/2016ApJS..222....8D} {222, 8}

\bibitem[\protect\citeauthoryear{{Draine} \& {Bertoldi}}{{Draine} \&
  {Bertoldi}}{1996}]{Draine:1996}
{Draine} B.~T.,  {Bertoldi} F.,  1996, \mn@doi [\apj] {10.1086/177689}, \href
  {https://ui.adsabs.harvard.edu/abs/1996ApJ...468..269D} {468, 269}

\bibitem[\protect\citeauthoryear{{Eggenberger}, {Meynet}, {Maeder}, {Hirschi},
  {Charbonnel}, {Talon}  \& {Ekstr{\"o}m}}{{Eggenberger}
  et~al.}{2008}]{Eggenberger:2008}
{Eggenberger} P.,  {Meynet} G.,  {Maeder} A.,  {Hirschi} R.,  {Charbonnel} C.,
  {Talon} S.,   {Ekstr{\"o}m} S.,  2008, \mn@doi [\apss]
  {10.1007/s10509-007-9511-y}, \href
  {https://ui.adsabs.harvard.edu/abs/2008Ap&SS.316...43E} {316, 43}

\bibitem[\protect\citeauthoryear{{Eldridge} \& {Stanway}}{{Eldridge} \&
  {Stanway}}{2022}]{Eldridge:2022}
{Eldridge} J.~J.,  {Stanway} E.~R.,  2022, \mn@doi [\araa]
  {10.1146/annurev-astro-052920-100646}, \href
  {https://ui.adsabs.harvard.edu/abs/2022ARA&A..60..455E} {60, 455}

\bibitem[\protect\citeauthoryear{{Fan} et~al.,}{{Fan} et~al.}{2006}]{Fan:2006}
{Fan} X.,  et~al., 2006, \mn@doi [\aj] {10.1086/504836}, \href
  {https://ui.adsabs.harvard.edu/abs/2006AJ....132..117F} {132, 117}

\bibitem[\protect\citeauthoryear{{Faucher-Gigu{\`e}re}, {Lidz}, {Zaldarriaga}
  \& {Hernquist}}{{Faucher-Gigu{\`e}re} et~al.}{2009}]{Faucher-Giguere:2009}
{Faucher-Gigu{\`e}re} C.-A.,  {Lidz} A.,  {Zaldarriaga} M.,   {Hernquist} L.,
  2009, \mn@doi [\apj] {10.1088/0004-637X/703/2/1416}, \href
  {https://ui.adsabs.harvard.edu/abs/2009ApJ...703.1416F} {703, 1416}

\bibitem[\protect\citeauthoryear{{Feng}, {Di-Matteo}, {Croft}, {Bird},
  {Battaglia}  \& {Wilkins}}{{Feng} et~al.}{2016}]{Feng:2016}
{Feng} Y.,  {Di-Matteo} T.,  {Croft} R.~A.,  {Bird} S.,  {Battaglia} N.,
  {Wilkins} S.,  2016, \mn@doi [\mnras] {10.1093/mnras/stv2484}, \href
  {https://ui.adsabs.harvard.edu/abs/2016MNRAS.455.2778F} {455, 2778}

\bibitem[\protect\citeauthoryear{{Fernandez}, {Bryan}, {Haiman}  \&
  {Li}}{{Fernandez} et~al.}{2014}]{Fernandez:2014}
{Fernandez} R.,  {Bryan} G.~L.,  {Haiman} Z.,   {Li} M.,  2014, \mn@doi
  [\mnras] {10.1093/mnras/stu230}, \href
  {https://ui.adsabs.harvard.edu/abs/2014MNRAS.439.3798F} {439, 3798}

\bibitem[\protect\citeauthoryear{{Fialkov}, {Barkana}, {Visbal},
  {Tseliakhovich}  \& {Hirata}}{{Fialkov} et~al.}{2013}]{Fialkov:2013}
{Fialkov} A.,  {Barkana} R.,  {Visbal} E.,  {Tseliakhovich} D.,   {Hirata}
  C.~M.,  2013, \mn@doi [\mnras] {10.1093/mnras/stt650}, \href
  {https://ui.adsabs.harvard.edu/abs/2013MNRAS.432.2909F} {432, 2909}

\bibitem[\protect\citeauthoryear{{Foreman-Mackey}, {Hogg}, {Lang}  \&
  {Goodman}}{{Foreman-Mackey} et~al.}{2013}]{Foreman-Mackey:2013}
{Foreman-Mackey} D.,  {Hogg} D.~W.,  {Lang} D.,   {Goodman} J.,  2013, \mn@doi
  [\pasp] {10.1086/670067}, \href
  {https://ui.adsabs.harvard.edu/abs/2013PASP..125..306F} {125, 306}

\bibitem[\protect\citeauthoryear{{Frebel}, {Johnson}  \& {Bromm}}{{Frebel}
  et~al.}{2007}]{Frebel:2007}
{Frebel} A.,  {Johnson} J.~L.,   {Bromm} V.,  2007, \mn@doi [\mnras]
  {10.1111/j.1745-3933.2007.00344.x}, \href
  {https://ui.adsabs.harvard.edu/abs/2007MNRAS.380L..40F} {380, L40}

\bibitem[\protect\citeauthoryear{{Fryer}, {Woosley}  \& {Heger}}{{Fryer}
  et~al.}{2001}]{Fryer:2001}
{Fryer} C.~L.,  {Woosley} S.~E.,   {Heger} A.,  2001, \mn@doi [\apj]
  {10.1086/319719}, \href
  {https://ui.adsabs.harvard.edu/abs/2001ApJ...550..372F} {550, 372}

\bibitem[\protect\citeauthoryear{{Galli} \& {Palla}}{{Galli} \&
  {Palla}}{2013}]{Galli:2013}
{Galli} D.,  {Palla} F.,  2013, \mn@doi [\araa]
  {10.1146/annurev-astro-082812-141029}, \href
  {https://ui.adsabs.harvard.edu/abs/2013ARA&A..51..163G} {51, 163}

\bibitem[\protect\citeauthoryear{{Garaldi}, {Kannan}, {Smith}, {Springel},
  {Pakmor}, {Vogelsberger}  \& {Hernquist}}{{Garaldi}
  et~al.}{2022}]{Garaldi:2022}
{Garaldi} E.,  {Kannan} R.,  {Smith} A.,  {Springel} V.,  {Pakmor} R.,
  {Vogelsberger} M.,   {Hernquist} L.,  2022, \mn@doi [\mnras]
  {10.1093/mnras/stac257}, \href
  {https://ui.adsabs.harvard.edu/abs/2022MNRAS.512.4909G} {512, 4909}

\bibitem[\protect\citeauthoryear{{Ge} \& {Wise}}{{Ge} \&
  {Wise}}{2017}]{Ge:2017}
{Ge} Q.,  {Wise} J.~H.,  2017, \mn@doi [\mnras] {10.1093/mnras/stx2074}, \href
  {https://ui.adsabs.harvard.edu/abs/2017MNRAS.472.2773G} {472, 2773}

\bibitem[\protect\citeauthoryear{{Gessey-Jones} et~al.,}{{Gessey-Jones}
  et~al.}{2022}]{Gessey-Jones:2022}
{Gessey-Jones} T.,  et~al., 2022, \mn@doi [\mnras] {10.1093/mnras/stac2049},
  \href {https://ui.adsabs.harvard.edu/abs/2022MNRAS.516..841G} {516, 841}

\bibitem[\protect\citeauthoryear{{Glover}}{{Glover}}{2015a}]{Glover:2015a}
{Glover} S. C.~O.,  2015a, \mn@doi [\mnras] {10.1093/mnras/stv1059}, \href
  {https://ui.adsabs.harvard.edu/abs/2015MNRAS.451.2082G} {451, 2082}

\bibitem[\protect\citeauthoryear{{Glover}}{{Glover}}{2015b}]{Glover:2015b}
{Glover} S. C.~O.,  2015b, \mn@doi [\mnras] {10.1093/mnras/stv1781}, \href
  {https://ui.adsabs.harvard.edu/abs/2015MNRAS.453.2901G} {453, 2901}

\bibitem[\protect\citeauthoryear{{Glover}}{{Glover}}{2016}]{Glover:2016}
{Glover} S. C.~O.,  2016, arXiv e-prints, \href
  {https://ui.adsabs.harvard.edu/abs/2016arXiv161005679G} {p. arXiv:1610.05679}

\bibitem[\protect\citeauthoryear{{Greif} \& {Bromm}}{{Greif} \&
  {Bromm}}{2006}]{Greif:2006}
{Greif} T.~H.,  {Bromm} V.,  2006, \mn@doi [\mnras]
  {10.1111/j.1365-2966.2006.11017.x}, \href
  {https://ui.adsabs.harvard.edu/abs/2006MNRAS.373..128G} {373, 128}

\bibitem[\protect\citeauthoryear{{Guzzo} et~al.,}{{Guzzo}
  et~al.}{2014}]{Guzzo:2014}
{Guzzo} L.,  et~al., 2014, \mn@doi [\aap] {10.1051/0004-6361/201321489}, \href
  {https://ui.adsabs.harvard.edu/abs/2014A&A...566A.108G} {566, A108}

\bibitem[\protect\citeauthoryear{{Haardt} \& {Madau}}{{Haardt} \&
  {Madau}}{2001}]{Haardt:2001}
{Haardt} F.,  {Madau} P.,  2001, in {Neumann} D.~M.,  {Tran} J.~T.~V.,  eds,
  Clusters of Galaxies and the High Redshift Universe Observed in X-rays. p.~64
  (\mn@eprint {arXiv} {astro-ph/0106018})

\bibitem[\protect\citeauthoryear{{Haardt} \& {Madau}}{{Haardt} \&
  {Madau}}{2012}]{Haardt:2012}
{Haardt} F.,  {Madau} P.,  2012, \mn@doi [\apj] {10.1088/0004-637X/746/2/125},
  \href {https://ui.adsabs.harvard.edu/abs/2012ApJ...746..125H} {746, 125}

\bibitem[\protect\citeauthoryear{{Haiman}, {Thoul}  \& {Loeb}}{{Haiman}
  et~al.}{1996}]{Haiman:1996}
{Haiman} Z.,  {Thoul} A.~A.,   {Loeb} A.,  1996, \mn@doi [\apj]
  {10.1086/177343}, \href
  {https://ui.adsabs.harvard.edu/abs/1996ApJ...464..523H} {464, 523}

\bibitem[\protect\citeauthoryear{{Haiman}, {Rees}  \& {Loeb}}{{Haiman}
  et~al.}{1997}]{Haiman:1997}
{Haiman} Z.,  {Rees} M.~J.,   {Loeb} A.,  1997, \mn@doi [\apj]
  {10.1086/303647}, \href
  {https://ui.adsabs.harvard.edu/abs/1997ApJ...476..458H} {476, 458}

\bibitem[\protect\citeauthoryear{{Haiman}, {Abel}  \& {Rees}}{{Haiman}
  et~al.}{2000}]{Haiman:2000}
{Haiman} Z.,  {Abel} T.,   {Rees} M.~J.,  2000, \mn@doi [\apj]
  {10.1086/308723}, \href
  {https://ui.adsabs.harvard.edu/abs/2000ApJ...534...11H} {534, 11}

\bibitem[\protect\citeauthoryear{{Harikane} et~al.,}{{Harikane}
  et~al.}{2023}]{Harikane:2023}
{Harikane} Y.,  et~al., 2023, \mn@doi [\apjs] {10.3847/1538-4365/acaaa9}, \href
  {https://ui.adsabs.harvard.edu/abs/2023ApJS..265....5H} {265, 5}

\bibitem[\protect\citeauthoryear{{Harris} et~al.,}{{Harris}
  et~al.}{2020}]{Harris:2020}
{Harris} C.~R.,  et~al., 2020, \mn@doi [\nat] {10.1038/s41586-020-2649-2},
  \href {https://ui.adsabs.harvard.edu/abs/2020Natur.585..357H} {585, 357}

\bibitem[\protect\citeauthoryear{{Hartwig}, {Bromm}, {Klessen}  \&
  {Glover}}{{Hartwig} et~al.}{2015a}]{Hartwig:2015b}
{Hartwig} T.,  {Bromm} V.,  {Klessen} R.~S.,   {Glover} S. C.~O.,  2015a,
  \mn@doi [\mnras] {10.1093/mnras/stu2740}, \href
  {https://ui.adsabs.harvard.edu/abs/2015MNRAS.447.3892H} {447, 3892}

\bibitem[\protect\citeauthoryear{{Hartwig}, {Clark}, {Glover}, {Klessen}  \&
  {Sasaki}}{{Hartwig} et~al.}{2015b}]{Hartwig:2015a}
{Hartwig} T.,  {Clark} P.~C.,  {Glover} S. C.~O.,  {Klessen} R.~S.,   {Sasaki}
  M.,  2015b, \mn@doi [\apj] {10.1088/0004-637X/799/2/114}, \href
  {https://ui.adsabs.harvard.edu/abs/2015ApJ...799..114H} {799, 114}

\bibitem[\protect\citeauthoryear{{Hartwig} et~al.,}{{Hartwig}
  et~al.}{2022}]{Hartwig:2022}
{Hartwig} T.,  et~al., 2022, \mn@doi [\apj] {10.3847/1538-4357/ac7150}, \href
  {https://ui.adsabs.harvard.edu/abs/2022ApJ...936...45H} {936, 45}

\bibitem[\protect\citeauthoryear{{Heger} \& {Woosley}}{{Heger} \&
  {Woosley}}{2002}]{Heger:2002}
{Heger} A.,  {Woosley} S.~E.,  2002, \mn@doi [\apj] {10.1086/338487}, \href
  {https://ui.adsabs.harvard.edu/abs/2002ApJ...567..532H} {567, 532}

\bibitem[\protect\citeauthoryear{{Heger} \& {Woosley}}{{Heger} \&
  {Woosley}}{2010}]{Heger:2010}
{Heger} A.,  {Woosley} S.~E.,  2010, \mn@doi [\apj]
  {10.1088/0004-637X/724/1/341}, \href
  {https://ui.adsabs.harvard.edu/abs/2010ApJ...724..341H} {724, 341}

\bibitem[\protect\citeauthoryear{{Hirano}, {Hosokawa}, {Yoshida}, {Omukai}  \&
  {Yorke}}{{Hirano} et~al.}{2015}]{Hirano:2015}
{Hirano} S.,  {Hosokawa} T.,  {Yoshida} N.,  {Omukai} K.,   {Yorke} H.~W.,
  2015, \mn@doi [\mnras] {10.1093/mnras/stv044}, \href
  {https://ui.adsabs.harvard.edu/abs/2015MNRAS.448..568H} {448, 568}

\bibitem[\protect\citeauthoryear{{Hirashita} \& {Ferrara}}{{Hirashita} \&
  {Ferrara}}{2002}]{Hirashita:2002}
{Hirashita} H.,  {Ferrara} A.,  2002, \mn@doi [\mnras]
  {10.1046/j.1365-8711.2002.05968.x}, \href
  {https://ui.adsabs.harvard.edu/abs/2002MNRAS.337..921H} {337, 921}

\bibitem[\protect\citeauthoryear{{Hunter}}{{Hunter}}{2007}]{Hunter:2007}
{Hunter} J.~D.,  2007, \mn@doi [Computing in Science and Engineering]
  {10.1109/MCSE.2007.55}, \href
  {https://ui.adsabs.harvard.edu/abs/2007CSE.....9...90H} {9, 90}

\bibitem[\protect\citeauthoryear{{Iliev}, {Scannapieco}, {Martel}  \&
  {Shapiro}}{{Iliev} et~al.}{2003}]{Iliev:2003}
{Iliev} I.~T.,  {Scannapieco} E.,  {Martel} H.,   {Shapiro} P.~R.,  2003,
  \mn@doi [\mnras] {10.1046/j.1365-8711.2003.06410.x}, \href
  {https://ui.adsabs.harvard.edu/abs/2003MNRAS.341...81I} {341, 81}

\bibitem[\protect\citeauthoryear{{Inayoshi} \& {Omukai}}{{Inayoshi} \&
  {Omukai}}{2011}]{Inayoshi:2011}
{Inayoshi} K.,  {Omukai} K.,  2011, \mn@doi [\mnras]
  {10.1111/j.1365-2966.2011.19229.x}, \href
  {https://ui.adsabs.harvard.edu/abs/2011MNRAS.416.2748I} {416, 2748}

\bibitem[\protect\citeauthoryear{{Inayoshi} \& {Tanaka}}{{Inayoshi} \&
  {Tanaka}}{2015}]{Inayoshi:2015}
{Inayoshi} K.,  {Tanaka} T.~L.,  2015, \mn@doi [\mnras] {10.1093/mnras/stv871},
  \href {https://ui.adsabs.harvard.edu/abs/2015MNRAS.450.4350I} {450, 4350}

\bibitem[\protect\citeauthoryear{{John}}{{John}}{1988}]{John:1988}
{John} T.~L.,  1988, \aap, \href
  {https://ui.adsabs.harvard.edu/abs/1988A&A...193..189J} {193, 189}

\bibitem[\protect\citeauthoryear{{Johnson}, {Dalla Vecchia}  \&
  {Khochfar}}{{Johnson} et~al.}{2013}]{Johnson:2013}
{Johnson} J.~L.,  {Dalla Vecchia} C.,   {Khochfar} S.,  2013, \mn@doi [\mnras]
  {10.1093/mnras/sts011}, \href
  {https://ui.adsabs.harvard.edu/abs/2013MNRAS.428.1857J} {428, 1857}

\bibitem[\protect\citeauthoryear{{Kannan}, {Garaldi}, {Smith}, {Pakmor},
  {Springel}, {Vogelsberger}  \& {Hernquist}}{{Kannan}
  et~al.}{2022}]{Kannan:2022}
{Kannan} R.,  {Garaldi} E.,  {Smith} A.,  {Pakmor} R.,  {Springel} V.,
  {Vogelsberger} M.,   {Hernquist} L.,  2022, \mn@doi [\mnras]
  {10.1093/mnras/stab3710}, \href
  {https://ui.adsabs.harvard.edu/abs/2022MNRAS.511.4005K} {511, 4005}

\bibitem[\protect\citeauthoryear{{Kennicutt}}{{Kennicutt}}{1998}]{Kennicutt:1998}
{Kennicutt} Robert~C. J.,  1998, \mn@doi [\araa]
  {10.1146/annurev.astro.36.1.189}, \href
  {https://ui.adsabs.harvard.edu/abs/1998ARA&A..36..189K} {36, 189}

\bibitem[\protect\citeauthoryear{{Kobayashi}, {Karakas}  \&
  {Lugaro}}{{Kobayashi} et~al.}{2020}]{Kobayashi:2020}
{Kobayashi} C.,  {Karakas} A.~I.,   {Lugaro} M.,  2020, \mn@doi [\apj]
  {10.3847/1538-4357/abae65}, \href
  {https://ui.adsabs.harvard.edu/abs/2020ApJ...900..179K} {900, 179}

\bibitem[\protect\citeauthoryear{{Komatsu} et~al.,}{{Komatsu}
  et~al.}{2009}]{Komatsu:2009}
{Komatsu} E.,  et~al., 2009, \mn@doi [\apjs] {10.1088/0067-0049/180/2/330},
  \href {https://ui.adsabs.harvard.edu/abs/2009ApJS..180..330K} {180, 330}

\bibitem[\protect\citeauthoryear{{Kulkarni}, {Visbal}  \& {Bryan}}{{Kulkarni}
  et~al.}{2021}]{Kulkarni:2021}
{Kulkarni} M.,  {Visbal} E.,   {Bryan} G.~L.,  2021, \mn@doi [\apj]
  {10.3847/1538-4357/ac08a3}, \href
  {https://ui.adsabs.harvard.edu/abs/2021ApJ...917...40K} {917, 40}

\bibitem[\protect\citeauthoryear{{Larkin}, {Gerasimov}  \&
  {Burgasser}}{{Larkin} et~al.}{2023}]{Larkin:2023}
{Larkin} M.~M.,  {Gerasimov} R.,   {Burgasser} A.~J.,  2023, \mn@doi [\aj]
  {10.3847/1538-3881/ac9b43}, \href
  {https://ui.adsabs.harvard.edu/abs/2023AJ....165....2L} {165, 2}

\bibitem[\protect\citeauthoryear{{Latif} \& {Khochfar}}{{Latif} \&
  {Khochfar}}{2019}]{Latif:2019}
{Latif} M.~A.,  {Khochfar} S.,  2019, \mn@doi [\mnras] {10.1093/mnras/stz2812},
  \href {https://ui.adsabs.harvard.edu/abs/2019MNRAS.490.2706L} {490, 2706}

\bibitem[\protect\citeauthoryear{{Latif}, {Bovino}, {Grassi}, {Schleicher}  \&
  {Spaans}}{{Latif} et~al.}{2015}]{Latif:2015}
{Latif} M.~A.,  {Bovino} S.,  {Grassi} T.,  {Schleicher} D.~R.~G.,   {Spaans}
  M.,  2015, \mn@doi [\mnras] {10.1093/mnras/stu2244}, \href
  {https://ui.adsabs.harvard.edu/abs/2015MNRAS.446.3163L} {446, 3163}

\bibitem[\protect\citeauthoryear{{Latif}, {Whalen}, {Khochfar}, {Herrington}
  \& {Woods}}{{Latif} et~al.}{2022a}]{Latif:2022b}
{Latif} M.~A.,  {Whalen} D.~J.,  {Khochfar} S.,  {Herrington} N.~P.,   {Woods}
  T.~E.,  2022a, \mn@doi [\nat] {10.1038/s41586-022-04813-y}, \href
  {https://ui.adsabs.harvard.edu/abs/2022Natur.607...48L} {607, 48}

\bibitem[\protect\citeauthoryear{{Latif}, {Whalen}  \& {Khochfar}}{{Latif}
  et~al.}{2022b}]{Latif:2022a}
{Latif} M.~A.,  {Whalen} D.,   {Khochfar} S.,  2022b, \mn@doi [\apj]
  {10.3847/1538-4357/ac3916}, \href
  {https://ui.adsabs.harvard.edu/abs/2022ApJ...925...28L} {925, 28}

\bibitem[\protect\citeauthoryear{{Lee}, {Giavalisco}, {Conroy}, {Wechsler},
  {Ferguson}, {Somerville}, {Dickinson}  \& {Urry}}{{Lee}
  et~al.}{2009}]{Lee:2009}
{Lee} K.-S.,  {Giavalisco} M.,  {Conroy} C.,  {Wechsler} R.~H.,  {Ferguson}
  H.~C.,  {Somerville} R.~S.,  {Dickinson} M.~E.,   {Urry} C.~M.,  2009,
  \mn@doi [\apj] {10.1088/0004-637X/695/1/368}, \href
  {https://ui.adsabs.harvard.edu/abs/2009ApJ...695..368L} {695, 368}

\bibitem[\protect\citeauthoryear{{Leitherer} et~al.,}{{Leitherer}
  et~al.}{1999}]{Leitherer:1999}
{Leitherer} C.,  et~al., 1999, \mn@doi [\apjs] {10.1086/313233}, \href
  {https://ui.adsabs.harvard.edu/abs/1999ApJS..123....3L} {123, 3}

\bibitem[\protect\citeauthoryear{{Liu} \& {Bromm}}{{Liu} \&
  {Bromm}}{2020}]{Liu:2020}
{Liu} B.,  {Bromm} V.,  2020, \mn@doi [\mnras] {10.1093/mnras/staa2143}, \href
  {https://ui.adsabs.harvard.edu/abs/2020MNRAS.497.2839L} {497, 2839}

\bibitem[\protect\citeauthoryear{{Lodato} \& {Natarajan}}{{Lodato} \&
  {Natarajan}}{2006}]{Lodato:2006}
{Lodato} G.,  {Natarajan} P.,  2006, \mn@doi [\mnras]
  {10.1111/j.1365-2966.2006.10801.x}, \href
  {https://ui.adsabs.harvard.edu/abs/2006MNRAS.371.1813L} {371, 1813}

\bibitem[\protect\citeauthoryear{{Lovell}, {Vijayan}, {Thomas}, {Wilkins},
  {Barnes}, {Irodotou}  \& {Roper}}{{Lovell} et~al.}{2021}]{Lovell:2021}
{Lovell} C.~C.,  {Vijayan} A.~P.,  {Thomas} P.~A.,  {Wilkins} S.~M.,  {Barnes}
  D.~J.,  {Irodotou} D.,   {Roper} W.,  2021, \mn@doi [\mnras]
  {10.1093/mnras/staa3360}, \href
  {https://ui.adsabs.harvard.edu/abs/2021MNRAS.500.2127L} {500, 2127}

\bibitem[\protect\citeauthoryear{{Luo}, {Ardaneh}, {Shlosman}, {Nagamine},
  {Wise}  \& {Begelman}}{{Luo} et~al.}{2018}]{Luo:2018}
{Luo} Y.,  {Ardaneh} K.,  {Shlosman} I.,  {Nagamine} K.,  {Wise} J.~H.,
  {Begelman} M.~C.,  2018, \mn@doi [\mnras] {10.1093/mnras/sty362}, \href
  {https://ui.adsabs.harvard.edu/abs/2018MNRAS.476.3523L} {476, 3523}

\bibitem[\protect\citeauthoryear{{Luo}, {Shlosman}, {Nagamine}  \&
  {Fang}}{{Luo} et~al.}{2020}]{Luo:2020}
{Luo} Y.,  {Shlosman} I.,  {Nagamine} K.,   {Fang} T.,  2020, \mn@doi [\mnras]
  {10.1093/mnras/staa153}, \href
  {https://ui.adsabs.harvard.edu/abs/2020MNRAS.492.4917L} {492, 4917}

\bibitem[\protect\citeauthoryear{{Lupi}, {Haiman}  \& {Volonteri}}{{Lupi}
  et~al.}{2021}]{Lupi:2021}
{Lupi} A.,  {Haiman} Z.,   {Volonteri} M.,  2021, \mn@doi [\mnras]
  {10.1093/mnras/stab692}, \href
  {https://ui.adsabs.harvard.edu/abs/2021MNRAS.503.5046L} {503, 5046}

\bibitem[\protect\citeauthoryear{{Machacek}, {Bryan}  \& {Abel}}{{Machacek}
  et~al.}{2001}]{Machacek:2001}
{Machacek} M.~E.,  {Bryan} G.~L.,   {Abel} T.,  2001, \mn@doi [\apj]
  {10.1086/319014}, \href
  {https://ui.adsabs.harvard.edu/abs/2001ApJ...548..509M} {548, 509}

\bibitem[\protect\citeauthoryear{{Madau} \& {Dickinson}}{{Madau} \&
  {Dickinson}}{2014}]{Madau:2014}
{Madau} P.,  {Dickinson} M.,  2014, \mn@doi [\araa]
  {10.1146/annurev-astro-081811-125615}, \href
  {https://ui.adsabs.harvard.edu/abs/2014ARA&A..52..415M} {52, 415}

\bibitem[\protect\citeauthoryear{{Madau} \& {Rees}}{{Madau} \&
  {Rees}}{2001}]{Madau:2001}
{Madau} P.,  {Rees} M.~J.,  2001, \mn@doi [\apjl] {10.1086/319848}, \href
  {https://ui.adsabs.harvard.edu/abs/2001ApJ...551L..27M} {551, L27}

\bibitem[\protect\citeauthoryear{{Maio}, {Ciardi}, {Dolag}, {Tornatore}  \&
  {Khochfar}}{{Maio} et~al.}{2010}]{Maio:2010}
{Maio} U.,  {Ciardi} B.,  {Dolag} K.,  {Tornatore} L.,   {Khochfar} S.,  2010,
  \mn@doi [\mnras] {10.1111/j.1365-2966.2010.17003.x}, \href
  {https://ui.adsabs.harvard.edu/abs/2010MNRAS.407.1003M} {407, 1003}

\bibitem[\protect\citeauthoryear{{Maio}, {Khochfar}, {Johnson}  \&
  {Ciardi}}{{Maio} et~al.}{2011}]{Maio:2011}
{Maio} U.,  {Khochfar} S.,  {Johnson} J.~L.,   {Ciardi} B.,  2011, \mn@doi
  [\mnras] {10.1111/j.1365-2966.2011.18455.x}, \href
  {https://ui.adsabs.harvard.edu/abs/2011MNRAS.414.1145M} {414, 1145}

\bibitem[\protect\citeauthoryear{{McLaughlin}, {Stancil}, {Sadeghpour}  \&
  {Forrey}}{{McLaughlin} et~al.}{2017}]{McLaughlin:2017}
{McLaughlin} B.~M.,  {Stancil} P.~C.,  {Sadeghpour} H.~R.,   {Forrey} R.~C.,
  2017, \mn@doi [Journal of Physics B Atomic Molecular Physics]
  {10.1088/1361-6455/aa6c1f}, \href
  {https://ui.adsabs.harvard.edu/abs/2017JPhB...50k4001M} {50, 114001}

\bibitem[\protect\citeauthoryear{{McLeod}, {McLure}, {Dunlop}, {Robertson},
  {Ellis}  \& {Targett}}{{McLeod} et~al.}{2015}]{McLeod:2015}
{McLeod} D.~J.,  {McLure} R.~J.,  {Dunlop} J.~S.,  {Robertson} B.~E.,  {Ellis}
  R.~S.,   {Targett} T.~A.,  2015, \mn@doi [\mnras] {10.1093/mnras/stv780},
  \href {https://ui.adsabs.harvard.edu/abs/2015MNRAS.450.3032M} {450, 3032}

\bibitem[\protect\citeauthoryear{{McLeod}, {McLure}  \& {Dunlop}}{{McLeod}
  et~al.}{2016}]{McLeod:2016}
{McLeod} D.~J.,  {McLure} R.~J.,   {Dunlop} J.~S.,  2016, \mn@doi [\mnras]
  {10.1093/mnras/stw904}, \href
  {https://ui.adsabs.harvard.edu/abs/2016MNRAS.459.3812M} {459, 3812}

\bibitem[\protect\citeauthoryear{{McLure} et~al.,}{{McLure}
  et~al.}{2013}]{McLure:2013}
{McLure} R.~J.,  et~al., 2013, \mn@doi [\mnras] {10.1093/mnras/stt627}, \href
  {https://ui.adsabs.harvard.edu/abs/2013MNRAS.432.2696M} {432, 2696}

\bibitem[\protect\citeauthoryear{{Mortlock} et~al.,}{{Mortlock}
  et~al.}{2011}]{Mortlock:2011}
{Mortlock} D.~J.,  et~al., 2011, \mn@doi [\nat] {10.1038/nature10159}, \href
  {https://ui.adsabs.harvard.edu/abs/2011Natur.474..616M} {474, 616}

\bibitem[\protect\citeauthoryear{{Nagamine}, {Choi}  \& {Yajima}}{{Nagamine}
  et~al.}{2010}]{Nagamine:2010}
{Nagamine} K.,  {Choi} J.-H.,   {Yajima} H.,  2010, \mn@doi [\apjl]
  {10.1088/2041-8205/725/2/L219}, \href
  {https://ui.adsabs.harvard.edu/abs/2010ApJ...725L.219N} {725, L219}

\bibitem[\protect\citeauthoryear{{Nakajima} \& {Maiolino}}{{Nakajima} \&
  {Maiolino}}{2022}]{Nakajima:2022}
{Nakajima} K.,  {Maiolino} R.,  2022, \mn@doi [\mnras]
  {10.1093/mnras/stac1242}, \href
  {https://ui.adsabs.harvard.edu/abs/2022MNRAS.513.5134N} {513, 5134}

\bibitem[\protect\citeauthoryear{{Neistein}, {Khochfar}, {Dalla Vecchia}  \&
  {Schaye}}{{Neistein} et~al.}{2012}]{Neistein:2012}
{Neistein} E.,  {Khochfar} S.,  {Dalla Vecchia} C.,   {Schaye} J.,  2012,
  \mn@doi [\mnras] {10.1111/j.1365-2966.2012.20584.x}, \href
  {https://ui.adsabs.harvard.edu/abs/2012MNRAS.421.3579N} {421, 3579}

\bibitem[\protect\citeauthoryear{{O'Shea} \& {Norman}}{{O'Shea} \&
  {Norman}}{2008}]{O'Shea:2008}
{O'Shea} B.~W.,  {Norman} M.~L.,  2008, \mn@doi [\apj] {10.1086/524006}, \href
  {https://ui.adsabs.harvard.edu/abs/2008ApJ...673...14O} {673, 14}

\bibitem[\protect\citeauthoryear{{O'Shea}, {Wise}, {Xu}  \& {Norman}}{{O'Shea}
  et~al.}{2015}]{O'Shea:2015}
{O'Shea} B.~W.,  {Wise} J.~H.,  {Xu} H.,   {Norman} M.~L.,  2015, \mn@doi
  [\apjl] {10.1088/2041-8205/807/1/L12}, \href
  {https://ui.adsabs.harvard.edu/abs/2015ApJ...807L..12O} {807, L12}

\bibitem[\protect\citeauthoryear{{Oesch} et~al.,}{{Oesch}
  et~al.}{2014}]{Oesch:2014}
{Oesch} P.~A.,  et~al., 2014, \mn@doi [\apj] {10.1088/0004-637X/786/2/108},
  \href {https://ui.adsabs.harvard.edu/abs/2014ApJ...786..108O} {786, 108}

\bibitem[\protect\citeauthoryear{{Oesch}, {Bouwens}, {Illingworth}, {Labb{\'e}}
   \& {Stefanon}}{{Oesch} et~al.}{2018}]{Oesch:2018}
{Oesch} P.~A.,  {Bouwens} R.~J.,  {Illingworth} G.~D.,  {Labb{\'e}} I.,
  {Stefanon} M.,  2018, \mn@doi [\apj] {10.3847/1538-4357/aab03f}, \href
  {https://ui.adsabs.harvard.edu/abs/2018ApJ...855..105O} {855, 105}

\bibitem[\protect\citeauthoryear{{Omukai}, {Tsuribe}, {Schneider}  \&
  {Ferrara}}{{Omukai} et~al.}{2005}]{Omukai:2005}
{Omukai} K.,  {Tsuribe} T.,  {Schneider} R.,   {Ferrara} A.,  2005, \mn@doi
  [\apj] {10.1086/429955}, \href
  {https://ui.adsabs.harvard.edu/abs/2005ApJ...626..627O} {626, 627}

\bibitem[\protect\citeauthoryear{{Omukai}, {Schneider}  \& {Haiman}}{{Omukai}
  et~al.}{2008}]{Omukai:2008}
{Omukai} K.,  {Schneider} R.,   {Haiman} Z.,  2008, \mn@doi [\apj]
  {10.1086/591636}, \href
  {https://ui.adsabs.harvard.edu/abs/2008ApJ...686..801O} {686, 801}

\bibitem[\protect\citeauthoryear{{Paardekooper}, {Khochfar}  \&
  {Dalla}}{{Paardekooper} et~al.}{2013}]{Paardekooper:2013}
{Paardekooper} J.~P.,  {Khochfar} S.,   {Dalla} C.~V.,  2013, \mn@doi [\mnras]
  {10.1093/mnrasl/sls032}, \href
  {https://ui.adsabs.harvard.edu/abs/2013MNRAS.429L..94P} {429, L94}

\bibitem[\protect\citeauthoryear{{Paardekooper}, {Khochfar}  \& {Dalla
  Vecchia}}{{Paardekooper} et~al.}{2015}]{Paardekooper:2015}
{Paardekooper} J.-P.,  {Khochfar} S.,   {Dalla Vecchia} C.,  2015, \mn@doi
  [\mnras] {10.1093/mnras/stv1114}, \href
  {https://ui.adsabs.harvard.edu/abs/2015MNRAS.451.2544P} {451, 2544}

\bibitem[\protect\citeauthoryear{{Park}, {Ricotti}  \& {Sugimura}}{{Park}
  et~al.}{2021}]{Park:2021}
{Park} J.,  {Ricotti} M.,   {Sugimura} K.,  2021, \mn@doi [\mnras]
  {10.1093/mnras/stab2999}, \href
  {https://ui.adsabs.harvard.edu/abs/2021MNRAS.508.6176P} {508, 6176}

\bibitem[\protect\citeauthoryear{{Peebles} \& {Dicke}}{{Peebles} \&
  {Dicke}}{1968}]{Peebles:1968}
{Peebles} P.~J.~E.,  {Dicke} R.~H.,  1968, \mn@doi [\apj] {10.1086/149811},
  \href {https://ui.adsabs.harvard.edu/abs/1968ApJ...154..891P} {154, 891}

\bibitem[\protect\citeauthoryear{{Perez} \& {Granger}}{{Perez} \&
  {Granger}}{2007}]{Perez:2007}
{Perez} F.,  {Granger} B.~E.,  2007, \mn@doi [Computing in Science and
  Engineering] {10.1109/MCSE.2007.53}, \href
  {https://ui.adsabs.harvard.edu/abs/2007CSE.....9c..21P} {9, 21}

\bibitem[\protect\citeauthoryear{{Phipps}, {Khochfar}, {Varri}  \& {Dalla
  Vecchia}}{{Phipps} et~al.}{2020}]{Phipps:2020}
{Phipps} F.,  {Khochfar} S.,  {Varri} A.~L.,   {Dalla Vecchia} C.,  2020,
  \mn@doi [\aap] {10.1051/0004-6361/202037884}, \href
  {https://ui.adsabs.harvard.edu/abs/2020A&A...641A.132P} {641, A132}

\bibitem[\protect\citeauthoryear{{Pillepich} et~al.,}{{Pillepich}
  et~al.}{2018}]{Pillepich:2018b}
{Pillepich} A.,  et~al., 2018, \mn@doi [\mnras] {10.1093/mnras/stx3112}, \href
  {https://ui.adsabs.harvard.edu/abs/2018MNRAS.475..648P} {475, 648}

\bibitem[\protect\citeauthoryear{{Qin}, {Mesinger}, {Park}, {Greig}  \&
  {Mu{\~n}oz}}{{Qin} et~al.}{2020}]{Qin:2020}
{Qin} Y.,  {Mesinger} A.,  {Park} J.,  {Greig} B.,   {Mu{\~n}oz} J.~B.,  2020,
  \mn@doi [\mnras] {10.1093/mnras/staa1131}, \href
  {https://ui.adsabs.harvard.edu/abs/2020MNRAS.495..123Q} {495, 123}

\bibitem[\protect\citeauthoryear{{Raiter}, {Schaerer}  \& {Fosbury}}{{Raiter}
  et~al.}{2010}]{Raiter:2010}
{Raiter} A.,  {Schaerer} D.,   {Fosbury} R.~A.~E.,  2010, \mn@doi [\aap]
  {10.1051/0004-6361/201015236}, \href
  {https://ui.adsabs.harvard.edu/abs/2010A&A...523A..64R} {523, A64}

\bibitem[\protect\citeauthoryear{{Regan}}{{Regan}}{2022}]{Regan:2022}
{Regan} J.,  2022, arXiv e-prints, \href
  {https://ui.adsabs.harvard.edu/abs/2022arXiv221004899R} {p. arXiv:2210.04899}

\bibitem[\protect\citeauthoryear{{Regan}, {Johansson}  \& {Haehnelt}}{{Regan}
  et~al.}{2014}]{Regan:2014a}
{Regan} J.~A.,  {Johansson} P.~H.,   {Haehnelt} M.~G.,  2014, \mn@doi [\mnras]
  {10.1093/mnras/stu068}, \href
  {https://ui.adsabs.harvard.edu/abs/2014MNRAS.439.1160R} {439, 1160}

\bibitem[\protect\citeauthoryear{{Regan}, {Johansson}  \& {Wise}}{{Regan}
  et~al.}{2016}]{Regan:2016}
{Regan} J.~A.,  {Johansson} P.~H.,   {Wise} J.~H.,  2016, \mn@doi [\mnras]
  {10.1093/mnras/stw1307}, \href
  {https://ui.adsabs.harvard.edu/abs/2016MNRAS.461..111R} {461, 111}

\bibitem[\protect\citeauthoryear{{Rosdahl} et~al.,}{{Rosdahl}
  et~al.}{2018}]{Rosdahl:2018}
{Rosdahl} J.,  et~al., 2018, \mn@doi [\mnras] {10.1093/mnras/sty1655}, \href
  {https://ui.adsabs.harvard.edu/abs/2018MNRAS.479..994R} {479, 994}

\bibitem[\protect\citeauthoryear{{Rossi}, {Salvadori}  \&
  {Sk{\'u}lad{\'o}ttir}}{{Rossi} et~al.}{2021}]{Rossi:2021}
{Rossi} M.,  {Salvadori} S.,   {Sk{\'u}lad{\'o}ttir} {\'A}.,  2021, \mn@doi
  [\mnras] {10.1093/mnras/stab821}, \href
  {https://ui.adsabs.harvard.edu/abs/2021MNRAS.503.6026R} {503, 6026}

\bibitem[\protect\citeauthoryear{{Salpeter}}{{Salpeter}}{1955}]{Salpeter:1955}
{Salpeter} E.~E.,  1955, \mn@doi [\apj] {10.1086/145971}, \href
  {https://ui.adsabs.harvard.edu/abs/1955ApJ...121..161S} {121, 161}

\bibitem[\protect\citeauthoryear{{Salumbides}, {Bagdonaite}, {Abgrall},
  {Roueff}  \& {Ubachs}}{{Salumbides} et~al.}{2015}]{Salumbides:2015}
{Salumbides} E.~J.,  {Bagdonaite} J.,  {Abgrall} H.,  {Roueff} E.,   {Ubachs}
  W.,  2015, \mn@doi [\mnras] {10.1093/mnras/stv656}, \href
  {https://ui.adsabs.harvard.edu/abs/2015MNRAS.450.1237S} {450, 1237}

\bibitem[\protect\citeauthoryear{{Sarmento} \& {Scannapieco}}{{Sarmento} \&
  {Scannapieco}}{2022}]{Sarmento:2022}
{Sarmento} R.,  {Scannapieco} E.,  2022, \mn@doi [\apj]
  {10.3847/1538-4357/ac815c}, \href
  {https://ui.adsabs.harvard.edu/abs/2022ApJ...935..174S} {935, 174}

\bibitem[\protect\citeauthoryear{{Saslaw} \& {Zipoy}}{{Saslaw} \&
  {Zipoy}}{1967}]{Saslaw:1967}
{Saslaw} W.~C.,  {Zipoy} D.,  1967, \mn@doi [\nat] {10.1038/216976a0}, \href
  {https://ui.adsabs.harvard.edu/abs/1967Natur.216..976S} {216, 976}

\bibitem[\protect\citeauthoryear{{Sassano}, {Schneider}, {Valiante},
  {Inayoshi}, {Chon}, {Omukai}, {Mayer}  \& {Capelo}}{{Sassano}
  et~al.}{2021}]{Sassano:2021}
{Sassano} F.,  {Schneider} R.,  {Valiante} R.,  {Inayoshi} K.,  {Chon} S.,
  {Omukai} K.,  {Mayer} L.,   {Capelo} P.~R.,  2021, \mn@doi [\mnras]
  {10.1093/mnras/stab1737}, \href
  {https://ui.adsabs.harvard.edu/abs/2021MNRAS.506..613S} {506, 613}

\bibitem[\protect\citeauthoryear{{Schaerer}}{{Schaerer}}{2002}]{Schaerer:2002}
{Schaerer} D.,  2002, \mn@doi [\aap] {10.1051/0004-6361:20011619}, \href
  {https://ui.adsabs.harvard.edu/abs/2002A&A...382...28S} {382, 28}

\bibitem[\protect\citeauthoryear{{Schauer}, {Glover}, {Klessen}  \&
  {Clark}}{{Schauer} et~al.}{2021}]{Schauer:2021}
{Schauer} A. T.~P.,  {Glover} S. C.~O.,  {Klessen} R.~S.,   {Clark} P.,  2021,
  \mn@doi [\mnras] {10.1093/mnras/stab1953}, \href
  {https://ui.adsabs.harvard.edu/abs/2021MNRAS.507.1775S} {507, 1775}

\bibitem[\protect\citeauthoryear{{Schaye} \& {Dalla Vecchia}}{{Schaye} \&
  {Dalla Vecchia}}{2008}]{Schaye:2008}
{Schaye} J.,  {Dalla Vecchia} C.,  2008, \mn@doi [\mnras]
  {10.1111/j.1365-2966.2007.12639.x}, \href
  {https://ui.adsabs.harvard.edu/abs/2008MNRAS.383.1210S} {383, 1210}

\bibitem[\protect\citeauthoryear{{Schaye} et~al.,}{{Schaye}
  et~al.}{2010}]{Schaye:2010}
{Schaye} J.,  et~al., 2010, \mn@doi [\mnras]
  {10.1111/j.1365-2966.2009.16029.x}, \href
  {https://ui.adsabs.harvard.edu/abs/2010MNRAS.402.1536S} {402, 1536}

\bibitem[\protect\citeauthoryear{{Schmidt}}{{Schmidt}}{1959}]{Schmidt:1959}
{Schmidt} M.,  1959, \mn@doi [\apj] {10.1086/146614}, \href
  {https://ui.adsabs.harvard.edu/abs/1959ApJ...129..243S} {129, 243}

\bibitem[\protect\citeauthoryear{{Shang}, {Bryan}  \& {Haiman}}{{Shang}
  et~al.}{2010}]{Shang:2010}
{Shang} C.,  {Bryan} G.~L.,   {Haiman} Z.,  2010, \mn@doi [\mnras]
  {10.1111/j.1365-2966.2009.15960.x}, \href
  {https://ui.adsabs.harvard.edu/abs/2010MNRAS.402.1249S} {402, 1249}

\bibitem[\protect\citeauthoryear{{Shapiro} \& {Kang}}{{Shapiro} \&
  {Kang}}{1987}]{Shapiro:1987}
{Shapiro} P.~R.,  {Kang} H.,  1987, \mn@doi [\apj] {10.1086/165350}, \href
  {https://ui.adsabs.harvard.edu/abs/1987ApJ...318...32S} {318, 32}

\bibitem[\protect\citeauthoryear{{Smith}, {Sigurdsson}  \& {Abel}}{{Smith}
  et~al.}{2008}]{Smith:2008}
{Smith} B.,  {Sigurdsson} S.,   {Abel} T.,  2008, \mn@doi [\mnras]
  {10.1111/j.1365-2966.2008.12922.x}, \href
  {https://ui.adsabs.harvard.edu/abs/2008MNRAS.385.1443S} {385, 1443}

\bibitem[\protect\citeauthoryear{{Smith}, {Wise}, {O'Shea}, {Norman}  \&
  {Khochfar}}{{Smith} et~al.}{2015}]{Smith:2015}
{Smith} B.~D.,  {Wise} J.~H.,  {O'Shea} B.~W.,  {Norman} M.~L.,   {Khochfar}
  S.,  2015, \mn@doi [\mnras] {10.1093/mnras/stv1509}, \href
  {https://ui.adsabs.harvard.edu/abs/2015MNRAS.452.2822S} {452, 2822}

\bibitem[\protect\citeauthoryear{{Smith}, {Kannan}, {Garaldi}, {Vogelsberger},
  {Pakmor}, {Springel}  \& {Hernquist}}{{Smith} et~al.}{2022}]{Smith:2022}
{Smith} A.,  {Kannan} R.,  {Garaldi} E.,  {Vogelsberger} M.,  {Pakmor} R.,
  {Springel} V.,   {Hernquist} L.,  2022, \mn@doi [\mnras]
  {10.1093/mnras/stac713}, \href
  {https://ui.adsabs.harvard.edu/abs/2022MNRAS.512.3243S} {512, 3243}

\bibitem[\protect\citeauthoryear{{Solomon}}{{Solomon}}{1965}]{Solomon:1965}
{Solomon} P.~M.,  1965, PhD thesis, THE UNIVERSITY OF WISCONSIN - MADISON.

\bibitem[\protect\citeauthoryear{{Spinoso}, {Bonoli}, {Valiante}, {Schneider}
  \& {Izquierdo-Villalba}}{{Spinoso} et~al.}{2023}]{Spinoso:2023}
{Spinoso} D.,  {Bonoli} S.,  {Valiante} R.,  {Schneider} R.,
  {Izquierdo-Villalba} D.,  2023, \mn@doi [\mnras] {10.1093/mnras/stac3169},
  \href {https://ui.adsabs.harvard.edu/abs/2023MNRAS.518.4672S} {518, 4672}

\bibitem[\protect\citeauthoryear{{Springel}}{{Springel}}{2005}]{Springel:2005}
{Springel} V.,  2005, \mn@doi [\mnras] {10.1111/j.1365-2966.2005.09655.x},
  \href {https://ui.adsabs.harvard.edu/abs/2005MNRAS.364.1105S} {364, 1105}

\bibitem[\protect\citeauthoryear{{Springel}, {Yoshida}  \& {White}}{{Springel}
  et~al.}{2001}]{Springel:2001}
{Springel} V.,  {Yoshida} N.,   {White} S. D.~M.,  2001, \mn@doi [\na]
  {10.1016/S1384-1076(01)00042-2}, \href
  {https://ui.adsabs.harvard.edu/abs/2001NewA....6...79S} {6, 79}

\bibitem[\protect\citeauthoryear{{Stacy}, {Bromm}  \& {Lee}}{{Stacy}
  et~al.}{2016}]{Stacy:2016}
{Stacy} A.,  {Bromm} V.,   {Lee} A.~T.,  2016, \mn@doi [\mnras]
  {10.1093/mnras/stw1728}, \href
  {https://ui.adsabs.harvard.edu/abs/2016MNRAS.462.1307S} {462, 1307}

\bibitem[\protect\citeauthoryear{{Stanway} \& {Eldridge}}{{Stanway} \&
  {Eldridge}}{2018}]{Stanway:2018}
{Stanway} E.~R.,  {Eldridge} J.~J.,  2018, \mn@doi [\mnras]
  {10.1093/mnras/sty1353}, \href
  {https://ui.adsabs.harvard.edu/abs/2018MNRAS.479...75S} {479, 75}

\bibitem[\protect\citeauthoryear{{Stecher} \& {Williams}}{{Stecher} \&
  {Williams}}{1967}]{Stecher:1967}
{Stecher} T.~P.,  {Williams} D.~A.,  1967, \mn@doi [\apjl] {10.1086/180047},
  \href {https://ui.adsabs.harvard.edu/abs/1967ApJ...149L..29S} {149, L29}

\bibitem[\protect\citeauthoryear{{Sugimura}, {Omukai}  \& {Inoue}}{{Sugimura}
  et~al.}{2014}]{Sugimura:2014}
{Sugimura} K.,  {Omukai} K.,   {Inoue} A.~K.,  2014, \mn@doi [\mnras]
  {10.1093/mnras/stu1778}, \href
  {https://ui.adsabs.harvard.edu/abs/2014MNRAS.445..544S} {445, 544}

\bibitem[\protect\citeauthoryear{{Sugimura}, {Coppola}, {Omukai}, {Galli}  \&
  {Palla}}{{Sugimura} et~al.}{2016}]{Sugimura:2016}
{Sugimura} K.,  {Coppola} C.~M.,  {Omukai} K.,  {Galli} D.,   {Palla} F.,
  2016, \mn@doi [\mnras] {10.1093/mnras/stv2655}, \href
  {https://ui.adsabs.harvard.edu/abs/2016MNRAS.456..270S} {456, 270}

\bibitem[\protect\citeauthoryear{{Tanaka}, {Perna}  \& {Haiman}}{{Tanaka}
  et~al.}{2012}]{Tanaka:2012}
{Tanaka} T.,  {Perna} R.,   {Haiman} Z.,  2012, \mn@doi [\mnras]
  {10.1111/j.1365-2966.2012.21539.x}, \href
  {https://ui.adsabs.harvard.edu/abs/2012MNRAS.425.2974T} {425, 2974}

\bibitem[\protect\citeauthoryear{{Tegmark}, {Silk}, {Rees}, {Blanchard}, {Abel}
   \& {Palla}}{{Tegmark} et~al.}{1997}]{Tegmark:1997}
{Tegmark} M.,  {Silk} J.,  {Rees} M.~J.,  {Blanchard} A.,  {Abel} T.,   {Palla}
  F.,  1997, \mn@doi [\apj] {10.1086/303434}, \href
  {https://ui.adsabs.harvard.edu/abs/1997ApJ...474....1T} {474, 1}

\bibitem[\protect\citeauthoryear{{Tornatore}, {Ferrara}  \&
  {Schneider}}{{Tornatore} et~al.}{2007}]{Tornatore:2007}
{Tornatore} L.,  {Ferrara} A.,   {Schneider} R.,  2007, \mn@doi [\mnras]
  {10.1111/j.1365-2966.2007.12215.x}, \href
  {https://ui.adsabs.harvard.edu/abs/2007MNRAS.382..945T} {382, 945}

\bibitem[\protect\citeauthoryear{{Trebitsch} et~al.,}{{Trebitsch}
  et~al.}{2021}]{Trebitsch:2021}
{Trebitsch} M.,  et~al., 2021, \mn@doi [\aap] {10.1051/0004-6361/202037698},
  \href {https://ui.adsabs.harvard.edu/abs/2021A&A...653A.154T} {653, A154}

\bibitem[\protect\citeauthoryear{{Trenti} \& {Stiavelli}}{{Trenti} \&
  {Stiavelli}}{2009}]{Trenti:2009}
{Trenti} M.,  {Stiavelli} M.,  2009, \mn@doi [\apj]
  {10.1088/0004-637X/694/2/879}, \href
  {https://ui.adsabs.harvard.edu/abs/2009ApJ...694..879T} {694, 879}

\bibitem[\protect\citeauthoryear{{Tseliakhovich} \& {Hirata}}{{Tseliakhovich}
  \& {Hirata}}{2010}]{Tseliakhovich:2010}
{Tseliakhovich} D.,  {Hirata} C.,  2010, \mn@doi [\prd]
  {10.1103/PhysRevD.82.083520}, \href
  {https://ui.adsabs.harvard.edu/abs/2010PhRvD..82h3520T} {82, 083520}

\bibitem[\protect\citeauthoryear{{Ubachs}, {Salumbides}, {Murphy}, {Abgrall}
  \& {Roueff}}{{Ubachs} et~al.}{2019}]{Ubachs:2019}
{Ubachs} W.,  {Salumbides} E.~J.,  {Murphy} M.~T.,  {Abgrall} H.,   {Roueff}
  E.,  2019, \mn@doi [\aap] {10.1051/0004-6361/201834782}, \href
  {https://ui.adsabs.harvard.edu/abs/2019A&A...622A.127U} {622, A127}

\bibitem[\protect\citeauthoryear{{Van Rossum} \& {Drake}}{{Van Rossum} \&
  {Drake}}{2009}]{vanRossum:2009}
{Van Rossum} G.,  {Drake} F.~L.,  2009, {Python 3 Reference Manual}.
CreateSpace, Scotts Valley, CA, \mn@doi{10.5555/1593511}

\bibitem[\protect\citeauthoryear{{Virtanen} et~al.,}{{Virtanen}
  et~al.}{2020}]{Virtanen:2020}
{Virtanen} P.,  et~al., 2020, \mn@doi [Nature Methods]
  {10.1038/s41592-019-0686-2}, \href
  {https://ui.adsabs.harvard.edu/abs/2020NatMe..17..261V} {17, 261}

\bibitem[\protect\citeauthoryear{{Visbal}, {Bryan}  \& {Haiman}}{{Visbal}
  et~al.}{2020}]{Visbal:2020}
{Visbal} E.,  {Bryan} G.~L.,   {Haiman} Z.,  2020, \mn@doi [\apj]
  {10.3847/1538-4357/ab994e}, \href
  {https://ui.adsabs.harvard.edu/abs/2020ApJ...897...95V} {897, 95}

\bibitem[\protect\citeauthoryear{{Vogelsberger}, {Marinacci}, {Torrey}  \&
  {Puchwein}}{{Vogelsberger} et~al.}{2020}]{Vogelsberger:2020}
{Vogelsberger} M.,  {Marinacci} F.,  {Torrey} P.,   {Puchwein} E.,  2020,
  \mn@doi [Nature Reviews Physics] {10.1038/s42254-019-0127-2}, \href
  {https://ui.adsabs.harvard.edu/abs/2020NatRP...2...42V} {2, 42}

\bibitem[\protect\citeauthoryear{{Wells} \& {Norman}}{{Wells} \&
  {Norman}}{2022}]{Wells:2022}
{Wells} A.~I.,  {Norman} M.~L.,  2022, \mn@doi [\apj]
  {10.3847/1538-4357/ac6c87}, \href
  {https://ui.adsabs.harvard.edu/abs/2022ApJ...932...71W} {932, 71}

\bibitem[\protect\citeauthoryear{{Welsh}, {Cooke}  \& {Fumagalli}}{{Welsh}
  et~al.}{2019}]{Welsh:2019}
{Welsh} L.,  {Cooke} R.,   {Fumagalli} M.,  2019, \mn@doi [\mnras]
  {10.1093/mnras/stz1526}, \href
  {https://ui.adsabs.harvard.edu/abs/2019MNRAS.487.3363W} {487, 3363}

\bibitem[\protect\citeauthoryear{{Welsh}, {Cooke}, {Fumagalli}  \&
  {Pettini}}{{Welsh} et~al.}{2022}]{Welsh:2022}
{Welsh} L.,  {Cooke} R.,  {Fumagalli} M.,   {Pettini} M.,  2022, \mn@doi [\apj]
  {10.3847/1538-4357/ac4503}, \href
  {https://ui.adsabs.harvard.edu/abs/2022ApJ...929..158W} {929, 158}

\bibitem[\protect\citeauthoryear{{Wiersma}, {Schaye}  \& {Smith}}{{Wiersma}
  et~al.}{2009}]{Wiersma:2009}
{Wiersma} R. P.~C.,  {Schaye} J.,   {Smith} B.~D.,  2009, \mn@doi [\mnras]
  {10.1111/j.1365-2966.2008.14191.x}, \href
  {https://ui.adsabs.harvard.edu/abs/2009MNRAS.393...99W} {393, 99}

\bibitem[\protect\citeauthoryear{{Wilkins} et~al.,}{{Wilkins}
  et~al.}{2022}]{Wilkins:2022}
{Wilkins} S.~M.,  et~al., 2022, \mn@doi [\mnras] {10.1093/mnras/stac3280},
  \href {https://ui.adsabs.harvard.edu/abs/2022MNRAS.tmp.3128W} {}

\bibitem[\protect\citeauthoryear{{Wise} \& {Abel}}{{Wise} \&
  {Abel}}{2005}]{Wise:2005}
{Wise} J.~H.,  {Abel} T.,  2005, \mn@doi [\apj] {10.1086/430434}, \href
  {https://ui.adsabs.harvard.edu/abs/2005ApJ...629..615W} {629, 615}

\bibitem[\protect\citeauthoryear{{Wise} \& {Abel}}{{Wise} \&
  {Abel}}{2007}]{Wise:2007}
{Wise} J.~H.,  {Abel} T.,  2007, \mn@doi [\apj] {10.1086/522876}, \href
  {https://ui.adsabs.harvard.edu/abs/2007ApJ...671.1559W} {671, 1559}

\bibitem[\protect\citeauthoryear{{Wise}, {Abel}, {Turk}, {Norman}  \&
  {Smith}}{{Wise} et~al.}{2012a}]{Wise:2012b}
{Wise} J.~H.,  {Abel} T.,  {Turk} M.~J.,  {Norman} M.~L.,   {Smith} B.~D.,
  2012a, \mn@doi [\mnras] {10.1111/j.1365-2966.2012.21809.x}, \href
  {https://ui.adsabs.harvard.edu/abs/2012MNRAS.427..311W} {427, 311}

\bibitem[\protect\citeauthoryear{{Wise}, {Turk}, {Norman}  \& {Abel}}{{Wise}
  et~al.}{2012b}]{Wise:2012a}
{Wise} J.~H.,  {Turk} M.~J.,  {Norman} M.~L.,   {Abel} T.,  2012b, \mn@doi
  [\apj] {10.1088/0004-637X/745/1/50}, \href
  {https://ui.adsabs.harvard.edu/abs/2012ApJ...745...50W} {745, 50}

\bibitem[\protect\citeauthoryear{{Wise}, {Regan}, {O'Shea}, {Norman}, {Downes}
  \& {Xu}}{{Wise} et~al.}{2019}]{Wise:2019}
{Wise} J.~H.,  {Regan} J.~A.,  {O'Shea} B.~W.,  {Norman} M.~L.,  {Downes}
  T.~P.,   {Xu} H.,  2019, \mn@doi [\nat] {10.1038/s41586-019-0873-4}, \href
  {https://ui.adsabs.harvard.edu/abs/2019Natur.566...85W} {566, 85}

\bibitem[\protect\citeauthoryear{{Wolcott-Green} \& {Haiman}}{{Wolcott-Green}
  \& {Haiman}}{2019}]{Wolcott-Green:2019}
{Wolcott-Green} J.,  {Haiman} Z.,  2019, \mn@doi [\mnras]
  {10.1093/mnras/sty3280}, \href
  {https://ui.adsabs.harvard.edu/abs/2019MNRAS.484.2467W} {484, 2467}

\bibitem[\protect\citeauthoryear{{Wolcott-Green}, {Haiman}  \&
  {Bryan}}{{Wolcott-Green} et~al.}{2011}]{Wolcott-Green:2011}
{Wolcott-Green} J.,  {Haiman} Z.,   {Bryan} G.~L.,  2011, \mn@doi [\mnras]
  {10.1111/j.1365-2966.2011.19538.x}, \href
  {https://ui.adsabs.harvard.edu/abs/2011MNRAS.418..838W} {418, 838}

\bibitem[\protect\citeauthoryear{{Wolcott-Green}, {Haiman}  \&
  {Bryan}}{{Wolcott-Green} et~al.}{2017}]{Wolcott-Green:2017}
{Wolcott-Green} J.,  {Haiman} Z.,   {Bryan} G.~L.,  2017, \mn@doi [\mnras]
  {10.1093/mnras/stx167}, \href
  {https://ui.adsabs.harvard.edu/abs/2017MNRAS.469.3329W} {469, 3329}

\bibitem[\protect\citeauthoryear{{Woods}, {Willott}, {Regan}, {Wise}, {Downes},
  {Norman}  \& {O'Shea}}{{Woods} et~al.}{2021}]{Woods:2021}
{Woods} T.~E.,  {Willott} C.~J.,  {Regan} J.~A.,  {Wise} J.~H.,  {Downes}
  T.~P.,  {Norman} M.~L.,   {O'Shea} B.~W.,  2021, \mn@doi [\apjl]
  {10.3847/2041-8213/ac2a45}, \href
  {https://ui.adsabs.harvard.edu/abs/2021ApJ...920L..22W} {920, L22}

\bibitem[\protect\citeauthoryear{{Xu}, {Wise}, {Norman}, {Ahn}  \&
  {O'Shea}}{{Xu} et~al.}{2016}]{Xu:2016}
{Xu} H.,  {Wise} J.~H.,  {Norman} M.~L.,  {Ahn} K.,   {O'Shea} B.~W.,  2016,
  \mn@doi [\apj] {10.3847/1538-4357/833/1/84}, \href
  {https://ui.adsabs.harvard.edu/abs/2016ApJ...833...84X} {833, 84}

\bibitem[\protect\citeauthoryear{{Yajima} \& {Khochfar}}{{Yajima} \&
  {Khochfar}}{2017}]{Yajima:2017}
{Yajima} H.,  {Khochfar} S.,  2017, \mn@doi [\mnras] {10.1093/mnrasl/slw249},
  \href {https://ui.adsabs.harvard.edu/abs/2017MNRAS.467L..51Y} {467, L51}

\bibitem[\protect\citeauthoryear{{Yoshida}, {Omukai}, {Hernquist}  \&
  {Abel}}{{Yoshida} et~al.}{2006}]{Yoshida:2006}
{Yoshida} N.,  {Omukai} K.,  {Hernquist} L.,   {Abel} T.,  2006, \mn@doi [\apj]
  {10.1086/507978}, \href
  {https://ui.adsabs.harvard.edu/abs/2006ApJ...652....6Y} {652, 6}

\bibitem[\protect\citeauthoryear{{Zackrisson}, {Rydberg}, {Schaerer},
  {{\"O}stlin}  \& {Tuli}}{{Zackrisson} et~al.}{2011}]{Zackrisson:2011}
{Zackrisson} E.,  {Rydberg} C.-E.,  {Schaerer} D.,  {{\"O}stlin} G.,   {Tuli}
  M.,  2011, \mn@doi [\apj] {10.1088/0004-637X/740/1/13}, \href
  {https://ui.adsabs.harvard.edu/abs/2011ApJ...740...13Z} {740, 13}

\bibitem[\protect\citeauthoryear{{Zammit} et~al.,}{{Zammit}
  et~al.}{2017}]{Zammit:2017}
{Zammit} M.~C.,  et~al., 2017, \mn@doi [\apj] {10.3847/1538-4357/aa9712}, \href
  {https://ui.adsabs.harvard.edu/abs/2017ApJ...851...64Z} {851, 64}

\bibitem[\protect\citeauthoryear{{Zammit} et~al.,}{{Zammit}
  et~al.}{2018}]{Zammit:2018}
{Zammit} M.~C.,  et~al., 2018, in Workshop on Astrophysical Opacities. p.~145

\bibitem[\protect\citeauthoryear{{da Silva}, {Fumagalli}  \& {Krumholz}}{{da
  Silva} et~al.}{2012}]{daSilva:2012}
{da Silva} R.~L.,  {Fumagalli} M.,   {Krumholz} M.~R.,  2012, \mn@doi [\apj]
  {10.1088/0004-637X/745/2/145}, \href
  {https://ui.adsabs.harvard.edu/abs/2012ApJ...745..145D} {745, 145}

\bibitem[\protect\citeauthoryear{{da Silva}, {Fumagalli}  \& {Krumholz}}{{da
  Silva} et~al.}{2014}]{daSilva:2014}
{da Silva} R.~L.,  {Fumagalli} M.,   {Krumholz} M.~R.,  2014, \mn@doi [\mnras]
  {10.1093/mnras/stu1688}, \href
  {https://ui.adsabs.harvard.edu/abs/2014MNRAS.444.3275D} {444, 3275}

\makeatother
\end{thebibliography}

\appendix

\section{SEDs examples}
\label{appen:example_spectra}

\begin{figure}
    \centering
    \includegraphics[width=\linewidth]{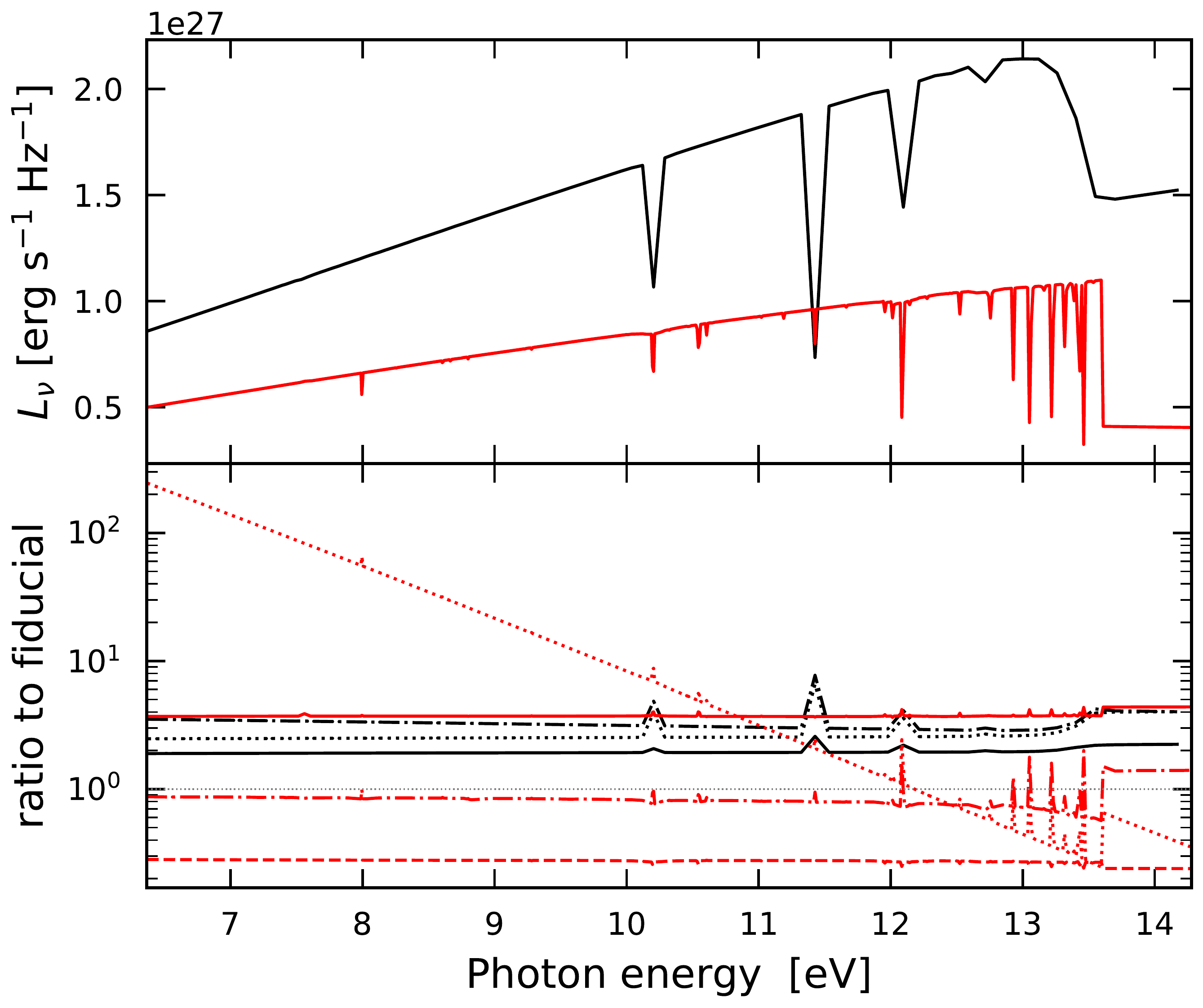}
    \vspace{-0.6cm}
    \caption{\textbf{\textit{Top panel}}: the UV section of the SEDs, relevant for the $\ce{H_2}$ dissociation, for a 1-Myr-old PopIII (black line) and a PopII (red line) population. Our fiducial choice of stellar models is shown here, hence Ygg2 for PopIII and BPASS\_Chab for PopII. The total mass is $10^6$ M$_\odot$ for each population. \textbf{\textit{Bottom panel}}: ratio of the other SEDs considered to the fiducial ones, again at an age of 1 Myr. Black lines represent PopIII models: Ygg1 (solid), Slug (dot-dashed) and BB5 (dotted). Red lines represent PopII models: BPASS\_TH (solid), BPASS\_BH (dashed), Slug (dot-dashed) and BB4 (dotted). Apart from the spectral features, that are influenced by the spectral resolution, the most noticeable feature is the completely different shape the all the PopII SEDs from the $10^4$ K black-body spectrum, that is commonly assumed in the literature to approximate them.}
    \vspace{-0.2cm}
\label{fig:example_spectra}
\end{figure}

In Fig.~\ref{fig:example_spectra} we compare the different stellar spectra employed in this work. The top panel reports the 1-Myr-old SEDs included in the fiducial setup: Ygg2 for PopIII stars in black and BPASS\_Chab for PopII stars in red. The UV spectral range shown in the x-axis is the relevant one for the \ce{H_2} dissociation rate. The additional SEDs described in Sec.~\ref{sec:SEDs} are shown in the bottom panel, as ratios relative to the corresponding fiducial choices. PopIII models are in black: Ygg1 (solid), Slug (dot-dashed) and BB5 (dotted). PopII models are in red: BPASS\_TH (solid), BPASS\_BH (dashed), Slug (dot-dashed) and BB4 (dotted). As expected, a young stellar population has a higher emission if a more top-heavy IMF is chosen (Ygg1 and BPASS\_TH) and the opposite is true for a bottom-heavy one (BPASS\_BH). The spectral shape seems consistent within the PopIII and PopII SEDs separately (ratios are mostly parallel to the horizontal line), apart from the BB4, i.e. the black-body spectrum with $T=10^4 \ \mathrm{K}$ that is commonly used in the literature to approximate PopII stellar emission \citep[e.g.][]{Johnson:2013,Glover:2015a}. The red dotted line in Fig.~\ref{fig:example_spectra} shows instead that the ratio between BB4 and BPASS\_Chab varies by $\sim3$ orders of magnitude in the 6-13.6 eV energy interval, hence implying a much softer spectral shape. This has important consequences in the $\ce{H^-}$ detachment rate, that is determined by photons in a wide energy range, from the UV to the IR, the reaction energy threshold being at 0.75 eV.

\section{SEDs choice}
\label{appen:SEDs}

\begin{figure}
    \centering
    \includegraphics[width=\linewidth]{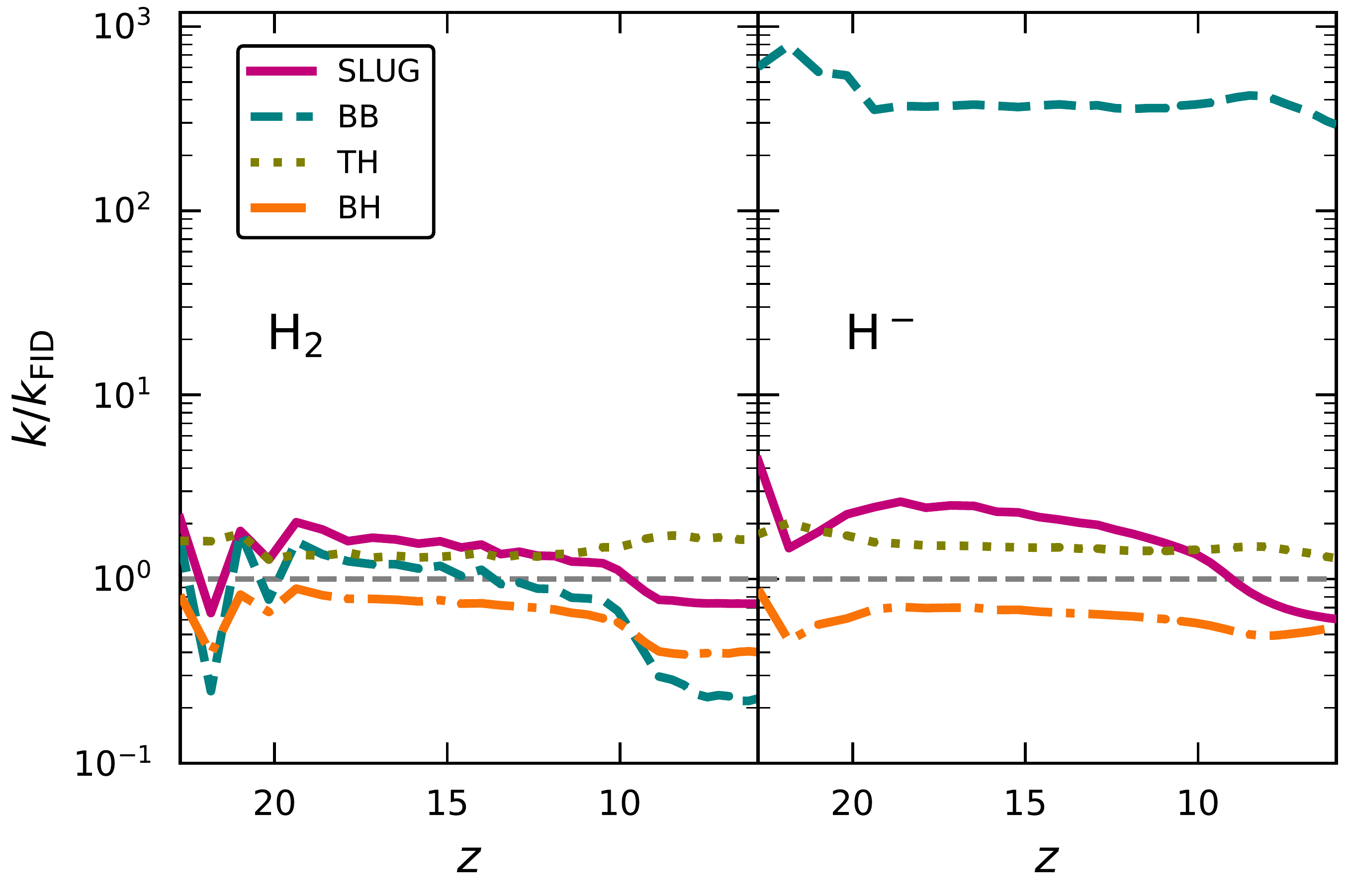}
    \vspace{-0.6cm}
    \caption{Ratio of the $\ce{H_2}$ dissociation (left panel) and $\ce{H^-}$ detachment rate (right panel) of the alternative SED combinations to the \textquotesingle FID\textquotesingle \ one.}
    \vspace{-0.2cm}
    \label{fig:rates_combs_Mbox}
\end{figure}

Throughout the paper we have been showing the results for our fiducial choice of stellar SEDs, although in Section~\ref{sec:SEDs} (Tables~\ref{Table:SEDpop3} and \ref{Table:SEDpop2}) we have listed all the additional SEDs included in the postprocessing algorithm. Here we show the potential, but limited, impact of a different choice of stellar emission models on some of our results. In particular, in Fig.~\ref{fig:rates_combs_Mbox} we report the ratio between the rates with the other combinations and the \textquotesingle FID\textquotesingle \ one, for the M simulation. The left panel shows this for the $\ce{H_2}$ dissociation rate: at early times the mean value is always within a factor of 2 from the fiducial case and is approximately constant with time (simply reflecting the small differences between PopIII SEDs), while, when PopII stars dominate at $z\lesssim10$, the ratio shows a different behaviour that is more prominent in the \textquotesingle BB\textquotesingle \ case, that by construction accounts only for very young stars ($<5$ Myr) emitting a constant black-body spectrum at $10^5$ and $10^4$ K for PopIII and PopII stars respectively. The latter case leads to underestimating the LWB by up to a factor of 5 at $z\sim8$.

The same ratios are shown for the $\ce{H^-}$ detachment rate in the right panel of Fig.~\ref{fig:rates_combs_Mbox}. \textquotesingle SLUG\textquotesingle, \textquotesingle TH\textquotesingle \ and \textquotesingle BH\textquotesingle \ cases all lie within a factor of $2-3$ above or below the fiducial case., while the \textquotesingle BB\textquotesingle \ spectra give a mean rate (dashed turquoise line) more than two orders of magnitude above the \textquotesingle FID\textquotesingle \ case. We have verified that this is due to the $10^4$ K black-body spectrum for PopII stars, normalised as suggested in \cite{Greif:2006}: such a soft spectrum, as already shown in \cite{Latif:2015} (their Figure 1) and highlighted in Appendix~\ref{appen:example_spectra}, when integrated over the wide wavelength range of the $\ce{H^-}$ cross section, gives a rate that is several orders of magnitude above the same rate for a harder spectrum with the same normalisation at the Lyman limit. Consequently, our results on the effective LW spectral shape (Sec.~\ref{sec:effective_spectral_shape}) do not change appreciably with a different SEDs choice, with the exception of \textquotesingle BB\textquotesingle \ that converges to $T_\mathrm{eff}=10^4 \ \mathrm{K}$ even before PopII stars dominate the LWB (Sec.~\ref{sec:age_contrib}).

In conclusion, the radiation background depends only mildly on the choice of the SEDs, as long as realistic stellar models are employed; on the other hand, approximations such as the \textquotesingle BB\textquotesingle \ case give different results that in turn can lead to inaccurate evaluations of the $\ce{H_2}$ abundance in the Early Universe.

\begin{figure}
    \centering
    \includegraphics[width=\linewidth]{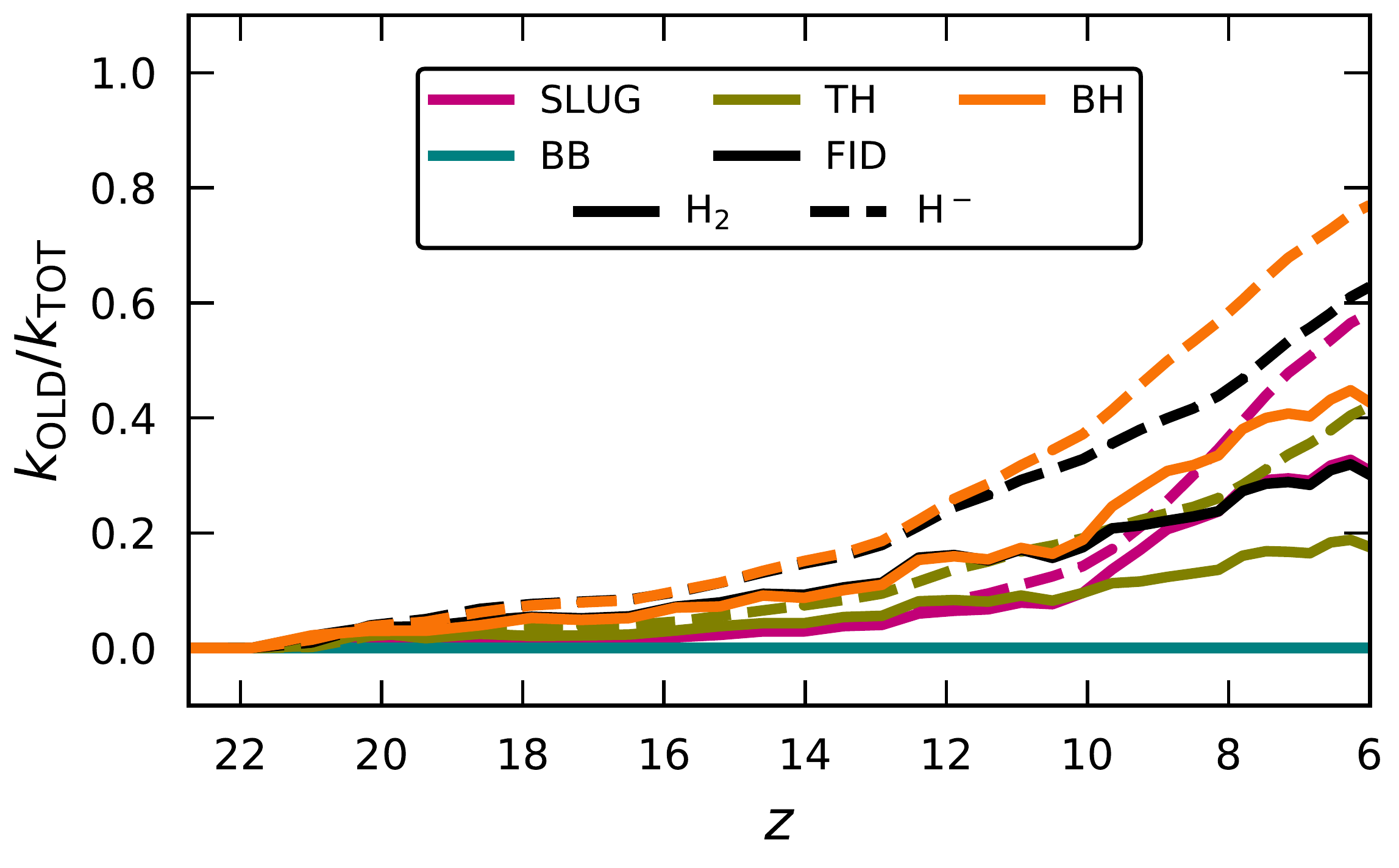}
    \vspace{-0.6cm}
    \caption{Similarly to Fig.~\ref{fig:agebins}, fraction of the $\ce{H_2}$ dissociation rate (solid lines) and $\ce{H^-}$ detachment rate (dashed lines) in the M FiBY simulation originated by stars older than 20 Myr. The SED choices are color-coded in the same way as in Fig.~\ref{fig:rates_combs_Mbox}.}
    \vspace{-0.2cm}
    \label{fig:frac_old_allcombs}
\end{figure}

The contribution from old stellar populations, normally neglected in the literature \citep{Ahn:2009,Johnson:2013,Wise:2012b}, also depends on the specific choice of stellar IMF and SED. In Fig.~\ref{fig:frac_old_allcombs} we show the fraction of the rates that is due to stellar populations older than 20 Myr, by combining the two \textquotesingle oldest\textquotesingle \ bins described in Section~\ref{sec:age_contrib}. The Figure again refers only to M, but the same results are valid for the other simulations, with only subtle variations depending on the specific star formation history. As already shown in Fig.~\ref{fig:agebins}, the contribution from old stars increases with time and is larger for the \ce{H^-} detachment rate (dashed lines). This is true for all the combinations that include old stars (namely, for everyone but the \textquotesingle BB\textquotesingle \ case) and the differences can be explained with the different number of low-mass stars in top-heavy (magenta - for PopIII - and gold lines) and bottom-heavy (orange and black lines) IMFs. Given the current uncertainties on the IMF in metal-free and metal-poor environments, old stellar populations can account for up to $\sim20\%-40\%$ of the $\ce{H_2}$ dissociation and $\sim40\%-80\%$ of the $\ce{H^-}$ detachment rate during the Epoch of Reionisation, while their contribution is limited to $\sim10\%-20\%$ at $z\lesssim14$.

\section{Normalised critical distance}
\label{appen:local_contribution}

\begin{figure}
    \centering
    \includegraphics[width=\linewidth]{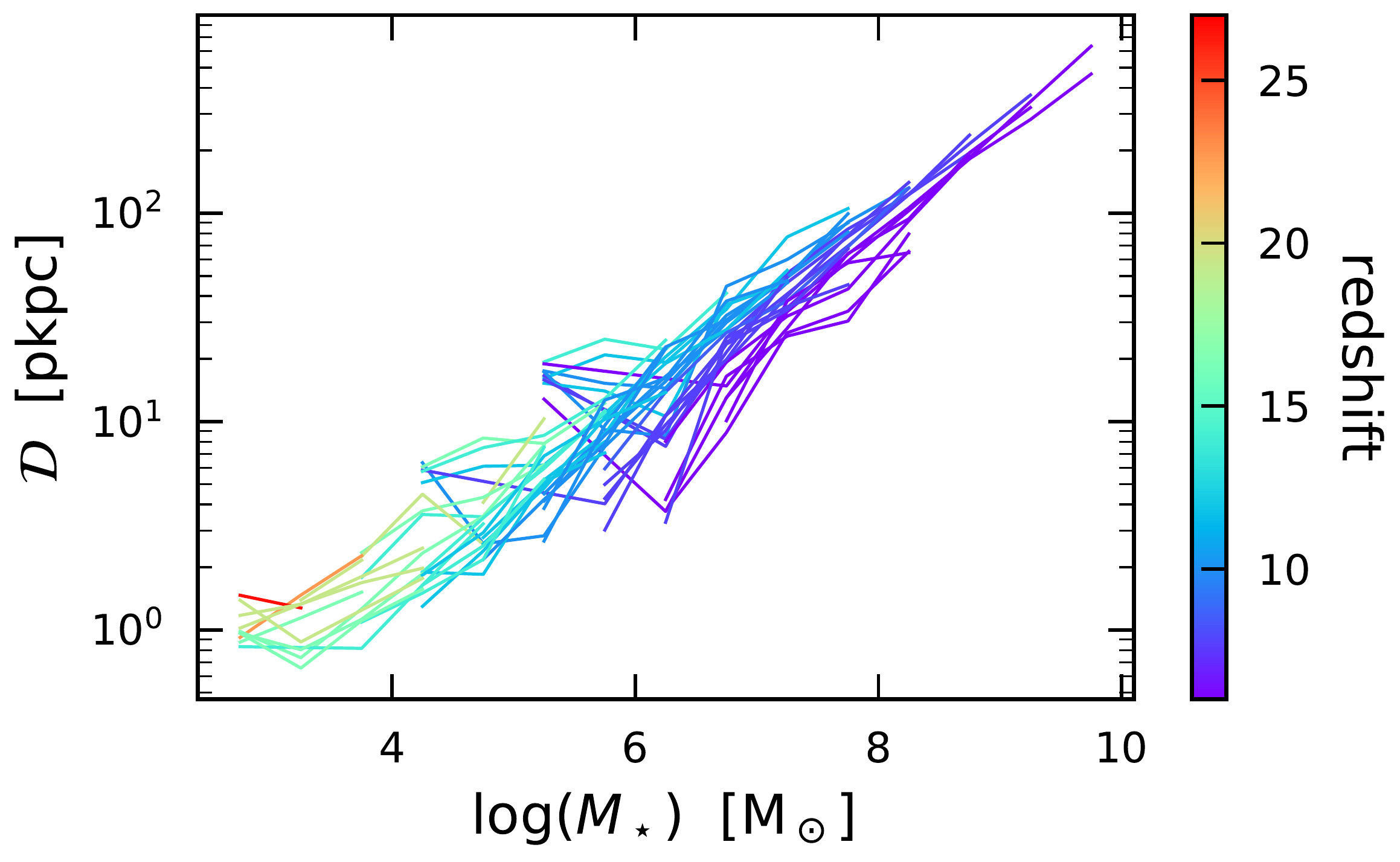}
    \vspace{-0.6cm}
    \caption{$\mathcal{D}(M_\star)$, i.e., the critical distance shown in Fig.~\ref{fig:local_contrib_SM} normalised by the mean LW radiation intensity, for all the FiBY simulations. Each line is colour-coded depending on the redshift. With the exception of the galaxies close to the resolution limit, $\mathcal{D}(M_\star)$ depicts a clear increasing trend with the galaxy stellar mass and hence can be fitted with Eq.~\ref{eq:fit_Dcrit}. The small evolution with redshift at a fixed $M_\star$ is explained by the lower UV emission per stellar mass from PopII stars with respect to PopIII-dominated galaxies.}
    \vspace{-0.2cm}
    \label{fig:Dcrit_normalised}
\end{figure}

As shown in Sec.~\ref{sec:local_contrib}, the radiation emitted by a single galaxy can exceed the LW radiation intensity in a volume whose size depends on the galaxy stellar mass and the mean LWB level. We show in Fig.~\ref{fig:Dcrit_normalised} the normalised critical distance $\mathcal{D}(M_\star)$ as defined in Eq.~\ref{eq:normalise_Dcrit}, that is then fitted as shown in Eq.~\ref{eq:fit_Dcrit}.

\bsp	
\label{lastpage}
\end{document}